\documentclass[aip,preprint,amsmath,amssymb]{revtex4-2} 
\usepackage{graphicx}
\usepackage{bm}
\usepackage{rotating}
\usepackage{amsmath}
\usepackage{amssymb}
\usepackage{txfonts}
\usepackage{url}
\usepackage{braket}
\usepackage{multirow}
\usepackage{here}
\usepackage[usenames,dvipsnames]{color}
\definecolor{Gray}{gray}{0.0}
\definecolor{lightGray}{gray}{0.35}
\usepackage[linesnumbered,ruled,vlined]{algorithm2e}
\usepackage[normalem]{ulem}
\usepackage{hyperref}

\makeatletter
\def\Hline{%
\noalign{\ifnum0=`}\fi\hrule \@height 1pt \futurelet
\reserved@a\@xhline}
\makeatother

\begin{document}
\title{jQMC: A JAX-based \emph{ab initio} quantum Monte Carlo package designed for GPU-accelerated computing}
%
%
\author{Kousuke Nakano}
\email{kousuke\_1123@icloud.com}
\affiliation{Center for Basic Research on Materials, National Institute for Materials Science (NIMS), 1-2-1 Sengen, Tsukuba, Ibaraki 305-0047, Japan}
\author{Michele Casula}
\affiliation{Institut de Min{\'e}ralogie, de Physique des Mat{\'e}riaux et de Cosmochimie (IMPMC), Sorbonne Universit{\'e}, CNRS UMR 7590, IRD UMR 206, MNHN, 4 Place Jussieu, 75252 Paris, France}


\date{\today}
\begin{abstract}
We present jQMC, a Python-based computational package for {\it ab initio} Quantum Monte Carlo (QMC) simulations, designed for modern GPU-accelerated computing environments.  
jQMC implements two well-established QMC algorithms: Variational Monte Carlo (VMC) and the lattice-regularized variant of Diffusion Monte Carlo (LRDMC).
The employed wave function is a Jastrow factor combined with the antisymmetrized geminal power with spin-singlet and spin-triplet pairings, which contains the single Slater determinant as its special lowest-rank case.
The wave function can be initialized from external Hartree-Fock/Density Functional Theory calculations through the TREX-IO library (a common wave-function format across electronic-structure packages) and optimized by stochastic reconfiguration and linear-method energy minimization.
One of the prominent features of jQMC is its use of JAX, which enables automatic differentiation for wave function optimization and atomic force calculations, and allows the main QMC algorithms to be Just-In-Time (JIT) compiled and portable across CPU and GPU. jQMC is vectorized over walkers at the top level of the QMC algorithms, providing efficient intra-GPU~(CPU) vectorization. The multi-GPU~(CPU) parallelization is also supported through MPI and JAX sharding.
To assess the practical performance of this implementation, we benchmarked jQMC performance on NVIDIA GPUs (A100 and H100) and analyzed CUDA kernels. 
For the test cases analyzed here, with system sizes up to 160 electrons, the current version of jQMC is faster than TurboRVB, a Fortran90 code implementing the same algorithms and wave functions, once jQMC is run on GPUs. In terms of wall-time, the gain can reach an order of magnitude for VMC, while it is more moderate for LRDMC.
Furthermore, to demonstrate the applications of jQMC, we focus on an ongoing debate in the community, namely the accuracy of QMC quantities compared with highly accurate Quantum Chemistry calculations such as coupled-cluster theory with single, double, and perturbative triple excitations [CCSD(T)]. We studied the accuracy of atomic forces computed by VMC and LRDMC for ethanol and malonaldehyde. We found that LRDMC forces achieve $\sim$ 1.5 kcal/mol/\AA\ accuracy in the mean absolute error (MAE) with respect to all-electron CCSD(T) forces for ethanol, while LRDMC and CCSD(T) forces are quite different ($\sim$ 5 kcal/mol/\AA\ in MAE) for malonaldehyde.
\end{abstract}
\maketitle

%
%

\section{Introduction}
\label{sec:introduction}
Solving the many-body Schr\"odinger equation remains a central challenge in computational chemistry and physics. For nearly a century, researchers have been developing many theoretical and numerical techniques to address this problem. Among these, Density Functional Theory (DFT) is one of the most widely adopted approaches. DFT maps the many-electron problem into a system of non-interacting electrons in an effective potential, called eXchange-Correlation (XC) functional~\cite{2004MAR}. Although DFT is formally exact, its practical application depends on the XC functional, and its exact form remains unknown. Progress in systematically improving this functional has been elusive~\cite{2017MED}. As a result, widely used approximations often fail in situations involving strong correlation effects or weak van der Waals interactions.

\vspace{1mm}
An alternative strategy for addressing the complexity of the many-electron Schr\"odinger equation is offered by \emph{ab initio} Quantum Monte Carlo (QMC) methods~\cite{2001FOU,2017BEC}. Unlike conventional deterministic approaches, QMC leverages stochastic sampling (i.e., Monte Carlo) to evaluate expectation values, thereby bypassing some of the limitations of deterministic methods. The theoretical foundation of QMC dates back to the Markov Chain Monte Carlo (MCMC) in the 1940s~\cite{1949ULA} and fixed-node Diffusion Monte Carlo (FN-DMC) in the 1980s~\cite{1980CEP, 1986CEP}, which have since become indispensable across diverse scientific communities.
QMC shows a favorable scaling with system size, albeit with a larger prefactor than in mean-field approaches. This prefactor, however, is mitigated by the nature of MCMC, which makes QMC suited for modern high-performance computing (HPC) architectures. The algorithmic structure allows for near-perfect weak scaling across thousands of cores, positioning QMC as a promising method for next-generation simulations of both molecular and extended systems.

\vspace{1mm}
To dates, several groups have implemented \emph{ab initio} QMC codes such as Amolqc, CASINO{~{\cite{2009NEE, 2020NEE}}}, CHAMP-EU\cite{CHAMP-EU}, CHAMP-US\cite{CHAMP-US}, CMQMC, PyQMC{~\cite{2023WHE}}, QMC=Chem{\cite{Scemama2013QMC}}, QMCPACK{~{\cite{2018KIM, 2020KEN}}}, QMeCha\cite{QMeCha}, QWalk\cite{qwalk}, and TurboRVB~{\cite{2020NAK, 2023NAK}} (named alphabetically). It is also worth highlighting recent efforts for standardizing data and functionalities required among QMC packages. For instance, TREX-IO~{\cite{2023POS}} is a library designed to standardize the wave function format among different QMC packages. QMCkl~{\cite{scemama2024qmckl}} provides common kernels for core QMC algorithms. They represent valuable efforts in the QMC community to consolidate their development efforts.

\vspace{2mm}
The recent shift of HPC facilities toward GPU-dominated infrastructures presents significant challenges for scientific computing and programming~{\cite{shinde2025shifting}}. Legacy codebases cannot fully exploit the computational power of GPU hardware without a comprehensive refactoring of the code because the suitable designs are often very different between them. Nearly all of the top-tier systems, such as those ranked in the TOP500 list~{\cite{strohmaier2015top500, TOP500}}, are equipped with CPU-GPU hybrid architectures, with very few exceptions such as Fugaku~{\cite{sato2022fugaku}}. 
There are several approaches to make codebases portable to GPU machines, such as using native programming languages (e.g., CUDA), directive-based programming models (e.g., OpenACC~{\cite{wienke2012openacc}}), or OpenMP offload, and exploiting GPU-optimized libraries such as NVIDIA's cuBLAS and cuSOLVER. A related and increasingly promising approach involves harnessing libraries, such as Kokkos~{\cite{2014_Kokkos, 2022_Kokkos3}}, TensorFlow~{\cite{tensorflow2015-whitepaper}}, PyTorch~{\cite{2019PyTorch}}, and JAX~{\cite{jax2018github}}, which are designed inherently with GPU execution. These frameworks offer a powerful abstraction for high-performance computing and are now being explored for \emph{ab initio} QMC applications. Notable examples of \emph{ab initio} QMC codes built on these modern frameworks include FermiNet~{\cite{pfau2020ferminet}}, PauliNet~{\cite{hermann2020deep}}, QMCTorch~{\cite{renaud2023qmctorch}}, and JaQMC~{\cite{JaQMC_ren_towards_2023}} which have demonstrated the potential of combining QMC with machine learning and neural network architectures.

\vspace{2mm}
Our developing \emph{ab initio} QMC code, named jQMC, relies on one of the libraries, JAX~{\cite{jax2018github}}. JAX has several features helping with making code fast and portable across different architectures. jQMC implements two ground-state QMC algorithms: variational Monte Carlo (VMC) and DMC{~\cite{2001FOU}} in its lattice regularized version. In VMC, the ground state is obtained by minimizing the total energy (or variance) with a parametrized many-body wave function ansatz. DMC is an imaginary-time projection technique{~\cite{2017BEC}} that filters out the ground-state wave function from a given trial wave function, typically combined with the so-called \emph{fixed-node} (FN) approximation. The nodal surface is constructed by a mean-field calculation, such as Hartree-Fock (HF) or DFT, or by an accurate variational optimization, and is usually kept frozen during the projection.

\vspace{2mm}
In particular, the jQMC package introduces several distinguishing features that collectively enhance its capability and flexibility for \emph{ab initio} quantum Monte Carlo simulations:
$\rm(\hspace{.18em}i\hspace{.18em})$ wave function ansatz:
jQMC supports the Jastrow Antisymmetrized Geminal Power (JAGP)~{\cite{2003CAS-AGP}}, a resonating valence bond (RVB)-type wave function. JAGP contains the single Slater determinant as a special case.
$\rm(\hspace{.18em}ii\hspace{.18em})$ wave function optimization:
The code implements the stochastic reconfiguration (SR) method{~\cite{1998SOR}} and the linear method (LM), robust and efficient optimization algorithms that enable simultaneous optimization of both wave function amplitudes and nodal surfaces.
$\rm(\hspace{.18em}iii\hspace{.18em})$ Projection method:
To perform DMC calculations with enhanced numerical control, jQMC includes an implementation of the lattice regularized DMC (LRDMC) method~{\cite{2005CAS}}, which provides improved stability over traditional DMC algorithms.
$\rm(\hspace{.18em}iv\hspace{.18em})$ Force evaluation:
By leveraging algorithmic differentiation with the JAX framework, jQMC enables efficient implementation of energy derivatives, such as atomic forces~{\cite{2010SOR}}, directly from the many-body wave function.
$\rm(\hspace{.18em}v\hspace{.18em})$ Software architecture:
Developed in Python, the codebase emphasizes ease of use for end-users and modular extensibility for developers. This makes it well-suited for both production simulations and the rapid prototyping of new QMC algorithms.
$\rm(\hspace{.18em}vi\hspace{.18em})$ Hardware acceleration:
Through just-in-time (JIT) compilation and vectorized execution (e.g., via jit and vmap) provided by JAX, jQMC achieves high computational performance across diverse hardware backends, including CPUs and GPUs.
$\rm(\hspace{.18em}vii\hspace{.18em})$ Parallel scalability:
Built-in MPI support (mpi4py~{\cite{2023mpi4py}}) and the native JAX sharding mechanism allow jQMC to scale efficiently on modern HPC platforms, enabling large-scale many-body simulations.
These components together position jQMC as a modern and versatile platform for both conducting accurate quantum Monte Carlo simulations and developing novel methodologies within the QMC framework.

\vspace{2mm}
This paper is organized as follows:
Sec.{{~\ref{sec:methods}}} briefly explains the QMC algorithms implemented in jQMC, MCMC, VMC, and LRDMC.
Sec.{~\ref{sec:wavefunction}} describes the RVB-family of wave functions combined with the Jastrow factor; 
Sec.{~\ref{sec:Coulomb-and-ECP}} focuses on the implementation of the bare Coulomb interactions and effective core potentials.
Sec.{~\ref{sec:derivatives}} details how energy derivatives are computed ({\it e.g.}, atomic forces) in jQMC;
Sec.{~\ref{sec:opt_wf}} reviews the wave function optimization techniques currently implemented in jQMC;
Sec.{~\ref{sec:implementations}} summarizes the implementation of jQMC and its typical QMC workflow;
Sec.{~\ref{sec:mixed-precision}} describes the mixed-precision support implemented in jQMC.
Sec.{~\ref{sec:vectorization}} explains the vectorization strategy employed in jQMC and its benchmark results measured on Miyabi (University of Tokyo) supercomputer;
Sec.~{\ref{sec:performance-analysis}} reports an analysis of the CUDA kernels in jQMC using NVIDIA Nsight Systems and NVIDIA Nsight Compute, and study the bottlenecks of the current implementation.
Sec.~{\ref{sec:wall-time-benchmark}} reports wall-time benchmarks of jQMC on CPU and GPU machines, and compares them with the TurboRVB QMC code, implementing the same algorithms and wave functions;
Sec.{~\ref{sec:weak-scaling-benchmark}} describes the parallelization strategy employed in jQMC and reports weak scaling results of jQMC, measured on Leonardo (CINECA) and Miyabi (University of Tokyo) supercomputers;
Sec.{~\ref{sec:verification}} reports a validation of jQMC by comparing the total energies of methane, water, and methane-water dimer, and its binding energy with other QMC codes;
Sec.~{\ref{sec:demonstrations}} demonstrates MCMC and DMC atomic-force calculations for the ethanol and malonaldehyde molecules, and compares them with all-electron coupled-cluster theory with single, double, and perturbative triple excitations [CCSD(T)].

%
%

\section{Methods}
\label{sec:methods}
jQMC implements two established Quantum Monte Carlo methods: Variational Monte Carlo (VMC) and Lattice Regularized Diffusion Monte Carlo (LRDMC). We summarize these methods in this Section. For further details, the reader is referred to more comprehensive reviews and textbooks on QMC~{\cite{2001FOU, 2017BEC}}. We notice that throughout this paper the formalism is presented assuming a real many-body wave function $\Psi(\mathbf{x})\in\mathbb{R}$, which is also the only case currently supported by the jQMC implementation (v.0.2.2). Extending the formalism to a complex wave function $\Psi(\mathbf{x})\in\mathbb{C}$ is left for future developments.

\subsection{Variational Monte Carlo}
\label{subsec:variational_monte_carlo}
Given the \emph{ab initio} Hamiltonian $\hat{\mathcal{H}}$, the expectation value of the energy can be computed by MCMC as:
\begin{equation}
{E_{{\text{MCMC}}}} = \frac{\left\langle \Psi | \hat{\mathcal{H}} | \Psi \right\rangle}{\left\langle \Psi | \Psi \right\rangle}  = \frac{{\int {d{\mathbf{x}}{\Psi ^2}\left( {\mathbf{x}} \right) \cdot \hat {\mathcal{H}} \Psi \left( {\mathbf{x}} \right)/\Psi \left( {\mathbf{x}} \right)} }}{{\int {d{\mathbf{x}}{\Psi ^2}\left( {\mathbf{x}} \right)} }} = \int {d{\mathbf{x}}{e_L}\left( {\mathbf{x}} \right)\pi \left( {\mathbf{x}} \right)},
\label{eq:expectation_energy}
\end{equation}
for a given wave function (WF) $\Psi$, where ${\mathbf{x}} = \left( {{{\mathbf{r}}_1}{\sigma _1},{{\mathbf{r}}_2}{\sigma _2}, \ldots {{\mathbf{r}}_N}{\sigma _N}} \right)$ represents the $N$ electron coordinates and their spins, and 
$$
{e_L}\left( {\mathbf{x}} \right) \equiv {\hat {\mathcal{H}}\Psi \left( {\mathbf{x}} \right) \over \Psi \left( {\mathbf{x}} \right)} 
\quad  \text{and} \quad 
\pi \left( {\mathbf{x}} \right) \equiv { {\Psi ^2}\left( {\mathbf{x}} \right) \over  \int {d{\mathbf{x'}}{\Psi ^2}\left( {\mathbf{x'}} \right)} },
$$ 
are the local energy and the probability distribution, respectively.
MCMC allows one to evaluate the multidimensional integration in Eq.~\ref{eq:expectation_energy} stochastically by generating a set of configurations $\left\{ {{{\mathbf{x}}_i}} \right\}$ according to the distribution $\pi \left( {\mathbf{x}} \right)$ and by averaging the accumulated local energies ${e_L}\left( {{{\mathbf{x}}_i}} \right)$:
\begin{equation}
{E_{{\text{MCMC}}}} = {\left\langle {{e_L}\left( {\mathbf{x}} \right)} \right\rangle _{\pi \left( {\mathbf{x}} \right)}} \approx \frac{1}{M}\sum\limits_{i = 1}^M {{e_L}\left( {{{\mathbf{x}}_i}} \right)},
\label{eq-eval-observable}
\end{equation}
The total energy and the mean of other local observables $O_L$ are always associated with a statistical error of $\sqrt{{\text{Var}[O_L(\mathbf{x}_i)] / \tilde M}}$, 
where $\text{Var}[O_L(\mathbf{x}_i)]$ is the variance of the sampled observables, and $\tilde M$ is the sampling size $M$ divided by the autocorrelation time. 
Then, assuming $\Psi$ is parametrized with a set of variational parameters $\alpha \equiv \left( {{\alpha _1},{\alpha _2}, \cdots,{\alpha _p}} \right)$, the corresponding variational energy reads:
\begin{equation}
{E_{{\text{MCMC}}}}\left( \alpha  \right) = \int {d{\mathbf{x}}{e_L}\left( {{\mathbf{x}},\alpha } \right)\pi \left( {{\mathbf{x}},\alpha } \right)}  \geqslant E_{0},
\label{eq:vmc_variational}
\end{equation}
where, according to the \emph{variational theorem}, ${E_{\text{MCMC}}}$ provides an upper bound to the exact ground-state energy ($E_0$).
This framework is called VMC.

\vspace{2mm}
One of the key implementations of VMC is the MCMC evaluation of  the Hamiltonian expectation value and the update of the parameters of the trial WF anstaz $\Psi_T$, to get the best variational representation of the ground state according to the variational principle. The possible WF parametrizations available in jQMC are discussed later. The MCMC implemented in jQMC relies on the generalized Metropolis algorithm (Metropolis--Hastings algorithm~{\cite{metropolis1953state, hastings1970monte}}). The acceptance of a proposed move ($\mathcal{A}(\mathbf{x} \rightarrow \mathbf{x}'$)) reads:
\begin{equation}
\mathcal{A}(\mathbf{x} \rightarrow \mathbf{x}') = \min \left[1,  \frac{\Psi^2(\mathbf{x}')}{\Psi^2(\mathbf{x})} \cdot \frac{T(\mathbf{x}' \rightarrow \mathbf{x})}{T(\mathbf{x} \rightarrow \mathbf{x}')} \right],
\label{eq:acceptance_ratio}
\end{equation}
where $\mathbf{x} = \{\cdots, \mathbf{r}_l, \cdots \}$ and $\mathbf{x}' = \{\cdots, \mathbf{r}'_l, \cdots \}$ differ by a single-electron position.
The WF ratios during the MCMC electron position update are efficiently computed via the fast-update algorithm, described in Subsec.~{\ref{subsec:fastupdate}}.
jQMC, first, detects the nearest neighbor nucleus $I$ of the target electron $l$ to compute the value $\sigma_{\rm prop}(\mathbf{r}_l)$ defined as:
\begin{equation}
\sigma_{\rm prop}(\mathbf{r}_l) = \frac{1}{Z_{I(l)}^2|\mathbf{r}_l - \mathbf{R}_{I(l)}|} \cdot \frac{1 + Z_{I(l)}^2|\mathbf{r}_l - \mathbf{R}_{I(l)}|}{1 + |\mathbf{r}_l - \mathbf{R}_{I(l)}|},
\label{eq:mcmc_step_size}
\end{equation}
then a single electron move is proposed:
\begin{equation}
\mathbf{x} \rightarrow \mathbf{x}': r'_{l, \gamma} = r_{l,\gamma} + \eta_l
\label{eq:mcmc_proposal_move}
\end{equation}
where $\gamma$ is a Cartesian label ($x,y,z$), which is chosen randomly, $\eta_l$ is a Gaussian random number with a mean of 0 and variance $\sigma_{\rm prop}(\mathbf{r}_l) \Delta t$, where $\Delta t$ is an input hyperparameter. The transition probability is defined as:
\begin{gather}
T(\mathbf{x} \rightarrow \mathbf{x}') = \cfrac{1}{\sqrt{2\pi (\sigma_{\rm prop}(\mathbf{r}_l) \Delta t)^2}} \exp \left[{- \frac{|\mathbf{r}'_l - \mathbf{r}_l|^2}{2 (\sigma_{\rm prop}(\mathbf{r}_l) \Delta t)^2} }\right] 
\label{eq:transition_probability}
\end{gather}
to satisfy the detailed-balance condition.
Thus, its ratio becomes:
\begin{equation}
\cfrac{T(\mathbf{x}' \rightarrow \mathbf{x})}{T(\mathbf{x} \rightarrow \mathbf{x}')} = \left(\cfrac{\sigma_{\rm prop}(\mathbf{r}_l)}{\sigma_{\rm prop}(\mathbf{r}'_l)}\right) \exp \left[{-|\mathbf{r}_l - \mathbf{r}'_l|^2 \left(\frac{1}{2 (\sigma_{\rm prop}(\mathbf{r}'_l) \Delta t)^2} - \frac{1}{2 (\sigma_{\rm prop}(\mathbf{r}_l) \Delta t)^2} \right)}\right].
\label{eq:transition_ratio}
\end{equation}
Following this transition probability, we repeatedly perform accept/reject steps to generate Monte Carlo samples and evaluate Eq.~{\ref{eq-eval-observable}}. Eqs.~\ref{eq:mcmc_step_size} and \ref{eq:transition_probability} automatically modulate the amplitude of the MCMC step, by reaching a higher sampling efficiency according to the distribution of the electronic clouds centered around each nucleus.

\vspace{1mm}
In the MCMC, probability distributions different from $\pi \left( {\mathbf{x}} \right)$ can also be employed. This is called the reweighting technique, or importance sampling. More specifically, one can use an arbitrary probability distribution function $\pi '\left( {\mathbf{x}} \right) = \Psi _{\text{G}}^2\left( {\mathbf{x}} \right)/\int {d{\mathbf{x}}\Psi _{\text{G}}^2\left( {\mathbf{x}} \right)}$, and estimate the expectation value of a local observable $O\left({\mathbf{x}} \right)$ by:
\begin{equation}
{{\bar O}_{{\text{MCMC}}}}
= \frac{{{{\left\langle {O\left( {{\mathbf{x'}}} \right)\mathcal{W}\left( {{\mathbf{x'}}} \right)} \right\rangle }_{\pi '\left(
{{\mathbf{x'}}} \right)}}}}{{{{\left\langle {\mathcal{W}\left( {{\mathbf{x'}}} \right)} \right\rangle }_{\pi '\left(
{\mathbf{x}'} \right)}}}}
\approx \frac{{\sum\nolimits_{i = 1}^{M'} {O\left( {{{{\mathbf{x'}}}_i}} \right)\mathcal{W}\left( {{{{\mathbf{x'}}}_i}}
\right)} }}{{\sum\nolimits_{i = 1}^{M'} {\mathcal{W}\left( {{{{\mathbf{x'}}}_i}} \right)} }},
\label{eq:reweighted_observable}
\end{equation}
where $\mathcal{W}\left( {{{{\mathbf{x}}}}} \right) = \left({\Psi }\left( {\mathbf{x}} \right)/\Psi _{\text{G}}\left( {\mathbf{x}}
\right)\right)^2$,
and the points $\mathbf{x}'_i$ are distributed according to $\pi'$. The reweighting technique is essential to compute atomic forces with finite variance, as discussed in Sec.~\ref{ionic_forces}.
jQMC implements the so-called Attaccalite and Sorella (AS) reweighting scheme~{\cite{2008ATT}} to avoid the divergence of derivatives in the vicinity of the nodal surface. 
In the AS scheme, the MCMC sampling is driven by a modified guiding function ${{\Psi _{\text{G}}}\left( {\mathbf{x}} \right)}$ defined by
\begin{equation}
{\Psi _{\text{G}}}\left( {\mathbf{x}} \right) = \frac{{{R_{\rm AS}^{\varepsilon_{\rm AS}} }\left( {\mathbf{x}} \right)}}{{R_{\rm AS}\left( {\mathbf{x}}
\right)}}{\Psi}\left( {\mathbf{x}} \right).
\label{guiding_function}
\end{equation}
${R_{\rm AS}(\mathbf{x})}$ is defined as
\begin{equation}
R_{\rm AS}\left( {\mathbf{x}} \right) = \left( s_{\rm AS} \sum\limits_{i,j} {{{\left| {G}^{ - 1}\right|}_{i,j}^2}}   \right)^{ - \theta_R}
\label{R_r}
\end{equation}
where $G$ is the so-called pairing matrix used for the anti-symmetric part of the many-body wave function, described later (see Sec.~\ref{subsec:AGP}),
and the scaling factor $s_{\rm AS}$ is defined as:
\begin{equation}
s_{\rm AS} = \min\left(\min_i \sum\limits_j |G_{ij}|^2,\, \min_j \sum\limits_i |G_{ij}|^2\right),
\label{AS_S}
\end{equation}
which ensures that both spin-up and spin-down electrons are treated on equal footing when approaching the nodal surface, and
$\theta_R=3/8$ is an empirically chosen value.
Finally, with the hyperparameter $\varepsilon_{\rm AS}$, ${R_{\rm AS}^{\varepsilon_{\rm AS}} }\left( {\mathbf{x}} \right)$ is defined as:
\begin{equation}
  {R_{\rm AS}^{\varepsilon_{\rm AS}} }\left( {\mathbf{x}} \right) = {\rm max}\left[R_{\rm AS}\left( {\mathbf{x}} \right), \varepsilon_{\rm AS}\right]
\label{regularization}
\end{equation}
By using the new probability, an observable with singular points in the vicinity of the nodal surface can be evaluated {\em with finite variance} as:
\begin{equation}
 {\left\langle {\mathcal{W} \left( {\mathbf{x}} \right){\mathbf O} \left( {\mathbf{x}} \right)} \right\rangle _{{\Pi _{\text{G}}}\left( {\mathbf{x}} \right)}}/{\left\langle {\mathcal{W}\left( {\mathbf{x}} \right)} \right\rangle _{{\Pi _{\text{G}}}\left( {\mathbf{x}} \right)}},
\label{reweighting}
\end{equation}
where the $ \mathcal{W} \left( {\mathbf{x}} \right)$ is the modified weight:
\begin{equation}
\mathcal{W}\left( {\mathbf{x}} \right) = {\left( {\Psi\left( {\mathbf{x}} \right)/{\Psi _{\text{G}}}\left( {\mathbf{x}} \right)} \right)^2} \equiv \left( {R_{\rm AS}\left( {\mathbf{x}} \right) \over {\rm max} \left[  R_{\rm AS}\left( {\mathbf{x}} \right) , \varepsilon_{\rm AS}\right] } \right)^2.
\label{eq:reweighting_factor}
\end{equation}
The value of $\varepsilon_{\rm AS}$ should be chosen manually in jQMC such that the average reweighting factor $\left\langle {\mathcal{W}\left( {\mathbf{x}} \right)} \right\rangle \simeq 0.8$, which is an empirically optimal setting.

\subsection{Lattice regularized Monte Carlo}
\label{subsec:lattice_regularized_monte_carlo}

\subsubsection{Formalism of Lattice regularized diffusion Monte Carlo (LRDMC)}
\label{subsubsec:lrdmc_formalism}
Lattice regularized diffusion Monte Carlo (LRDMC)~{\cite{2005CAS}} is a method built to filter out the ground state WF ${\left| {{\Upsilon _0}} \right\rangle }$ from a given guidance WF $\left| {{\Psi _{\text{T}}}} \right\rangle$. This method is based on Green's function Monte Carlo (GFMC){~\cite{1995TEN, 1998BUO, 2000SOR}}, and thus, it is sometimes called ``power method''. In the GFMC formalism, the ground-state WF can be obtained by iteratively applying ${\left( {\lambda  - \hat {\mathcal{H}}} \right)^M}$ to a given trial wave function:
\begin{equation}
\begin{split}
\left| {{\Upsilon _0}} \right\rangle &\propto \mathop {\lim }\limits_{M \to \infty } {\left( {\lambda  - \hat {\mathcal{H}}} \right)^M}\left| {{\Psi _{\text{T}}}} \right\rangle  \\
& = \mathop {\lim }\limits_{M \to \infty } {\left( {\lambda  - {E_0}} \right)^M}\left[ {{a_0}\left| {{\Upsilon _0}} \right\rangle  + \sum\limits_{n \ne 0} {{{\left( {\frac{{\lambda  - {E_n}}}{{\lambda  - {E_0}}}} \right)}^M}{a_n}\left| {{\Upsilon _n}} \right\rangle } } \right],
\end{split}
\label{eq:power-method}
\end{equation}
where $\lambda$ is a large positive number, ${{E_n}}$ is $n$-th eigenvalue of $\hat {\mathcal{H}}$, and ${a_n}$ is the coefficient for the $n$-th eigenvectors (${{\Upsilon _n}}$) (i.e. $\left| {{\Psi _{\text{T}}}} \right\rangle  = \sum\limits_n {{a_n}\left| {{\Upsilon _n}} \right\rangle}$). 
The above projection filters out the ground state WF ${{\Upsilon _0}}$ from a given guidance WF $\left| {{\Psi _{\text{T}}}} \right\rangle$ unless the trial WF is orthogonal to the ground state ({\it i.e.}, ${a_0} \equiv \braket{{{\Psi _{\text{T}}}}|{{\Upsilon _0}}} \ne 0$) and $|{\lambda  - {E_n}}|/|{\lambda  - {E_0}}| \ge 1$.

\vspace{1mm}
For fermions (electrons), LRDMC is always associated with the fixed-node (FN) approximation{~\cite{2017BEC}} because the Green's function it not always positive for fermions. This is referred to as the sign problem. 
Therefore, in LRDMC, the Hamiltonian is modified by setting to zero the ``wrong'' sign terms and summing them up into the spin-flip term ${\mathcal{V}_{{\rm{sf}}}}\left( \mathbf{x} \right) = \sum\limits_{\mathbf{x}':{s_{\mathbf{x}',\mathbf{x}}} > 0}^{} {{{\mathcal{H}}_{\mathbf{x}',\mathbf{x}}}{\Psi _\text{T}}\left( \mathbf{x}' \right)} /{\Psi _\text{T}}\left( \mathbf{x} \right)$. This defines an effective FN Hamiltonian $\hat{\mathcal{H}}^{\text{FN}}$, which reads as:
\begin{equation}
{\mathcal{H}}_{\mathbf{x}',\mathbf{x}}^{\text{FN}} = 
 \begin{cases}
  {{\mathcal{H}}_{\mathbf{x},\mathbf{x}}} + {\mathcal{V}_{{\rm{SF}}}}\left( \mathbf{x} \right)\,\,\,\,{\rm{for}}\,\,\,\,\mathbf{x}' = \mathbf{x},\\
  {{\mathcal{H}}_{\mathbf{x}',\mathbf{x}}}\,\,\,\,\,\,\,\,\,\,\,\,\,\,\,\,\,\,\,\,\,\,\,\,\,\,\,\,{\rm{for}}\,\,\,\,\mathbf{x}' \ne \mathbf{x},{s_{\mathbf{x}',\mathbf{x}}} < 0,\\
  0\,\,\,\,\,\,\,\,\,\,\,\,\,\,\,\,\,\,\,\,\,\,\,\,\,\,\,\,\,\,\,\,\,\,\,\,\,\,{\rm{for}}\,\,\,\,\mathbf{x}' \ne \mathbf{x},{s_{\mathbf{x}',\mathbf{x}}} > 0,
\end{cases}
\label{eq:fn_hamiltonian}
\end{equation}
where ${s_{\mathbf{x}',\mathbf{x}}} = {\Psi _\text{T}}\left( \mathbf{x} \right){{\mathcal{H}}_{\mathbf{x}',\mathbf{x}}}{\Psi _\text{T}}\left( \mathbf{x}' \right)$ (i.e, ${s_{\mathbf{x}',\mathbf{x}}} >0$ indicates that the $\mathbf{x} \rightarrow \mathbf{x}'$ crosses the nodal surface of the guiding function, or simply contributes to the Green's function $\lambda - \hat{\mathcal{H}}$ with the ``wrong'' sign). The fixed-node Green's function with the guiding function is then:
\begin{equation}
{\mathcal{G}}_{\mathbf{x}',\mathbf{x}}^{{\text{FN}}} = \left( {\lambda \delta_{\mathbf{x}', \mathbf{x}}  - {\mathcal{H}}_{\mathbf{x}',\mathbf{x}}^{{\text{FN}}}} \right)\frac{{\Psi _{\text{T}}^{}\left( \mathbf{x}' \right)}}{{\Psi _{\text{T}}^{}\left( \mathbf{x} \right)}} = {\lambda} \delta_{\mathbf{x}', \mathbf{x}}  - \frac{{\Psi _{\text{T}}^{}\left( \mathbf{x}' \right)}}{{\Psi _{\text{T}}^{}\left( \mathbf{x} \right)}} {\mathcal{H}}_{\mathbf{x}',\mathbf{x}}^{{\text{FN}}}
\label{eq:def-greenfunction}
\end{equation}
prevents walkers from crossing the nodal surface.
%

\vspace{2mm}
In LRDMC, the original continuous Hamiltonian is discretized with electron hopping with step size $a$. The corresponding Hamiltonian ${{\hat{\mathcal{H}}}^a}$ satisfies ${{{\hat {\mathcal{H}}}^a}} \to {\hat {\mathcal{H}}}$ for $a \to 0$. In other words, the kinetic and potential parts are approximated by a discretized form, ${{{\hat {\mathcal{H}}}^a}} \equiv \hat{K}^a + \hat{V}^a$. The discretized kinetic term is computed via the finite-difference Laplacian $\nabla_a^2$ acting on a function $f\left( {{x_i},{y_i},{z_i}} \right)$ as
\begin{equation}
\nabla^2_{a,i}f\left( {{x_i},{y_i},{z_i}} \right) = \frac{1}{{{a^2}}}\left\{ {\left[ {f\left( {{x_i} + a} \right) - f\left( {{x_i}} \right)} \right] + \left[ {f\left( {{x_i} - a} \right) - f\left( {{x_i}} \right)} \right]} \right\} \leftrightarrow {y_i} \leftrightarrow {z_i}.
\label{eq:discretized_laplacian}
\end{equation}
More specifically, the discretized kinetic operator $\hat{K}^a \equiv -\cfrac{1}{2}\sum_i \nabla^{2}_{i,a}$ acts on $\ket{x}$ as:
\begin{eqnarray}
\hat{K}^a \ket{x} \equiv - \cfrac{1}{2} \sum_{i=1}^{N_e} \nabla^2_{a,i} \ket{x} = \cfrac{3N_e}{a^2} \ket{x} - \cfrac{1}{2a^2} \sum_{j=1}^{6N_e} \ket{x_j},
\label{eq:discretized_kinetic}
\end{eqnarray}
leading to
\begin{eqnarray}
\hat{K}^a_{\mathbf{x}', \mathbf{x}} \frac{{\Psi _\text{T}}\left( \mathbf{x}' \right)}{{\Psi _\text{T}}\left( \mathbf{x} \right)} = \cfrac{3N_e}{a^2} \delta_{\mathbf{x}', \mathbf{x}} - \cfrac{1}{2a^2} \frac{{\Psi _\text{T}}\left( \mathbf{x}' \right)}{{\Psi _\text{T}}\left( \mathbf{x} \right)} \delta^{ad}_{\mathbf{x}', \mathbf{x}},
\label{eq:kinetic_ratio}
\end{eqnarray}
where $\delta^{ad}_{\mathbf{x}', \mathbf{x}} =1$ only if $\mathbf{x}'$ is adjacent to $\mathbf{x}$, otherwise 0. Here, the point is that one should consider {\it only} $6N_e$ $\mathbf{x}'$ adjacent to $\mathbf{x}$, making the GFMC approach applicable to the continuum \emph{ab initio} Hamiltonian. 
The 6$N_e$ wave function ratios ($\Psi_\text{T}\left( \mathbf{x}' \right)/{\Psi _\text{T}}\left( \mathbf{x} \right)$) are computed in practice using the fast update method, described in Subsec.~{\ref{subsec:fastupdate}}, which is critically important for a fast calculation of LRDMC.
Once the kinetic term is discretized, the potential term $\hat{V}^a$ is defined in such a way that the local energy of $\hat {\mathcal{H}}^a$ computed on the trial WF coincides with the local energy of the continuous Hamiltonian $\hat {\mathcal{H}}$. In practice $\hat{V}^a$ is divided into local and non-local terms, ${\hat{V}^a} \equiv \hat{V}_{\rm loc}^a + \hat{V}_{\rm nl}$, where only the local term depends on the discretized mesh $a$. The non-local potential operator $\hat{V}_{\rm nl}$ exists only in the calculations with effective core potentials (ECPs). The details of the non-local potential operator are described in Sec.~{\ref{sec:Coulomb-and-ECP}}.

\subsubsection{Implementation of LRDMC algorithms}
\label{subsubsec:lrdmc_implementation}
Since several papers have already described the details of the LRDMC implementation~\cite{casula2005PhD,2005CAS,2017BEC,2020NAK,2020_Nakano_LRDMC,nakano2025-loadbalance-lrdmc}, we do not repeat them here. jQMC implements two LRDMC algorithms: the original (conventional) LRDMC scheme~\cite{2005CAS} and the recently developed load-balanced scheme~\cite{nakano2025-loadbalance-lrdmc}. In the conventional algorithm, for a given projection time $\tau$, the electrons are moved a number of times corresponding to the application of the projector $\exp(-\tau \hat{H})$, and branching (reconfiguration) is then performed. Because this number is determined stochastically, the number of electron moves varies from walker to walker. As a result, in large-scale parallel calculations, load imbalance develops among walkers, which degrades weak scaling~\cite{nakano2025-loadbalance-lrdmc}. To remove this problem, we recently proposed an alternative algorithm in which all walkers perform the same number of projection steps, and demonstrated that this substantially improves weak scaling~\cite{nakano2025-loadbalance-lrdmc}. Both algorithms are implemented in jQMC.

\vspace{2mm}
Here, we focus on one implementation aspect that has received less attention in the literature, namely the branching (reconfiguration) step~\cite{2017BEC}, especially its MPI implementation involving inter-node communication. In LRDMC, regardless of whether the conventional or load-balanced algorithm is used, branching (reconfiguration)~\cite{2017BEC} is carried out as follows. Every $N_{\rm proj}$ projections, the code performs a branching step:
\begin{enumerate}
\item[(1)] New walkers are selected from the previous population with probability proportional to their weights before branching:
\begin{equation}
p_{\alpha,n}=\frac{w_{\alpha,n}}{\sum_\beta w_{\beta,n}}.
\label{eq:branching-2}
\end{equation}
\item[(2)] The weights of all new walkers are reset to the average weight of the previous population:
\begin{equation}
w_{\alpha,n+1}=\bar{w}_n\equiv\frac{1}{N_w}\sum_\beta w_{\beta,n}\qquad \forall \alpha.
\label{eq:branching-1}
\end{equation}
\end{enumerate}
In the multi-walker formalism, the walker state therefore carries an additional walker index $\alpha$, namely $(x_{\alpha,n},w_{\alpha,n})$.
The selection of new configurations is implemented using an efficient reconfiguration scheme in which the surviving walkers are chosen from $N_w$ \emph{correlated} random numbers~\cite{2017BEC}:
\begin{equation}
Z_\alpha=\frac{\xi+\alpha-1}{N_w},\qquad \alpha=1,\ldots,N_w,
\label{eq:correlated_random_numbers}
\end{equation}
where $\xi$ is a uniform random number in $[0,1)$. These numbers are compared with the cumulative normalized weights to determine the new walker population.
In our calculations, the survival rate is typically around 90\%, and most walkers survive unchanged after each branching step. Therefore, when walker replication and deletion are communicated across MPI ranks, all-to-all communication is unnecessary. Instead, point-to-point communication is sufficient, and this is essential for achieving good weak scaling. It is also important to perform communications simultaneously with the other computations associated with reconfiguration, that is, to use asynchronous communication so that data transfer does not become a bottleneck.

\subsection{Error analysis}
\label{subsec:error-analysis}
To estimate the mean and variance of an observable $\hat{O}$ from a Monte Carlo simulation, one accumulates its local values along the MCMC samples. Statistical efficiency is gained by running multiple {\it independent} MCMC chains in parallel. Let $i \in \{1,\ldots, M\}$, where $M$ is the chain length, be the MCMC step index in a chain, and $n \in \{ 1, \ldots, N\}$ the index for an independent chain; $X_{i,n}$ is the electron configuration at the $i$-th step of the $n$-th chain, and $O_{i,n} \equiv \hat{O}(X_{i,n})$ are the accumulated local values. The autocorrelation of MCMC samples within a chain causes the naive independent and identically distributed (iid) variance estimator to underestimate the true variance of the sample mean, leading to underestimation of error bars. To mitigate this problem, one usually uses two techniques, reblocking and jackknife resampling~\cite{2017BEC}. 

\vspace{2mm}
Let $b$ be the block (bin) size and $M_b = \lfloor M/b \rfloor$ the resulting number of binned samples per chain. One defines the binned samples $O_{j,n}$ with $j \in \{1, \ldots, M_{b} \}$ as $O_{j,n} = (1/b) \sum_{i \in \mathcal{B}_j} O_{i,n}$, where $\mathcal{B}_j = \{(j-1)b+1, \dots, jb\}$ is the index set of MCMC steps belonging to the $j$-th block. When the block size $b$ exceeds the autocorrelation time, the binned samples become effectively independent, so naive iid-based variance estimators recover the correct standard error of the mean. We notice that samples from different chains are already mutually independent, so blocking is performed only along the in-chain direction $i$ and never across the chain direction $n$.

\vspace{2mm}
When the statistical estimator is a nonlinear function of the Monte Carlo samples, such as a ratio estimator like the DMC/LRDMC mixed average, resampling techniques are required to estimate its uncertainty accurately. jQMC employs the so-called jackknife method to estimate the mean and variance of the accumulated observables. First, one flattens the binned samples $\{O_{j,n}\}$ into a single sequence $\{O_l\}_{l=1}^{N_{J}}$ with $N_{J} = M_b \cdot N$, and defines the $m$-th leave-one-out jackknife sample as $O_{m}^{\rm{J}} = \cfrac{\sum_{l \ne m} O_{l}}{N_{J}-1}$ for $m \in \{1,\ldots, N_{J}\}$. Then, the mean and variance are computed:
\begin{equation}
\mu = \mu^{J} = \cfrac{1}{N_{J}} \sum_{m=1}^{N_{J}} O_{m}^{\rm{J}}
\label{eq:jackknife_mean}
\end{equation}
and
\begin{equation}
\sigma^2 = (N_{J} - 1) (\sigma^{J})^2 = \cfrac{N_{J} - 1}{N_{J}} \sum_{m=1}^{N_{J}} \left( O_{m}^{\rm{J}} - \mu^{\rm{J}} \right)^2.
\label{eq:jackknife_variance}
\end{equation}

\vspace{2mm}
The jackknife (with reblocking) can also be applied to estimate the mean and variance of the accumulated {\it weighted} observables. This is needed for MCMC with the reweighting technique or for LRDMC. The weighted average is defined as $\bar{O} = \cfrac{\sum_l w_{l}O_{l}}{\sum_l w_{l}}$, with $l$ running over all MCMC samples. The $m$-th leave-one-out jackknife sample is $O_{m}^{\rm{J}} = \cfrac{\sum_{l \ne m} w_{l}O_{l}}{\sum_{l \ne m}w_{l}} = \cfrac{\sum_{l} w_{l}O_{l} - w_{m}O_{m}}{\sum_{l }w_{l} - w_{m}}$. If $\mathcal{I}_\beta$ denotes the set of MCMC step indices belonging to block $\beta$, the leave-one-block-out jackknife sample is then $O_{\beta}^{\rm{J}} = \cfrac{\sum_{l \notin \mathcal{I}_\beta} w_{l}O_{l}}{\sum_{l \notin \mathcal{I}_\beta} w_{l}} = \cfrac{\sum_{l} w_{l}O_{l} - \sum_{l \in \mathcal{I}_\beta} w_{l}O_{l}}{\sum_{l}w_{l} - \sum_{l \in \mathcal{I}_\beta} w_{l}}$. Then, the mean and variance are computed in the same way for the $O_{m}^{\rm{J}}$ quantities as described by Eqs.~\ref{eq:jackknife_mean} and \ref{eq:jackknife_variance}, respectively.

%
%

\section{Wave functions}
\label{sec:wavefunction}

\vspace{1mm}
jQMC employs a many-body wavefunction ansatz $\Psi$ composed of the Jastrow factor and the antisymmetric (AS) part:
\begin{equation} 
\Psi  =  \exp(J)  \times \Phi _\text{AS} \,,
\label{eq:wf_ansatz}
\end{equation}
where $\exp(J)$ is symmetric under electron exchange, while $\Phi _\text{AS}$ is antisymmetric.
jQMC currently offers three choices of $\Phi_\text{AS}$:
$\rm(\hspace{.18em}i\hspace{.18em})$ 
the Antisymmetrized Geminal Power (AGP),
$\rm(\hspace{.08em}ii\hspace{.08em})$ 
the AGP with a constrained number of molecular orbitals (AGPn), and
$\rm(i\hspace{-.08em}i\hspace{-.08em}i)$ 
the single Slater determinant.
In this Section, we will describe the functional form of the Jastrow factor (sec.~\ref{subsec:Jastrow}), the AGP (sec.~\ref{subsec:AGP}), the AGPn (sec.~\ref{subsec:AGPn}), the SD (sec.~\ref{subsec:SD}), their fast-update implementations (sec.~\ref{subsec:fastupdate}), and the basis set and atomic orbitals used in the Jastrow and AS parts (sec.~\ref{subsec:atomic_orbitals}).

\subsection{Jastrow factor}\label{subsec:Jastrow}
The Jastrow factor, $\exp(J)$, is required to include electronic correlation and satisfy Kato's cusp conditions~\cite{1957KATO_cusp}. In jQMC, the Jastrow term is written as the sum of one-body, two-body, and three- or four-body contributions: $J = J_1 + J_2 + J_{3/4}$. The one-body and two-body terms play a role in satisfying the electron-ion and electron-electron cusp conditions, respectively. The three- and four-body terms account for additional electron correlation.                                                               
The one-body Jastrow factor $J_1$ is the sum of two parts, the homogeneous part for the electron-ion cusp:   
\begin{equation}
J_1^{\rm h} \left( \mathbf{r}_1,\ldots,\mathbf{r}_N \right) = \sum_{i=1}^{N_{\rm{e}}} \sum_{I=1}^{N_\text{a}} \left( { { -
{{\left( {2{Z_{I}}} \right)}^{3/4}}u\left( {(2{Z_{I}})^{1/4}\left| {{\mathbf{r}_i} - {{\mathbf{R}}_{I}}}
\right|} \right)} } \right),
\label{onebody_J_hom}
\end{equation}
and the inhomogeneous part:
\begin{equation}
{J_1^{\rm inh}}\left( {{{\mathbf{r}}_1}, \ldots, {{\mathbf{r}}_N}} \right) =  \sum_{i=1}^{N_{\rm{e}}}
\sum_{I=1}^{N_\text{a}} \left( {\sum\limits_{l} {L_{I,l} \chi_{I,l}\left( {{{\mathbf{r}}_i}} \right)} } \right)
,
\label{onebody_J_inhom}
\end{equation}
where ${{{\mathbf{r}}_i}}$ are the electron positions, ${{{\mathbf{R}}_{I}}}$ are the atomic positions with their atomic number $Z_{I}$, $l$ runs over atomic orbitals $\chi _{I,l}$ centered on the atom $I$, $\{ L_{I,l}\}$  are variational parameters, and ${u\left( r \right)}$ is a simple bounded function. In jQMC, two functional forms are available for $u$: the exponential form (default option),
\begin{equation}
u^{\rm exp}\left( r \right) = \frac{ 1 }{2 b_{\text{ei}}} \left( {1 - {e^{ - r b_{\text{ei}}}}} \right) \,,
\label{onebody_u_exp}
\end{equation}
and the Pad\'{e} form,
\begin{equation}
u^{\rm Pad\acute{e}}\left( r \right) = \frac{r}{2} \left(1 + b_{\text{ei}} \, r \right)^{-1} \,,
\label{onebody_u_pade}
\end{equation}
both depending on a single variational parameter $ b_{\text{ei}}$. Both forms satisfy the electron-ion cusp condition and are bounded for $r \to \infty$.

\vspace{1mm}
The two-body Jastrow factor is defined as:
\begin{equation}
{J_2}\left( {{{\mathbf{r}}_1}, \ldots, {{\mathbf{r}}_N}} \right) =  {\sum\limits_{i < j} {{v_{{\sigma _i},{\sigma _j}}}\left(
{\left| {{{\mathbf{r}}_i} - {{\mathbf{r}}_j}} \right|} \right)} },
\label{twobody_jastrow}
\end{equation}
where
$v_{{\sigma _i},{\sigma _j}}$
is another simple bounded function.
In jQMC, two functional forms are implemented: the Pad\'{e} form (default),
\begin{equation}
{v^{\rm Pad\acute{e}}}\left( {{r_{i,j}}} \right) =
\frac{{{r_{i,j}}}}{2} {\left( {1 + b_{\rm{ee}} {{r_{i,j}}}} \right)^{ - 1}} \,,
\label{twobody_v_pade}
\end{equation}
and the exponential form,
\begin{equation}
{v^{\rm exp}}\left( {{r_{i,j}}} \right) =
\frac{1}{2 b_{\rm{ee}}} \left( 1 - e^{-b_{\rm{ee}} \, r_{i,j}} \right) \,,
\label{twobody_v_exp}
\end{equation}
where  ${r_{i,j}} = \left| {{{\mathbf{r}}_i} - {{\mathbf{r}}_j}} \right|$,
and $b_{\rm{ee}}$ is the variational parameter. In principle, $v_{{\sigma _i},{\sigma _j}}$ depends on spin, if both like-spin and unlike-spin cusp conditions are fulfilled. In jQMC, we have implemented so far only the spin-independent version, fulfilling the electron-electron cusp conditions for opposite spins.

\vspace{1mm}
The three/four-body Jastrow factor reads:
\begin{equation}
J_{3/4}\left( {{{\mathbf{r}}_1}, \ldots, {{\mathbf{r}}_N}} \right) = 
\sum_{i < j}
\left(  
\sum_{l} \sum_{l'}
M_{l,l'}
\chi _{l}( \mathbf{r}_i )
\chi _{l'}( \mathbf{r}_j )
\right),
\label{threebody_jastrow}
\end{equation}
where the indices $l$ and $l'$ indicate different orbitals and $\{ M_{l,l'} \}$ are variational parameters.
$\chi$ are Gaussian-type atomic orbitals (GTOs), described later.

\subsection{Antisymmetrized Geminal Power (AGP)}
\label{subsec:AGP}
jQMC implements the antisymmetrized geminal power (AGP) ansatz for the fermionic antisymmetric part of the trial wave function. The AGP ansatz was first applied to {\it ab initio} quantum Monte Carlo calculations by Casula and Sorella~\cite{2003CAS-AGP}, and has since been adopted in several QMC codes. For simplicity, we first consider a system with an even number $N$ of electrons. The AGP wave function is written as an antisymmetrized product of two-electron geminal $g$,
\begin{equation}
\Phi_{\rm AS}(\mathbf{1},\ldots,\mathbf{N})
=
{\cal A}
\left[
g(\mathbf{1},\mathbf{2})
g(\mathbf{3},\mathbf{4})
\cdots
g(\mathbf{N-1},\mathbf{N})
\right],
\label{eq:agp_definition}
\end{equation}
where $\mathbf{i}\equiv(\mathbf r_i,\sigma_i)$ denotes the spatial and spin coordinates of electron $i$. The antisymmetrization operator is defined as
\begin{equation}
{\cal A}
=
{1\over N!}
\sum_{P\in S_N}
\epsilon_P \hat P,
\label{eq:antisymmetrization_operator}
\end{equation}
where $S_N$ is the permutation group of $N$ elements, $\hat P$ is the operator associated with the permutation $P$, and $\epsilon_P$ is the sign.
Let $G$ be the $N\times N$ matrix whose elements are
\begin{equation}
G_{ij}=g(\mathbf{i},\mathbf{j}).
\label{eq:full_geminal_matrix}
\end{equation}
Because of fermionic antisymmetry, the geminal satisfies
\begin{equation}
g(\mathbf{i},\mathbf{j})=-g(\mathbf{j},\mathbf{i}),
\qquad
G_{ij}=-G_{ji},
\label{eq:geminal_antisymmetry}
\end{equation}
and hence $G$ is skew-symmetric. In this general spin-coordinate representation, the AGP amplitude is therefore proportional to the Pfaffian of $G$. The overall proportionality factor is irrelevant for variational Monte Carlo because the wave function is used up to normalization.
In the current implementation, jQMC supports only opposite-spin pairings.
Although the full geminal $g(\mathbf{i},\mathbf{j})$ is defined in the spin-coordinate space, only the spin up--spin down block is evaluated explicitly in the opposite-spin implementation. With the electrons ordered such that all spin-up are followed by all spin-down electrons, the full skew-symmetric geminal matrix has the block form
\begin{equation}
G=
\begin{pmatrix}
0 & F \\
-F^T & 0
\end{pmatrix},
\end{equation}
where
\begin{equation}
F_{ij}=f(\mathbf r_i^\uparrow,\mathbf r_j^\downarrow).
\label{eq:up_down_pairing_matrix}
\end{equation}
The spin down--spin up block is therefore fixed by antisymmetry and does not need to be computed independently. For $N_e^{\uparrow}=N_e^{\downarrow}$, the Pfaffian of $G$ is proportional to $\det F$, and the irrelevant overall sign is absorbed in the normalization convention. Thus, jQMC evaluates only the determinant of the spin up--spin down pairing matrix $F$:
\begin{equation}
\Phi_{\rm AS}(\mathbf{x}^{\uparrow},\mathbf{x}^{\downarrow})
=
\det F,
\label{eq:agp_det_form}
\end{equation}
where $\mathbf{x}^{\uparrow}=(\mathbf r_1^\uparrow,\ldots,\mathbf r_{N_\uparrow}^\uparrow)$ and $\mathbf{x}^{\downarrow}=(\mathbf r_1^\downarrow,\ldots,\mathbf r_{N_\downarrow}^\downarrow)$, with the matrix $F$ defined in Eq.~\ref{eq:up_down_pairing_matrix}.
The spatial pairing function is expanded on a chosen one-particle basis as
\begin{equation}
f(\mathbf r_i^\uparrow,\mathbf r_j^\downarrow)
=
\sum_{lm}
\Lambda^{\uparrow\downarrow}_{lm}
\psi_l^\uparrow(\mathbf r_i^\uparrow)
\psi_m^\downarrow(\mathbf r_j^\downarrow),
\label{eq:agp_pairing_expansion}
\end{equation}
where $\Lambda^{\uparrow\downarrow}$ is the geminal coefficient matrix in the chosen orbital basis. The orbitals $\psi_l^\uparrow$ and $\psi_m^\downarrow$ can be either atomic or molecular.
The spin character of the pair is especially transparent when the spatial orbitals are spin independent,
\begin{equation}
\psi_l^\uparrow(\mathbf r)=\psi_l^\downarrow(\mathbf r)\equiv \psi_l(\mathbf r).
\label{eq:spin_independent_orbitals}
\end{equation}
In this case, the opposite-spin geminal can be written as
\begin{equation}
g(\mathbf{i},\mathbf{j})
=
f(\mathbf r_i,\mathbf r_j)
\chi_\uparrow(\sigma_i)\chi_\downarrow(\sigma_j)
-
f(\mathbf r_j,\mathbf r_i)
\chi_\downarrow(\sigma_i)\chi_\uparrow(\sigma_j).
\label{eq:opposite_spin_geminal}
\end{equation}
Introducing the symmetric and skew-symmetric parts of the spatial pairing function,
\begin{align}
f_S(\mathbf r_i,\mathbf r_j)
&=
{1\over 2}
\left[
f(\mathbf r_i,\mathbf r_j)+f(\mathbf r_j,\mathbf r_i)
\right],
\\
f_A(\mathbf r_i,\mathbf r_j)
&=
{1\over 2}
\left[
f(\mathbf r_i,\mathbf r_j)-f(\mathbf r_j,\mathbf r_i)
\right],
\label{eq:pairing_symmetric_decomposition}
\end{align}
the geminal becomes
\begin{align}
g(\mathbf{i},\mathbf{j})
=&\;
f_S(\mathbf r_i,\mathbf r_j)
\left[
\chi_\uparrow(\sigma_i)\chi_\downarrow(\sigma_j)
-
\chi_\downarrow(\sigma_i)\chi_\uparrow(\sigma_j)
\right]
\nonumber\\
&+
f_A(\mathbf r_i,\mathbf r_j)
\left[
\chi_\uparrow(\sigma_i)\chi_\downarrow(\sigma_j)
+
\chi_\downarrow(\sigma_i)\chi_\uparrow(\sigma_j)
\right].
\label{eq:singlet_triplet_decomposition}
\end{align}
Thus, the symmetric part of the spatial pairing function is associated with a spin-singlet pair, whereas the skew-symmetric part is associated with the $M_S=0$ component of a spin-triplet pair. Equivalently, when the same spin-independent orbital basis is used for both spins, the decomposition is controlled by
\begin{equation}
\Lambda^{\uparrow\downarrow}
=
\Lambda^S+\Lambda^A,
\qquad
\Lambda^S={1\over 2}
\left[
\Lambda^{\uparrow\downarrow}
+
(\Lambda^{\uparrow\downarrow})^T
\right],
\qquad
\Lambda^A={1\over 2}
\left[
\Lambda^{\uparrow\downarrow}
-
(\Lambda^{\uparrow\downarrow})^T
\right].
\label{eq:lambda_symmetric_skew_decomposition}
\end{equation}
A symmetric $\Lambda^{\uparrow\downarrow}$ therefore gives a singlet AGP component, while a skew-symmetric $\Lambda^{\uparrow\downarrow}$ gives the $M_S=0$ triplet AGP component. A general matrix $\Lambda^{\uparrow\downarrow}$ contains both components. jQMC preserves a given symmetry of the geminal matrix during the WF optimization.

\vspace{2mm}
The AGP ansatz can also be generalized to spin-polarized systems. Without loss of generality, we can assume $N_{e}^{\uparrow} > N_e^{\downarrow}$ and the system contains $N_e^{\downarrow}$ opposite-spin pairs with $k = N_e^{\uparrow} - N_e^{\downarrow}$ unpaired spin-up electrons. In this case, the determinant form is obtained by concatenating the spin up--spin down pairing matrix with $k$ unpaired orbitals:
\begin{equation}
\Phi_{\rm AS}
=
\det \tilde F,
\label{eq:agp_polarized_det}
\end{equation}
where
\begin{equation}
\tilde F
=
\left[
F^{\uparrow\downarrow}
\;\middle|\;
U^\uparrow
\right].
\label{eq:agp_augmented_pairing_matrix}
\end{equation}
Here, $F^{\uparrow\downarrow}$ is an $N_e^{\uparrow} \times N_e^{\downarrow}$ matrix describing opposite-spin pairings, and $U^\uparrow$ is an $N_e^{\uparrow} \times k$ matrix describing the unpaired orbitals. Explicitly,
\begin{align}
F^{\uparrow\downarrow}_{ij}
&=
\sum_{lm}
\Lambda^{\uparrow\downarrow}_{lm}
\psi_l^\uparrow(\mathbf r_i^\uparrow)
\psi_m^\downarrow(\mathbf r_j^\downarrow),
\\
U^\uparrow_{ia}
&=
\sum_l
\Lambda^\uparrow_{la}
\psi_l^\uparrow(\mathbf r_i^\uparrow),
\qquad
a=1,\ldots,k .
\label{eq:paired_unpaired_blocks}
\end{align}
The resulting matrix $\tilde F$ is square, with dimension $N_e^{\uparrow} \times N_e^{\uparrow}$, and its determinant gives the antisymmetric part of the wave function for the spin-polarized system.

\vspace{2mm}
We now describe the matrix form used in the implementation. Let $\Psi^\uparrow$ and $\Psi^\downarrow$ be the orbital-value matrices,
\begin{equation}
\Psi^\uparrow_{li}
=
\psi_l^\uparrow(\mathbf r_i^\uparrow),
\qquad
\Psi^\downarrow_{mj}
=
\psi_m^\downarrow(\mathbf r_j^\downarrow).
\label{eq:orbital_value_matrices}
\end{equation}
For an atomic orbital (AO) basis of size $N_{ao}$, the dimensions of these matrices are ($N_{ao}$, $N_e^{\uparrow}$) and ($N_{ao}$, $N_e^{\downarrow}$), respectively. The paired block is assembled as a matrix--matrix product,
\begin{equation}
F^{\uparrow\downarrow}
=
(\Psi^\uparrow)^T
\Lambda^{\uparrow\downarrow}
\Psi^\downarrow,
\label{eq:paired_block_matrix_product}
\end{equation}
where $\Lambda^{\uparrow\downarrow}$ has dimension ($N_{ao}$, $N_{ao}$). The unpaired block is similarly written as
\begin{equation}
U^\uparrow
=
(\Psi^\uparrow)^T
\Lambda^\uparrow,
\label{eq:unpaired_block_matrix_product}
\end{equation}
where $\Lambda^\uparrow$ has dimension ($N_{ao}$, $k$). Therefore,
\begin{equation}
\tilde F
=
\left[
(\Psi^\uparrow)^T
\Lambda^{\uparrow\downarrow}
\Psi^\downarrow
\;\middle|\;
(\Psi^\uparrow)^T
\Lambda^\uparrow
\right].
\label{eq:full_augmented_matrix_product}
\end{equation}
For convenience, the variational parameters may be stored as the concatenated matrix
\begin{equation}
\tilde\Lambda
=
\left[
\Lambda^{\uparrow\downarrow}
\;\middle|\;
\Lambda^\uparrow
\right],
\label{eq:concatenated_lambda}
\end{equation}
whose dimension is ($M$, $M+k$). The first $M$ columns define the opposite-spin pairing block, while the remaining $k$ columns define the unpaired orbitals.

\vspace{2mm}
The same formalism applies whether the orbital basis is chosen to be AOs or molecular orbitals (MOs). Let $\Psi^\sigma$ denote the AO-value matrix and $\Phi^\sigma$ the MO-value matrix for spin $\sigma$. We use the convention
\begin{equation}
\Phi^\uparrow
=
C^\uparrow \Psi^\uparrow,
\qquad
\Phi^\downarrow
=
C^\downarrow \Psi^\downarrow,
\label{eq:mo_ao_relation}
\end{equation}
where $C^\uparrow$ and $C^\downarrow$ are the MO coefficient matrices, whose dimensions are ($N_{mo}^{\uparrow}$, $N_{ao}$) and ($N_{mo}^{\downarrow}$, $N_{ao}$), respectively. In spin-polarized calculations, the MO coefficients may differ between the two spin sectors. If $\Lambda_M^{\uparrow\downarrow}$ and $\Lambda_M^\uparrow$ are the geminal and unpaired-orbital coefficient matrices in the MO representation, respectively, the corresponding AO representation is
\begin{align}
\Lambda_A^{\uparrow\downarrow}
&=
(C^\uparrow)^T
\Lambda_M^{\uparrow\downarrow}
C^\downarrow,
\\
\Lambda_A^\uparrow
&=
(C^\uparrow)^T
\Lambda_M^\uparrow .
\label{eq:lambda_ao_mo_relation}
\end{align}
With these definitions, the concatenated pairing matrix is invariant under the AO--MO transformation:
\begin{align}
\tilde F^{\rm AO}
&=
\left[
(\Psi^\uparrow)^T
\Lambda_A^{\uparrow\downarrow}
\Psi^\downarrow
\;\middle|\;
(\Psi^\uparrow)^T
\Lambda_A^\uparrow
\right],
\\
\tilde F^{\rm MO}
&=
\left[
(\Phi^\uparrow)^T
\Lambda_M^{\uparrow\downarrow}
\Phi^\downarrow
\;\middle|\;
(\Phi^\uparrow)^T
\Lambda_M^\uparrow
\right],
\\
\tilde F^{\rm AO}
&=
\tilde F^{\rm MO}.
\label{eq:ao_mo_equivalence}
\end{align}
Thus, the AGP wave functions expanded in AOs and MOs are evaluated by the same matrix operations in jQMC.

\subsection{AGP with constrained number of molecular orbital (AGPn)}\label{subsec:AGPn}
If one expands the AGP pairing function with molecular orbitals, and neglects the non-diagonal terms of the pairing matrix or diagonalizes the pairing function to express it as a function of AGP natural orbitals, the resultant pairing function reads:
\begin{equation}
\tilde{\mathbf{F}}^{MO}(\mathbf{x}^{\uparrow}, \mathbf{x}^{\downarrow}) = [(\bm{\Phi}^{\uparrow})^T \bm{\Lambda}_{M, {\rm diag}}^{\uparrow, \downarrow} \bm{\Phi}^{\downarrow} | (\bm{\Phi}^{\uparrow})^T \bm{\Lambda}_M^{\uparrow} ],
\label{eq:agpn_pairing}
\end{equation}
where $\Lambda_{M, {\rm diag}}^{\uparrow, \downarrow}$ is the diagonal matrix (i.e., the non-diagonal terms are set to zero, neglecting couplings between different MO indices). This is what we call the AGPn ansatz.

\subsection{Single Slater determinant (SD)}\label{subsec:SD}
The AGP ansatz is reduced to the Slater Determinant if one restricts $M_{mo} = N_{e}^{\uparrow}$ or, equivalently, if the rank of the $\Lambda^{\uparrow, \downarrow}$ matrix is equal to $N_{e}^{\downarrow}$, with $N_{e}^{\uparrow}-N_{e}^{\downarrow}$ independent MOs. In this limit, we can neglect the non-diagonal elements of  $\Lambda^{\uparrow, \downarrow}$ and $\Lambda^{\uparrow}$, i.e., $\lambda_{i,j}^{\uparrow, \downarrow} = \delta_{i,j}$  (dim.: $N^{\downarrow}_{e}, N^{\downarrow}_{e}$) and $\lambda_{l,k}^{\uparrow} = \delta_{l,k+N_{e}^{\downarrow}}$ (dim: $N_{e}^{\uparrow}, N_{e}^{\uparrow} - N_{e}^{\downarrow}$). A typical starting point of a VMC calculation from HF/DFT is the SD anstaz with randomized (or zero) Jastrow factor parameters.

\subsection{Fast update of wave function}
\label{subsec:fastupdate}
jQMC employs a single-electron move algorithm in MCMC, where only one electron position is updated in each proposed MCMC step. The same single-electron update is also used in LRDMC, as the algorithm requires. Furthermore, both MCMC and LRDMC require evaluating the wave function at specific spatial grid points for the non-local ECP term and for the discretized kinetic-energy operator (LRDMC only). These evaluations involve moving only one electron at the time from the original electron configuration.
In such cases, one does not need to evaluate the wave function with all electron coordinates from scratch, but one needs just to compute its variation through the single-electron move. In this situation, one should exploit the so-called fast-update technique to drastically save the computational cost for both the Jastrow factor and determinant parts. 
More specifically, MCMC or LRDMC requires the evaluation of the wave function ratio $\Psi(\mathbf{x}')/\Psi(\mathbf{x})$. The ratio is decomposed into the determinant part $D(\mathbf{x}')/D(\mathbf{x})$ and $e^{J(\mathbf{x}')}/e^{J(\mathbf{x})}$, and computed separately.

\vspace{2mm}
The Jastrow ratio is computed as $e^{J(\mathbf{x}')}/e^{J(\mathbf{x})} \equiv e^{\Delta J}$, where $\Delta J = J(\mathbf{x}') - J(\mathbf{x})$. Most contributions to $J$ cancel exactly since only a single electron has been moved. As already described in Sec.~{\ref{subsec:Jastrow}}, the Jastrow factor is made of three different terms. Each of them requires a different implementation for fast updates.

\vspace{2mm}
For the determinant ratio, let $\tilde{F}^{\rm old}$ be the spatial pairing matrix [the same $\tilde{F}$ as in Eq.~(\ref{eq:agp_augmented_pairing_matrix})] before a proposed move and $\tilde{F}^{\rm new}$ the matrix after updating electron $l$. A single-electron move modifies only one row or one column of $\tilde{F}$:
a spin-up move updates row $l$, whereas a spin-down move updates column $l$.
Therefore, the new matrix can always be written as a rank-1 update,
\begin{equation}
\tilde{F}^{\rm new}=\tilde{F}^{\rm old}+\mathbf u_l \mathbf v_l^T,
\label{eq:rank1_update}
\end{equation}
with
\begin{equation}
(\mathbf u_l,\mathbf v_l)=
\begin{cases}
\left(\mathbf e_l,\; [\tilde{F}^{\rm new}_{l,:}-\tilde{F}^{\rm old}_{l,:}]^T\right), & \text{spin-up move},\\[1mm]
\left(\tilde{F}^{\rm new}_{:,l}-\tilde{F}^{\rm old}_{:,l},\; \mathbf e_l\right), & \text{spin-down move},
\end{cases}
\label{eq:rank1_vectors}
\end{equation}
where $\mathbf e_l$ is the $l$-th Cartesian basis vector.
The Slater determinant is treated in the same way, since the single-determinant ansatz is a special case of AGP.
Using the matrix determinant lemma,
\begin{equation}
\det[\tilde{F}^{\rm new}]
=
\det[\tilde{F}^{\rm old}+\mathbf u_l \mathbf v_l^T]
=
\left(1+\mathbf v_l^T (\tilde{F}^{\rm old})^{-1}\mathbf u_l\right)\det[\tilde{F}^{\rm old}],
\label{eq:det_lemma}
\end{equation}
and thus the determinant ratio is
\begin{equation}
\rho_{\rm det} \equiv \frac{\det[\tilde{F}^{\rm new}]}{\det[\tilde{F}^{\rm old}]}
=
1+\mathbf v_l^T (\tilde{F}^{\rm old})^{-1}\mathbf u_l.
\label{det_ratio_fast_update}
\end{equation}
Thus, the acceptance ratio can be evaluated without recomputing $\det[\tilde{F}^{\rm new}]$ from scratch.
After an accepted move, the inverse matrix is updated using the Sherman-Morrison formula,
\begin{equation}
(\tilde{F}^{\rm new})^{-1}
=
(\tilde{F}^{\rm old}+\mathbf u_l\mathbf v_l^T)^{-1}
=
(\tilde{F}^{\rm old})^{-1}
-
(\tilde{F}^{\rm old})^{-1}\mathbf u_l
\left(1+\mathbf v_l^T (\tilde{F}^{\rm old})^{-1}\mathbf u_l\right)^{-1}
\mathbf v_l^T (\tilde{F}^{\rm old})^{-1}.
\label{eq:sherman_morrison}
\end{equation}
The cost of this standard procedure is dominated by matrix-vector operations. To reduce this overhead, McDaniel et al.~{\cite{2017_McDaniel_delayed_update}} proposed the delayed-update scheme, but this is not yet implemented in jQMC.

\subsection{Atomic orbitals for the pairing function and the Jastrow factor}\label{subsec:atomic_orbitals}

One of the most common choices for atomic orbitals is atom-centered Gaussian-type orbitals, denoted as GTOs. A GTO is composed of the radial and angular parts. There are two conventions for the angular part: One is the solid harmonics, and the other is the polynomial functions in the Cartesian basis. jQMC implements both of them.

\subsubsection{Gaussian-Type Atomic orbitals with Solid-harmonics basis functions}
\label{subsubsec:gto_solid_harmonics}
Primitive atomic orbitals ($\psi_{l, m, I}(\mathbf{r})$) with the (regular) solid harmonic basis functions are given in the following Gaussian form:
\begin{equation}
\psi_{l, m, I}(\mathbf{r})  = \mathcal{N}^{\rm solid}_{l,m,I} \mathcal{R}_{I}(\mathbf{r}) \mathcal{S}_{l,m,I}(\mathbf{r}),
\label{eq:gto_solid}
\end{equation}
which can be evaluated using Cartesian coordinates (i.e., $x,y,z$) because there is no singular point, where the radial part and solid harmonics are defined as:
\begin{equation}
\mathcal{R}_{I}(\mathbf{r}) = e^{-Z_I \cdot |\mathbf{R}_I - \mathbf{r}|^2},
\label{eq:gto_radial}
\end{equation}
and
\begin{equation}
\mathcal{S}_{l,m,I}(\mathbf{r}) = \sqrt{\frac{2l + 1}{4\pi}} \cdot |\mathbf{R}_{I} - \mathbf{r}|^l \cdot \mathcal{Y}_{l,m}(\theta_{I}, \phi_{I}),
\label{eq:solid_harmonics}
\end{equation}
where $\sqrt{\cfrac{2l + 1}{4\pi}}$ in $\mathcal{S}_{l,m,I}(\mathbf{r})$ is the Racah's normalization factor for the solid harmonics part, resulting in $
\int_{0}^{\pi}\int_{0}^{2\pi} \sin\theta d\theta d\phi \mathcal{S}_{l,m,I}^*(\mathbf{r}) \mathcal{S}_{l,m,I}(\mathbf{r}) = \cfrac{4 \pi}{2l +1} r^{2l}$.
The normalization factor, $\mathcal{N}^{\rm solid}_{l,m,I}$, is defined as
\begin{equation}
\mathcal{N}_{l,I}^{\rm{solid}} = \sqrt{\frac{(2Z_{I}/\pi)^{3/2}(4Z_{I})^{l}}{(2l-1)!!}}.
\label{eq:norm_solid}
\end{equation}
The GTO can also be written with real spherical harmonics as:
\begin{equation}
\psi_{l, m, I}(\mathbf{r}) = \mathcal{N}_{l,m,I}^{\rm{sphe}} \cdot
            |\mathbf{R}_I - \mathbf{r}|^l \cdot \mathcal{R}_{I}(\mathbf{r}) \cdot \mathcal{Y}_{l,m}(\theta_{I}, \phi_{I}),
\label{eq:gto_spherical}
\end{equation}
where $\theta_{I}$ and $\phi_{I}$ are polar ($0 \le \theta < \pi$) and azimuthal ($0 \le \phi < 2\pi$) angles in the spherical coordinate whose origin is $\mathbf{R}_{I}$, $\mathcal{Y}_{l,m}(\theta, \phi)$ is the real spherical harmonics, and $\mathcal{N}^{\rm sphe}_{l,I}$ is the normalization factor:
\begin{equation}
\mathcal{N}^{\rm sphe}_{l,I} = \sqrt{\frac{2^{2l+3}(l+1)!(2Z_I)^{l+\frac{3}{2}}}{(2l+2)!\sqrt{\pi}}}.
\label{eq:norm_spherical}
\end{equation}
Notice that the relation between the normalizations factors is $\cfrac{\mathcal{N}^{\rm solid}_{l,m,I}}{\mathcal{N}^{\rm sphe}_{l,m,I}} = \sqrt{\cfrac{2l + 1}{4\pi}}$.

\subsubsection{Gaussian-Type Atomic orbitals with Cartesian basis functions}
\label{subsubsec:gto_cartesian}
Primitive atomic orbitals ($\psi_{l, m, I}(\mathbf{r})$) with Cartesian basis functions are given in the following Gaussian form:
\begin{equation}
\psi_{n_x, n_y, n_z, I}(\mathbf{r}) = \mathcal{N}_{n_x, n_y, n_z,I}^{\rm{cart}} \cdot e^{-Z_I \cdot |\mathbf{R}_I - \mathbf{r}|^2} \cdot \mathcal{P}_{n_x, n_y, n_z}(x,y,z),
\label{eq:gto_cartesian}
\end{equation}
where $\mathcal{P}_{n_x, n_y, n_z}(x,y,z) = x^{n_x} y^{n_y} x^{n_z}$, $\mathcal{N}_{n_x, n_y, n_z,I}^{\rm{cart}}$ is the normalization factor for an orbital $(n_x, n_y, n_z,I)$ discussed later. The normalization $\mathcal{N}_{n_x, n_y, n_z,I}^{\rm{cart}}$ is computed as:
\begin{equation}
\mathcal{N}_{n_x, n_y, n_z,I}^{\rm{cart}} = \sqrt{\frac{(2Z_{I}/\pi)^{3/2}(4Z_{I})^{n_x + n_y + n_z}}{(2n_x-1)!!(2n_y-1)!!(2n_z-1)!!}} \equiv \sqrt{\cfrac{(2Z_{I}/\pi)^{3/2}(8Z_{I})^{n_x + n_y + n_z} n_x! n_y! n_z! }{(2n_x)!(2n_y)!(2n_z)!}}.
\label{eq:norm_cartesian}
\end{equation}
The total angular momentum $l$ is defined as $l = n_x + n_y + n_z$. jQMC stores the polynomial functions in the canonical (or alphabetical) ordering, as in TREX-IO~{\cite{2023POS}}, e.g., $l=0 \rightarrow s$, $l=1 \rightarrow p_x,p_y,p_z$, $l=2 \rightarrow d_{xx}, d_{xy}, d_{xz}, d_{yy}, d_{yz}, d_{zz}$.
Notice that the above equation implies that the normalization factor is not common for all orbitals with the same $l$ except for $s$ and $p$ orbitals. For instance, for $d_{xx}$ and $d_{xy}$ orbitals,
\begin{equation}
\mathcal{N}_{2, 0, 0,I}^{\rm{cart}} = \sqrt{\frac{(2Z_{I}/\pi)^{3/2}(4Z_{I})^{2}}{3}},
\label{eq:norm_cart_dxx}
\end{equation}
while
\begin{equation}
\mathcal{N}_{1, 1, 0,I}^{\rm{cart}} = \sqrt{(2Z_{I}/\pi)^{3/2}(4Z_{I})^{2}}.
\label{eq:norm_cart_dxy}
\end{equation}
%

%
%

\section{Coulomb interaction and Effective Core Potential}
\label{sec:Coulomb-and-ECP}
The potential term is decomposed into two parts, bare Coulomb interactions and effective core potential parts, $V(\mathbf{x}) = V^{\rm bare}(\mathbf{x}) + V^{\rm ecp}(\mathbf{x})$. The bare Coulomb part is further decomposed into three components. i.e., $V^{\rm bare}(\mathbf{x}) = V^{\rm ei}(\mathbf{x}) + V^{\rm ee}(\mathbf{x}) + V^{\rm ii}(\mathbf{x})$. They are:
\begin{gather}
V^{\rm ei}(\mathbf{x}) = -\sum_{l, I} \frac{Z_{I}}{|\mathbf{r}_l - \mathbf{R}_{I}|},
\label{eq:coulomb_ei}
\end{gather}
\begin{gather}
V^{\rm ee}(\mathbf{x}) = \sum_{l' > l} \frac{1}{|\mathbf{r}_l - \mathbf{r}_{l'}|},
\label{eq:coulomb_ee}
\end{gather}
\begin{gather}
V^{\rm ii}(\mathbf{x}) = \sum_{I' > I} \frac{Z_{I'}Z_{I}}{|\mathbf{R}_{I'} - \mathbf{R}_{I}|}.
\label{eq:coulomb_ii}
\end{gather}
Notice that the electron-ion potential  is often computed with semi-local potentials when ECPs are employed. In the latter case, the atomic number $Z_{I}$ is replaced by the effective charge, or pseudoatomic number, $Z^{\rm eff}_{I}$, such that the electron-ion potential reads $V^{\rm ei}(\mathbf{x}) = -\sum_{l, I} Z^{\rm eff}_{I}/|\mathbf{r}_l - \mathbf{R}_{I}|$. Nevertheless, in jQMC the electron-ion potential is separated from the semi-local potential part for the sake of clarity. Thus, $V^{\rm bare}(\mathbf{x})$ is defined for both all-electron and ECP cases, with $Z_{I}$ replaced by $Z^{\rm eff}_{I}$ in Eqs.~\ref{eq:coulomb_ei} and \ref{eq:coulomb_ii} if ECP are used.

\vspace{2mm}
In the presence of ECPs, one should treat their non-local contributions. The semi-local effective core potential is usually employed in QMC calculations, which for nucleus $I$ is parametrized as follows:
\begin{equation}
\hat{V}_{\rm{ecp}}^{i, I} = \hat{V}_{\rm{eloc}}^{i, I} + \sum_{l=0}^{l_{\rm{max}}}\hat{V}_{l}^{i, I} \hat{P}_l^{i, I},
\label{eq:ecp_semilocal}
\end{equation}
where $i$ and $I$ are electron and nucleus indices, $\hat{P}_l^{i,I} = \sum_{m=-l}^l |Y_{lm}\rangle_{i,I}\langle Y_{lm}|_{i,I}$ projects the angular part of electron $i$'s coordinates relative to nucleus $I$ onto the angular momentum $l$, and $V^{i,I}_{\rm eloc}$, $V^{i,I}_l$ are radial functions of $r_{i,I} \equiv |\mathbf{r}_i - \mathbf{R}_I|$.
Both $\hat{V}_{\rm{eloc}}^{i, I} \equiv V_{\rm{eloc}}^{I}(\mathbf{r}_i)$ and $\hat{V}_{l}^{i, I} \equiv V_{l}^{I}(\mathbf{r}_i)$ are parametrized with three parameters, $a$, $b$, and $n$:
\begin{gather}
 V_{\rm{eloc}}^{I}(\mathbf{r}_i) =  \frac{1}{|\mathbf{r}_i - \mathbf{R}_{I}|^2} \sum_k^{N_{I}} a_{I, k} \cdot |\mathbf{r}_i - \mathbf{R}_{I}|^{n_{I, k}} \exp(-b_{I, k} |\mathbf{r}_i - \mathbf{R}_{I}|^2), \\
 V_{l}^{I}(\mathbf{r}_i) = \frac{1}{|\mathbf{r}_i - \mathbf{R}_{I}|^2} \sum_k^{N_{I, l}} a_{I, k,l} \cdot |\mathbf{r}_i - \mathbf{R}_{I}|^{n_{I, k,l}} \exp(-b_{I, k,l} |\mathbf{r}_i - \mathbf{R}_{I}|^2),
\label{eq:ecp_parametrization}
\end{gather}
where $N_{I}$ and $N_{I, l}$ are the number of ECP terms  of the local contribution and the non-local part with angular momentum $l$ for nucleus $I$, respectively.
Following the standard treatment of the semi-local form~\cite{2003GOT,1991MIT,2017BEC,casula2005PhD}, the action of $\hat V_l^{i,I}\hat P_l^{i,I}$ on a many-body wave function $\Psi$ reduces to an angular integration at fixed electron-nucleus distance $r_{i,I}$:
\begin{eqnarray}
\frac{\braket{x|\hat{V}_{l}^{i, I}\hat{P}_l^{i, I}|\Psi}}{\braket{x|\Psi}} &=& \int d\mathbf{x}'\braket{x|\hat{V}_{l}^{i, I}\hat{P}_l^{i, I}|x'}
\frac{\braket{x'|\Psi}}{\braket{x|\Psi}} \\
&=& V_l^{\,i,I}(r_{i,I})
  \sum_{m=-l}^l Y_{lm}(\Omega_{i,I})
  \int d\Omega_{i,I}'\;Y_{lm}^*(\Omega_{i,I}')\,
  \frac{\Psi(\dots,\mathbf{r}_i',\dots)}
       {\Psi(\dots,\mathbf{r}_i,\dots)} \\
&=& V_l^{\,i,I}(r_{i,I})\,\frac{2l+1}{4\pi}
  \int d\Omega_{i,I}'\;P_l\!\left(\mathbf{s}_{i,I} \cdot \mathbf{s}'_{i,I}\right)\,
  \frac{\Psi(\dots,\mathbf{r}_i',\dots)}{\Psi(\dots,\mathbf{r}_i,\dots)},
\label{eq:ecp_projection}
\end{eqnarray}
where $\mathbf{s}_{i,I} \equiv \cfrac{\mathbf{r}_i - \mathbf{R}_I}{|\mathbf{r}_i - \mathbf{R}_I|}$ and $ \mathbf{s}'_{i,I} \equiv \cfrac{\mathbf{r}_i' - \mathbf{R}_I}{|\mathbf{r}_i' - \mathbf{R}_I|}$.
The last identity of the above equation follows from a well-known property of the spherical harmonics~\cite{2003GOT}, called the addition theorem, $\sum_{m=-l}^l Y_{lm}(\Omega)\,Y_{lm}^*(\Omega') = \cfrac{2l+1}{4\pi}\,P_l(\mathbf{s}_{i,I} \cdot \mathbf{s}'_{i,I})$ where $P_{l}(z)$ is the $l$-th Legendre polynomial.

\vspace{2mm}
jQMC implements the integration scheme proposed in Ref.~{\onlinecite{1991MIT}} to perform a numerical quadrature with tetrahedrical ($N_V=4$), octahedrical ($N_V = 6$, $N_V = 18$), or icosahedrical ($N_V = 12$) symmetry. The quadrature meshes are defined as $\mathbf{R}_{I} + \bm{\delta}_m \cdot |\mathbf{r}_i - \mathbf{R}_{I}|$. Since the numerical quadrature is not exact for a general wave function, the direction of $z$ axis of the spherical coordinate is selected randomly, $\theta$ ($ \in [0, \pi[$) and $\phi$ ($ \in [0, 2\pi[$) are selected randomly, to mitigate the bias. With the mesh, the projection is rewritten as a sum of $N_V$ terms:
\begin{eqnarray}
\frac{\braket{x|\hat{V}_{l}^{i, I}\hat{P}_l^{i, I}|\Psi}}{\braket{x|\Psi}} &=& \int d\mathbf{x}'\braket{x|\hat{V}_{l}^{i, I}\hat{P}_l^{i, I}|x'}
\frac{\braket{x'|\Psi}}{\braket{x|\Psi}} \\
&\sim& \sum_{m=1}^{N_v}\braket{x|\hat{V}_{l}^{i, I}\hat{P}_l^{i, I}|x_m}
\frac{\braket{x_m|\Psi}}{\braket{x|\Psi}} \\
&=& \sum_{m=1}^{N_v} (2l+1) \cdot V_l^{\,i,I}(r_{i,I}) \cdot w_m \cdot P_{l}\left(\mathbf{s}_{i,I} \cdot \bm{\Delta}_m \right) \frac{\Psi(\cdots, \mathbf{R}_{I} + \bm{\delta}_m \cdot |\mathbf{r}_i - \mathbf{R}_{I}|, \cdots)}{\Psi(\cdots \mathbf{r}_i \cdots)},
\label{eq:ecp_quadrature}
\end{eqnarray}
where $a_m$ are the weights of the quadrature, and $\bm{\delta}^m$ are the points of the mesh, and $\bm{\Delta}_m $ is the normalized one: $\bm{\Delta}_m  \equiv \bm{\delta}_m /|\bm{\delta}_m|$. The grid points are tabulated in Ref.~{\onlinecite{1991MIT}}, which are hard-coded in jQMC.
The WF ratios on the grid points are efficiently computed via the fast-update algorithm, described in Sec.~\ref{subsec:fastupdate}, which is critically important for a fast computation for both MCMC and GFMC.

\vspace{2mm}
The non-local contribution arising from the application of non-local operators is naturally considered in GFMC, and more specifically in LRDMC. Indeed, in the LRDMC framework, the non-local potentials defined as $\hat{V}^{nl}_{\mathbf{x}, \mathbf{x}'} \cfrac{{\Psi _\text{T}}\left( \mathbf{x}' \right)}{{\Psi _\text{T}}\left( \mathbf{x} \right)}$ is nothing but $\braket{x|\hat{V}_{l}^{i, I}\hat{P}_l^{i, I}|x_m}
\cfrac{\braket{x_m|\Psi_\text{T}}}{\braket{x|\Psi_\text{T}}}$ for nucleus $I$ and electron $i$. Therefore, within the discretized summation implementation, one needs to consider only the non-diagonal hoppings $\mathbf{x} \rightarrow \mathbf{x}'$ satisfying $|\mathbf{r}'_i - \mathbf{R}_{I}| = |\mathbf{r}_i - \mathbf{R}_{I}|$. Indeed, one should considers only $N_V \cdot N_a \cdot N_e$ non-diagonal matrix elements at maximum for the non-local PP term, even if, in practice, we can consider only a few neighboring atoms $I$ to the target electron index $i$ since the potential decays exponentially, resulting in reducing the mesh size that should be considered to $\sim N_V \cdot N_e$.

\vspace{2mm}
With the semi-local ECPs, the fixed-node Hamiltonian (Eq.~{\ref{eq:fn_hamiltonian}}) corresponds to the T-move approximation of the standard DMC framework~{\cite{2006CAS-Tmove}}. If one replaces ${{\mathcal{V}^{\rm nl}_{\mathbf{x},\mathbf{x}'}}{\Psi _\text{T}}\left( \mathbf{x}' \right)} /{\Psi _\text{T}}\left( \mathbf{x} \right)$ with ${{\mathcal{V}^{\rm nl}_{\mathbf{x},\mathbf{x}'}}{D _\text{T}}\left( \mathbf{x}' \right)} /{D _\text{T}}\left( \mathbf{x} \right)$, where $D _\text{T}$ represents the determinant part of the trial wave function $\Psi _\text{T}$, the resultant fixed-node Hamiltonian corresponds to T-move with determinant locality approximation (DTM) in the standard DMC framework~{\cite{2019ZEN-DLA}}. jQMC implements the two fixed-node Hamiltonians for calculations with ECPs.

%
%

\section{Derivatives of total energy}
\label{sec:derivatives}
The total-energy derivatives with respect to the variational parameters are a key component in order to optimize the many-body wave function. Similarly, those with respect to atomic coordinates (i.e., atomic forces) are essential for tasks such as geometry optimization, phonon dispersion calculations, and for making machine-learning interatomic potentials.
Algorithmic Differentiation or Automatic Differentiation (AD) offers a powerful solution to this problem by systematically applying the chain rule~{\cite{2010SOR}}. AD comes in two primary forms: forward-mode AD (FAD), which propagates derivatives from the inputs to the outputs, and adjoint-mode AD (AAD), which starts from the outputs to the inputs. They correspond to Jacobian-Vector products (JVPs) and Vector-Jacobian products (VJPs) in the JAX terminology~{\cite{jax2018github}}, respectively. When the number of input parameters significantly exceeds the number of outputs, AAD is more efficient, as it enables the evaluation of all gradients in a computational time comparable to that of a single function evaluation (e.g., the energy or wave function).
This computational advantage makes AAD particularly well-suited for QMC simulations involving wave functions with many variational parameters. AAD was first introduced in the QMC context for atomic force calculations in Ref.~\onlinecite{2010SOR}. Our implementation, jQMC, leverages AAD provided by JAX (e.g. grad) to efficiently compute derivatives of the local energy and the logarithm of the wave function.

\vspace{1mm}
\subsection{Derivatives with respect to variational parameters (MCMC)}
\label{subsec:derivatives_with_respect_to_variational}
The energy derivative with respect to a variational parameter ${\alpha_k}$ is represented as a generalized \emph{force}:
\begin{equation}
{f_k} =- \frac{{\partial E\left( \bm{\alpha}  \right)}}{{\partial {\alpha_k}}} =  - \frac{\partial }{{\partial {\alpha_k}}}\frac{{\braket{{\Psi ^{\bm{\alpha}} }|\hat {\mathcal{H}}|{\Psi ^{\bm{\alpha}} }}}}{{\braket{{\Psi ^{\bm{\alpha}} }|{\Psi ^{\bm{\alpha}} }}}}.
\label{eq_general_force}
\end{equation}
In MCMC, the derivative can be evaluated using $M$ configurations of electron coordinates{~\cite{2017BEC}}:
\begin{equation}
\begin{split}
{f_k}
&=  - 2\,\cfrac{\sum_{\mathbf{x}} e_L(\mathbf{x})\left( O_k(\mathbf{x}) - \bar O_k \right) |\Psi^{\bm{\alpha}}(\mathbf{x})|^2}{\sum_{\mathbf{x}} |\Psi^{\bm{\alpha}}(\mathbf{x})|^2} \\
&\approx  - 2\,\cfrac{\sum_{i = 1}^M w_i\, e_L(\mathbf{x}_i)\left( O_k(\mathbf{x}_i) - \bar O_k \right)}{\sum_{i = 1}^M w_i},
\end{split}
\label{eq:generalized_force_mc}
\end{equation}
where ${e_L \left( {{{\mathbf{x}}}} \right)}$ is the local energy,
${O_k}\left( {\mathbf{x}} \right)$ is the logarithmic derivative of the WF, {\it i.e.}, ${O_k}\left( {\mathbf{x}} \right) = \cfrac{\partial}{{\partial {\alpha_k}}} \ln {\Psi ^{\bm{\alpha}} }\left( {\mathbf{x}} \right)$), and ${{\bar O}_k}$ is its average over $M$ samples, {\it {i.e.}}, ${{\bar O}_k} =  \sum\limits_{i = 1}^M w_i{O_k}\left( {{{\mathbf{x}}_i}} \right) / \sum\limits_{i = 1}^M w_i$. 
In jQMC, the logarithmic derivative ${O_k}\left( {\mathbf{x}} \right)$ is computed via JAX's automatic differentiation (grad) functionality. In jQMC, the derivatives with respect to the wave function parameters are implemented only for MCMC at present; in the future, they will be implemented also for LRDMC.

\subsection{Derivatives with respect to nuclear coordinates: Atomic forces (MCMC)}
\label{ionic_forces}
jQMC employs the differential form of atomic force calculations with the space-warp coordinate transformation (SWCT) technique to reduce the statistical error~{\cite{1989UMR-SWCT}}:
\begin{equation}
{{\mathbf{F}}_I} =  -\left\langle {\frac{d}{{d{{\mathbf{R}}_{I}}}}{e_L}} \right\rangle  + 2\left[ {\left\langle {{e_L}} \right\rangle \left\langle {\frac{d}{{d{{\mathbf{R}}_{I}}}}\log \left( {{{\mathcal{J}}^{\frac{1}{2}}}\Psi } \right)} \right\rangle  - \left\langle {{e_L}\,\frac{d}{{d{{\mathbf{R}}_{I}}}}\log \left( {{{\mathcal{J}}^{\frac{1}{2}}}\Psi } \right)} \right\rangle } \right]
\label{forces:aad}
\end{equation}
where ${\mathcal{J}}$ is the Jacobian of the SWCT.
The SWCT is used to follow the displacement of charges around the nucleus,
\begin{equation}
{{\mathbf{\bar r}}_i} = {{\mathbf{r}}_i} + d {{\mathbf{R}}_{I}}{\omega _{I}}\left( {{{\mathbf{r}}_i}} \right),
\label{eq:swct_displacement}
\end{equation}
\begin{equation}
{\omega _{I}}\left( {\mathbf{r}} \right) = \frac{{ \kappa \left( {\left| {{\mathbf{r}} - {{\mathbf{R}}_{I}}} \right|} \right)}}{{\sum\nolimits_{{J} = 1}^{N_\text{a}} {\kappa\left( {\left| {{\mathbf{r}} - {{\mathbf{R}}_{J}}} \right|} \right)} }},
\label{eq:swct_omega}
\end{equation}
where ${\kappa\left( r \right)}$ is a function that decays sufficiently fast for large $r$. The original\cite{1989UMR-SWCT} choice, $\kappa\left( r \right) = 1/{r^4}$, is adopted in jQMC.
The parameter derivatives are computed with the partial derivatives of the local energy and those of the logarithm of the wavefunction{~\cite{2010SOR}}:
\begin{eqnarray}
\mathbf{F}_{I}^{\rm VMC} = - \frac{dE}{d\mathbf{R}_{I}} = &-& \braket{\frac{\partial E_{\rm L}}{\partial \mathbf{R}_{I}} + \sum\limits_i^{} {{\omega}\left({\mathbf{R}}_{I},  {{{\mathbf{r}}_i}} \right)\frac{\partial E_{\rm L}}{{\partial {{\mathbf{r}}_i}}}}} \nonumber \\ &-&2\braket{(E_{\rm L} - E) \left(\frac{\partial \log \Psi }{\partial \mathbf{R}_{I}} + \sum\limits_i^{} {\left[ {{\omega}\left({\mathbf{R}}_{I},  {{{\mathbf{r}}_i}} \right)\frac{\partial \log \Psi}{{\partial {{\mathbf{r}}_i}}} + \frac{1}{2}\frac{\partial {\omega} \left( {{{\mathbf{R}}_{I}, {\mathbf{r}}_i}} \right)}{{\partial {{\mathbf{r}}_i}}}} \right]} \right)}.
\label{eq:force_vmc_matrix}
\end{eqnarray}
Here, 6$N_e$ + 6$N_{\rm a}$ components have to be evaluated, i.e., $\left\langle {\cfrac{\partial {E_L}}{{\partial {{\mathbf{r}}_i}}}} \right\rangle$, $\left\langle {\cfrac{\partial \log \Psi}{{\partial {{\mathbf{r}}_i}}} } \right\rangle$, $\left\langle {\cfrac{\partial {E_L}}{{\partial {{\mathbf{R}}_I}}}} \right\rangle $, $\left\langle {\cfrac{\partial \log \Psi }{{\partial {{\mathbf{R}}_I}}}} \right\rangle ${~{\cite{2010SOR}}}.
The dimensions of $\cfrac{\partial E_{\rm L}}{\partial \mathbf{R}_{I}}$, $\cfrac{\partial E_{\rm L}}{{\partial {{\mathbf{r}}_i}}}$, $\cfrac{\partial \log \Psi}{\partial \mathbf{R}_{I}}$, $\cfrac{\partial \log \Psi}{{\partial {{\mathbf{r}}_i}}}$, and $\cfrac{\partial {\omega} \left( {{{\mathbf{R}}_{I}, {\mathbf{r}}_i}} \right)}{{\partial {{\mathbf{r}}_i}}}$ are $(N_{a}, 3)$, $(N_{e}, 3)$, $(N_{a}, 3)$, $(N_{e}, 3)$, and $(N_{a}, N_{e}, 3)$, respectively. ${\omega}\left({\mathbf{R}}_{I},  {{{\mathbf{r}}_i}} \right)$ is always a scalar value.
One can define the following matrices to compute the force components with a given electronic configuration $\mathbf{x}$:
\begin{equation}
  \begin{cases}
  \partial E^R \equiv \cfrac{\partial E_{\rm L}}{\partial \mathbf{R}_{I}} & dim. = (N_{a}, 3), \\
  \partial E^{r, \uparrow} \equiv \cfrac{\partial E_{\rm L}}{{\partial {{\mathbf{r}}_i^{\uparrow}}}} & dim. = (N_{e}^{\uparrow}, 3), \\
  \partial E^{r, \downarrow} \equiv \cfrac{\partial E_{\rm L}}{{\partial {{\mathbf{r}}_i^{\downarrow}}}} & dim. = (N_{e}^{\downarrow}, 3), \\
  \partial \Psi^R \equiv \cfrac{\partial \log \Psi}{\partial \mathbf{R}_{I}} & dim. = (N_{a}, 3), \\
  \partial \Psi^{r,\uparrow} \equiv \cfrac{\partial \log \Psi}{{\partial {{\mathbf{r}}_i^{\uparrow}}}} & dim. = (N_{e}^{\uparrow}, 3), \\
  \partial \Psi^{r,\downarrow} \equiv \cfrac{\partial \log \Psi}{{\partial {{\mathbf{r}}_i^{\downarrow}}}} & dim. = (N_{e}^{\downarrow}, 3), \\
  \partial \Omega^{r,\uparrow} \equiv \sum_i \cfrac{\partial {\omega} \left( {{{\mathbf{R}}_{I}, {\mathbf{r}}_i^{\uparrow}}} \right)}{{\partial {{\mathbf{r}}_i^{\uparrow}}}} & dim. = (N_{a}, 3), \\
  \partial \Omega^{r,\downarrow} \equiv \sum_i \cfrac{\partial {\omega} \left( {{{\mathbf{R}}_{I}, {\mathbf{r}}_i^{\downarrow}}} \right)}{{\partial {{\mathbf{r}}_i^{\downarrow}}}} & dim. = (N_{a}, 3), \\
  \Omega^{\uparrow} \equiv {\omega}\left({\mathbf{R}}_{I},  {{{\mathbf{r}}_i^{\uparrow}}} \right) & dim. = (N_a, N_{e}^{\uparrow}).\\
  \Omega^{\downarrow} \equiv {\omega}\left({\mathbf{R}}_{I},  {{{\mathbf{r}}_i^{\downarrow}}} \right) & dim. = (N_a, N_{e}^{\downarrow}).\\
  \end{cases}
\label{eq:force_components}
\end{equation}
Once the objects in Eq.~\ref{eq:force_components} are evaluated and stored, the $(N_{\rm a}, 3)$ force matrix ${F}^{\rm VMC}$ can be obtained via matrix-matrix operations, according to the following contributions:
\begin{equation}
{F}^{\rm VMC} = {F}^{\rm HF} + {F}^{\rm Pulay},
\label{eq:force_vmc_full}
\end{equation}
where the Hellmann--Feynman (HF) term is
\begin{equation}
{F}^{\rm HF} = -\braket{\partial E^R + \Omega^{\uparrow} \partial E^{r,\uparrow} + \Omega^{\downarrow} \partial E^{r,\downarrow}},
\label{eq:force_hf}
\end{equation}
and the Pulay contribution reads
\begin{eqnarray}
{F}^{\rm Pulay} &=&  - 2\braket{(E_{\rm L} - E)(\partial \Psi^R + \Omega^{\uparrow} \partial \Psi^{r,\uparrow} + \Omega^{\downarrow} \partial \Psi^{r,\downarrow} + \cfrac{1}{2}(\partial \Omega^{r,\uparrow} + \partial \Omega^{r,\downarrow}))}.
\label{eq:force_pulay}
\end{eqnarray}

\vspace{2mm}
We validated the force implementation through the PES calculation of the Hydrogen and Nitrogen dimer. To test the implementation, we verified that the the following conditions are satisfied: (a) The derivative of the PES with respect to the bond distance should be consistent with the computed atomic forces as long as all the parameters are variationally optimized, and (b), with the SWCT, the atomic forces acting on the left and right atoms of a dimer should have exactly the same absolute values~{\cite{2022NAK-SWCT}}.

\subsection{Derivatives with respect to nuclear coordinates: Atomic forces (LRDMC)}
\label{ionic_forces_lrdmc}                                                            
In LRDMC, the atomic force computation differs from the MCMC case (Sec.~\ref{ionic_forces}) in three respects: (i) the SWCT is not recommended because it leads to a bias, (ii) the Reynolds approximation~\cite{1986_Reynolds_DMC_force} is introduced to replace the inaccessible derivative $\partial \Psi_0 / \partial \mathbf{R}_{I}$ with the available $\partial \Psi_T / \partial \mathbf{R}_{I}$, and (iii) the Pathak-Wagner~{\cite{Pathak2020_PW_reweight}} regularization is applied to suppress the divergence of the local energy derivative near the nodal surface.

\vspace{2mm}
Since the SWCT is typically not used in LRDMC, all $\Omega$ terms vanish ($\omega_I = 0$), and the force expression simplifies considerably. Using the same shorthand notation defined in Sec.~\ref{ionic_forces}, the force on atom $I$ sampled from the mixed distribution $\Psi_{\rm T}\Psi_0$ reads:
\begin{equation}
{F}_{I}^{\rm mixed} = {F}^{\rm HF}_{\rm mixed} + {F}^{\rm Pulay}_{\rm mixed},
\label{eq:force_lrdmc_mixed}
\end{equation}
where
\begin{equation}
{F}^{\rm HF}_{\rm mixed} = -\braket{\partial E^R}_{\Psi_{\rm T}\Psi_0},
\label{eq:force_lrdmc_hf}
\end{equation}
\begin{equation}
{F}^{\rm Pulay}_{\rm mixed} = -2\braket{(E_{\rm L} - E)\,\partial \Psi^R}_{\Psi_{\rm T}\Psi_0}.
\label{eq:force_lrdmc_pulay}
\end{equation}
Here, $\braket{\cdot}_{\Psi_{\rm T}\Psi_0}$ denotes the average over the mixed distribution sampled by the LRDMC walker population, in contrast to the MCMC average over $\Psi_{\rm T}^2$.
The mixed estimator ${F}_{I}^{\rm mixed}$ is biased because the sampling distribution $\Psi_{\rm T}\Psi_0$ differs from the pure distribution $\Psi_0^2$. This approximation becomes exact when $\Psi_{\rm T} = \Psi_0$.

\vspace{2mm}
Near the nodal surface of $\Psi_{\rm T}$, the local energy and its derivatives can diverge, leading to large statistical fluctuations in the force estimator. To mitigate this issue, we employ the Pathak--Wagner regularization~{\cite{Pathak2020_PW_reweight}}. According to the reweighting method, the bracket $\braket{O}$ is evaluated using the regularized estimator with a function $R_{\eta}(\mathbf{x})$:
\begin{equation}
\bar{O} = \braket{O(\mathbf{x})\, R_{\eta}(\mathbf{x})}.
\label{eq:regularized_estimator}
\end{equation}
Here,
\begin{equation}
R_{\eta}(\mathbf{x}) =
\begin{cases}
7\bigl|\bm{\zeta}/\eta\bigr|^6
- 15\bigl|\bm{\zeta}/\eta\bigr|^4
+ 9\bigl|\bm{\zeta}/\eta\bigr|^2,
& \bigl|\bm{\zeta}/\eta\bigr| \le 1, \\
1, & \bigl|\bm{\zeta}/\eta\bigr| > 1,
\end{cases}
\label{eq:f_epsilon}
\end{equation}
with
\begin{equation}
\bm{\zeta} = \cfrac{\Psi(\mathbf{x})\nabla\Psi(\mathbf{x})}{|\nabla\Psi(\mathbf{x})|^2}.
\label{eq:zeta_nodal_distance}
\end{equation}
Here, $\nabla$ denotes the many-body gradient, collecting all particle-coordinate derivatives into a single vector. Defining
\begin{equation}
Q_{i,k} \equiv \frac{\partial}{\partial r_{i,k}} \Psi(\mathbf{x}),
\label{eq:grad_psi_matrix}
\end{equation}
one may regard $\mathbf{Q}$ as an $(N_e,3)$ matrix. Then,
\begin{equation}
|\nabla\Psi(\mathbf{x})|^2 = \|\mathbf{Q}\|_F^2,
\label{eq:grad_psi_norm}
\end{equation}
and
\begin{equation}
\bigl|\bm{\zeta}/\eta\bigr|
=
\frac{|\Psi(\mathbf{x})|}{\eta \,\|\mathbf{Q}\|_F},
\label{eq:zeta_eta_ratio}
\end{equation}
where $\|\cdot\|_F$ denotes the Frobenius norm. The quantity $\bm{\zeta}$ represents the normal distance between $\mathbf{x}$ and the nearest nodal point of $\Psi$, while $\eta$ controls the distance from the nodal surface within which the regularization is applied.

Each expectation values entering the total force are regularized as follows:
\begin{equation}
{F}^{\rm HF}_{\rm mixed} = -\braket{\partial E^R \cdot R_\eta(\mathbf{x})}_{\Psi_{\rm T}\Psi_0},
\label{eq:force_lrdmc_hf_reg}
\end{equation}
\begin{equation}
{F}^{\rm Pulay}_{\rm mixed} = -2\braket{(E_{\rm L} - E)\,\partial \Psi^R \cdot R_\eta(\mathbf{x})}_{\Psi_{\rm T}\Psi_0},
\label{eq:force_lrdmc_pulay_reg}
\end{equation}
where $R_\eta(\mathbf{x})$ is the regularization function defined in Eq.~(\ref{eq:f_epsilon}), which smoothly suppresses contributions from walkers within a distance $\eta$ of the nodal surface.

%

\section{Optimization of WF\texorpdfstring{\MakeLowercase{s}}{s}}
\label{sec:opt_wf}
Optimizing the variational parameters of the many-body wave function is the most challenging task in \emph{ab initio} QMC calculations. Since the QMC energy (and sometimes the variance) is evaluated stochastically and is strongly nonlinear in the variational parameters, a straightforward optimization method such as gradient descent does not work and is extremely inefficient. The wave-function optimization in the context of \emph{ab initio} QMC has been studied for a long time. Stochastic Reconfiguration (SR) {~\cite{1998SOR}} developed by Sorella, the Linear method (LM) {~{\cite{2007_Umrigar_LM}}} developed by Umrigar et al., and their variants {~{\cite{2007_Umrigar_LM, 2007_Toulouse_LM, 2007_Sorella_adaptive_SR}}} are the most widely adopted optimization methods in the QMC community. jQMC implements both of them.

\vspace{2mm}
In both SR and LR, the {\it metric} for the parameter space (${\bm{\mathcal{S}}}$), which measures the distance of the underlying normalized wavefunction~{\cite{2012MAZ}}, plays a central role.
The normalized wavefunction at the given variational parameters $\alpha$ is defined as
\begin{equation}
\ket{\tilde{\Psi}^{\alpha}}=\ket{\Psi^\alpha}/\sqrt{\braket{\Psi^\alpha|\Psi^\alpha}},
\label{eq:normalized_wavefunction}
\end{equation}
while its derivatives with respect to the variational parameters define
a \emph{semi-orthogonal} basis set:
\begin{equation}
\ket{\tilde{\Psi}_{k}^{\alpha}}
=\frac{\partial}{\partial \alpha_k}\ket{\tilde{\Psi}^{\alpha}}
=(\hat{O}_k-\bar{O}_k)\ket{\tilde{\Psi}^{\alpha}},
\qquad k=1,\ldots,P,
\label{eq:lm_semi_orthogonal_basis}
\end{equation}
where $P$ is the total number of variational parameters, $\hat{O}_k$ is the logarithmic derivative operator, whose
sample-wise representation, introduced in Sec.~\ref{subsec:derivatives_with_respect_to_variational}, is $O_k(\mathbf{x})=\partial\log\Psi^\alpha(\mathbf{x})/\partial\alpha_k$, 
and $\bar{O}_k=\braket{\tilde{\Psi}^{\alpha}|\hat{O}_k|\tilde{\Psi}^{\alpha}}$
is its expectation value.
Then, the {\it metric} for the parameter space (${\bm{\mathcal{S}}}$), which is based on the overlap matrix between the  \emph{semi-orthogonal} basis components, is computed as:
\begin{equation}
\mathcal{S}_{k, k'} = \braket{\tilde{\Psi}_{k}^{\alpha} | \tilde{\Psi}_{k'}^{\alpha}} \equiv \braket{\tilde{\Psi}^{\alpha} |(\hat{O}_k-\bar{O}_k) (\hat{O}_{k'}-\bar{O}_{k'}) | \tilde{\Psi}^{\alpha}} = \frac{\braket{{\Psi}^{\alpha} |(\hat{O}_k-\bar{O}_k) (\hat{O}_{k'}-\bar{O}_{k'}) | {\Psi}^{\alpha}}}{\braket{{\Psi}^{\alpha}|{\Psi}^{\alpha}}}.
\label{eq:S_metrix}
\end{equation}

\subsection{Stochastic Reconfiguration (a.k.a. natural gradient optimization)}
\label{subsec:stochastic_reconfiguration_a_k_a_natural}
SR is a nonlinear optimization method for many-body wave functions developed by Sorella\cite{1998SOR}. It uses a positive-definite preconditioning matrix ${\bm{\mathcal{S}}}$ which represents the metric of the parameter space (Eq.~\ref{eq:S_metrix}), and the generalized force vector $\mathbf{f}$ (Eq.~\ref{eq_general_force}), such that during an optimization step the WF parameters are updated as follows: 
\begin{equation}\label{eq:SRupdate}
{\alpha_k} \to {\alpha_k} + \Delta  \cdot {\left( {{{\bm{\mathcal{S}}}^{ - 1}}{\mathbf{f}}} \right)_k}.
\end{equation}
%
The matrix ${\bm{\mathcal{S}}}$ is stochastically evaluated using $M$ samples ${\mathbf{x}} = \left\{ {{{\mathbf{x}}_1},{{\mathbf{x}}_2}, \ldots {{\mathbf{x}}_M}} \right\}$ (for the simple MCMC) with weights ${w} = \left\{ w_1, w_2, \ldots w_M \right\}$ (for the weighted MCMC):
\begin{equation}
{{\mathcal{S}}_{k,k'}} = \frac{{\sum\limits_{i = 1}^M w_{i} {\left( {{O_k}\left( {{{\mathbf{x}}_i}} \right) - {{\bar O}_k}} \right) \left( {{O_{k'}}\left( {{{\mathbf{x}}_i}} \right) - {{\bar O}_{k'}}} \right)} }}{\sum_{i=1}^{M} w_i},
\label{eq:sr_matrix}
\end{equation}
where ${O_k}\left( {{{\mathbf{x}}_i}} \right) = \cfrac{\partial}{{\partial {\alpha_k}}} {\log \Psi^{\alpha} \left( {{{\mathbf{x}}_i}} \right)}$, ${{\bar O}_k} = \sum\limits_{i = 1}^M {w_i {O_k}\left( {{{\mathbf{x}}_i}} \right)}/\sum\limits_{i = 1}^M w_i$, and $\mathbf{f}$ is computed as in Eq.~\ref{eq:generalized_force_mc}
of Sec.~\ref{subsec:derivatives_with_respect_to_variational}.
We notice that the natural gradient method{~\cite{1998AMA}} developed by Amari et al. in a different context and used for machine learning in recent years has a direct correspondence with the SR.

\vspace{2mm}
The native SR method often suffers from numerical instabilities because the statistical noise makes the matrix ${\bm{\mathcal{S}}}$ ill-conditioned{~\cite{2007_Sorella_adaptive_SR}}. jQMC mitigates the problem by the following numerical technique. First, to simplify the computation~{\cite{rende2024simple}} and to stabilize the solution, jQMC introduces the quantities
\begin{equation}
X_{k, i}  \equiv X_{k}(x_i) = \frac{ \sqrt{w_i} (O_{k}(x_i) - \bar{O}_{k})}{\sqrt{\sum_{i = 1}^M w_i}},
\label{eq:sr_X_matrix}
\end{equation}
and
\begin{equation}
e_{i} \equiv e(x_i) = \frac{ -2 \sqrt{w_i} ({e_L}(x_i) - {\bar{e}_{L}} )}{\sqrt{\sum_{i = 1}^M w_i}},
\label{eq:sr_e_vector}
\end{equation}
which give $\bm{\mathcal{S}} = X X^T$ and $\mathbf{f} = X e$. To make $\bm{\mathcal{S}}$ scale-invariant~{\cite{2017BEC}}, $X$ is rescaled row-wise by the inverse square root of the diagonal of $\bm{\mathcal{S}}$:
\begin{equation}
X_{k, i} \leftarrow X_{k, i} /\sqrt{\mathcal{S}_{k, k}}.
\label{eq:sr_X_normalization}
\end{equation}
Notice that $e$ does not need to be rescaled separately, since the rescaling of $\mathbf{f}$ is considered via $\mathbf{f} = X e$. In this normalized basis, the regularized SR update reads
\begin{equation}
\delta \theta = \tau (X X^T + \varepsilon_{\rm SR} I_P)^{-1} X e,
\label{eq:sr_delta_theta}
\end{equation}
where $P$ is the total number of variational parameters. After solving for $\delta\theta$ in the normalized basis, the natural-gradient direction is rescaled back via $\delta\theta_k \leftarrow \delta\theta_k / \sqrt{\mathcal{S}_{k, k}}$.

\vspace{2mm}
The above reformulation also enables the use of the well-known push-through identity technique~{\cite{henderson1981deriving, Petersen2012matrixcookbook}}, namely $(AB+\lambda I)^{-1}A = A(BA+\lambda I)^{-1}$:
\begin{equation}
\delta \theta = \tau X(X^T X + \varepsilon_{\rm SR} I_{M})^{-1} e,
\label{eq:sr_push_through}
\end{equation}
which is more efficient when the number of samples $M$ is smaller than the number of parameters $P$.

\vspace{2mm}
Either with or without the push-through identity technique, jQMC does not compute the inverse of the matrix $XX^T$ or $X^TX$. Instead, it solves the inversion problem using the conjugate gradient (CG) method. Let each MPI rank hold a local block $X^{(d)} \in \mathbb{R}^{P \times M_d}$, where $P$ is the number of variational parameters and $M_d$ is the number of MCMC samples held by rank $d$. The global $X$ is then
$
X = \bigl[\,X^{(1)},\,X^{(2)},\,\dots,\,X^{(D)}\bigr]\in\mathbb{R}^{P\times M},\quad
M=\sum_{d=1}^{D} M_d.
$
In the CG mode, jQMC never assembles $\bm{\mathcal{S}} \equiv X X^T$ $\in\mathbb{R}^{P\times P}$ or $X^T X$ $\in\mathbb{R}^{M\times M}$; it only stores the distributed blocks $X^{(d)}$ and auxiliary vectors. More specifically, jQMC implements a matrix-free CG method to solve
$
\bigl(\bm{\mathcal{S}} + \varepsilon_{\rm SR} I_P\bigr)\,\theta = \mathbf{f}
$
(or its dual form), where each rank $d$ independently computes only its local matrix-vector contributions.
Indeed, before starting the CG iterations, the right-hand side is constructed as $\mathbf{f}=\sum_{d=1}^{D}X^{(d)}e^{(d)}$, where all ranks first compute their local contributions ($X^{(d)}e^{(d)}$), summed them up using a global all-reduce.
Then, during the CG iterations, each rank repeatedly computes its local contribution,
$
v \mapsto X^{(d)}\left(X^{(d)T}v\right),
$
and a global all-reduce is then used to obtain
$
v \mapsto
\sum_{d=1}^{D}
X^{(d)}\left(X^{(d)T}v\right)
+
\varepsilon_{\rm SR}v.
$
This naturally distributes both memory and compute workloads among CPUs or GPUs, yielding excellent weak-scaling behavior without storing the full SR matrix, as demonstrated in Sec.~{\ref{sec:weak-scaling-benchmark}}.

\vspace{2mm}
In practice, during an optimization, jQMC monitors the variational energy ($E$) and the maximum value of the signal to noise ratio among all  the force components:
\begin{equation} \label{eq:devmax}
\verb| signal-to-noise ratio| \equiv \max_k \left( {\left| {\frac{{{f_k}}}{{{\sigma _{{f_k}}}}}} \right|} \right)
\end{equation}
where  ${\sigma _{{f_k}}}$ represents the estimated error bar of the force ${f_k}$. This value  is used in jQMC as one of the convergence criteria of the WF optimization.

\subsection{The Linear method and its variants}
\label{subsec:linear_method}
The LM starts from the linear expansion of the normalized variational wave function with respect to the variational parameters $\bm{\alpha}$:
\begin{equation}
\ket{\tilde{\Psi}_{0}^{\alpha+\delta\bm{\alpha}}}
=
\ket{\tilde{\Psi}_{0}^{\alpha}}
+
\sum_{k=1}^{p}\delta\alpha_k\ket{\tilde{\Psi}_{k}^{\alpha}}
\equiv
\sum_{k=0}^{p} z_k \ket{\tilde{\Psi}_{k}^{\alpha}},
\label{eq:lm_linear_expansion}
\end{equation}
where $z_0=1$ and $z_k=\delta\alpha_k$ for $k\ge 1$. Here, $\ket{\tilde{\Psi}_{0,\alpha}} \equiv \ket{\tilde{\Psi}_{\alpha}}$ is the normalized wavefunction defined in Eq.~{\ref{eq:normalized_wavefunction}}, and $\ket{\tilde{\Psi}_{k,\alpha}}$ is the \emph{semi-orthogonal} basis defined in Eq.~{\ref{eq:lm_semi_orthogonal_basis}}.
By construction, this basis satisfies
$\braket{\tilde{\Psi}_{0}^{\alpha}|\tilde{\Psi}_{k}^{\alpha}}=0$ for all
$k\ge 1$.
Using this expansion, the variational energy is written as
\begin{equation}
E_{\alpha+\delta\bm{\alpha}}
=
\frac{\braket{\tilde{\Psi}_{0}^{\alpha+\delta\bm{\alpha}}|\hat{H}|\tilde{\Psi}_{0}^{\alpha+\delta\bm{\alpha}}}}
{\braket{\tilde{\Psi}_{0}^{\alpha+\delta\bm{\alpha}}|\tilde{\Psi}_{0}^{\alpha+\delta\bm{\alpha}}}}
=
\frac{\sum_{k,k'=0}^{p} z_k z_{k'} \braket{\tilde{\Psi}_{k}^{\alpha}|\hat{H}|\tilde{\Psi}_{k'}^{\alpha}}}
{\sum_{k,k'=0}^{p} z_k z_{k'} \braket{\tilde{\Psi}_{k}^{\alpha}|\tilde{\Psi}_{k'}^{\alpha}}}.
\label{eq:lm_energy_functional}
\end{equation}
Stationarity with respect to $z_l$ leads to the generalized eigenvalue problem
\begin{equation}
\mathbf{H}\mathbf{z}=E\bm{\mathcal{S}}\mathbf{z},
\label{eq:lm_generalized_evp}
\end{equation}
with
\begin{equation}
H_{k,k'}=\braket{\tilde{\Psi}_{k}^{\alpha}|\hat{H}|\tilde{\Psi}_{k'}^{\alpha}},
\qquad
\mathcal{S}_{k,k'}=\braket{\tilde{\Psi}_{k}^{\alpha}|\tilde{\Psi}_{k'}^{\alpha}}.
\label{eq:lm_matrix_elements}
\end{equation}
The matrix elements can be evaluated efficiently by Monte Carlo sampling over $\pi(\mathbf{x})\propto|\Psi^\alpha(\mathbf{x})|^2$. In particular,
\begin{equation}
\braket{\tilde{\Psi}_{k}^{\alpha}|\hat{H}|\tilde{\Psi}_{k'}^{\alpha}}
=
\left\langle
(O_k(\mathbf{x})-\bar{O}_k)\,W_{k'}(\mathbf{x})
\right\rangle_{\pi(\mathbf{x})},
\label{eq:lm_H_mc}
\end{equation}
where
\begin{equation}
W_k(\mathbf{x})
=
\frac{\braket{x|\hat{H}(\hat{O}_k-\bar{O}_k)|\Psi^\alpha}}{\braket{x|\Psi^\alpha}}.
\label{eq:lm_W_function}
\end{equation}
This quantity is related to the derivatives of the local energy:
\begin{equation}
\frac{\partial E_{\rm L}(\mathbf{x})}{\partial \alpha_k}
=
\frac{\braket{x|\hat{H}(\hat{O}_k-\bar{O}_k)|\tilde{\Psi}_{0,\alpha}}}{\braket{x|\tilde{\Psi}_{0,\alpha}}}
-
E_{\rm L}(\mathbf{x})(O_k(\mathbf{x})-\bar{O}_k),
\label{eq:local_energy_deriv}
\end{equation}
so that
\begin{equation}
W_k(\mathbf{x})
=
\frac{\partial E_{\rm L}(\mathbf{x})}{\partial \alpha_k}
+
E_{\rm L}(\mathbf{x})(O_k(\mathbf{x})-\bar{O}_k).
\label{eq:lm_W_expression}
\end{equation}
We notice that if the local energy is real, then $\left\langle \partial E_{\rm L}(\mathbf{x})/\partial\alpha_k \right\rangle = 0$ (see Ref.~\onlinecite{2005_Sorella_LM}).
Substituting Eq.~(\ref{eq:lm_W_expression}) into Eq.~(\ref{eq:lm_H_mc}), the LM Hamiltonian matrix is naturally decomposed as $\mathbf{H} = \mathbf{K} + \mathbf{B}$, with
\begin{align}
K_{k,k'} &= \langle\, \delta O_k(\mathbf{x})\,\delta O_{k'}(\mathbf{x})\,E_{\rm L}(\mathbf{x})\,\rangle_{\pi(\mathbf{x})}, \label{eq:lm_K}\\
B_{k,k'} &= \langle\, \delta O_k(\mathbf{x})\,\frac{\partial E_{\rm L}(\mathbf{x})}{\partial \alpha_{k'}}\,\rangle_{\pi(\mathbf{x})}, \label{eq:lm_B}
\end{align}
where $\delta O_k(\mathbf{x}) = O_k(\mathbf{x}) - \bar O_k$. The $K$ matrix is symmetric in $k$ and $k'$, whereas $B$ inherits the asymmetry of the $W$ derivation. Since the exact $\mathbf{H}$ is hermitian (symmetric for the real wave functions considered in this study), $\mathbf{K} + \mathbf{B}$ is symmetrized post-hoc as $\mathbf{H} \leftarrow (\mathbf{H} + \mathbf{H}^T)/2$ in jQMC to avoid the numerical instaibility.

\vspace{2mm}
We note that the elements of the LM overlap ($\mathcal{S}$) and Hamiltonian ($H$) matrices involving the zeroth basis state are given as follows. Since the derivative basis is semi-orthogonal to the current normalized wave function, we have
$\mathcal{S}_{00}=1$ and $\mathcal{S}_{0k}=\mathcal{S}_{k0}=0 \quad (k\ge 1)$.
The corresponding Hamiltonian matrix elements are
$
H_{00}
=
\braket{\tilde{\Psi}_{0}^{\alpha}|\hat{H}|\tilde{\Psi}_{0}^{\alpha}}
=
E_{\alpha},
$
$
H_{k0}
=
\braket{\tilde{\Psi}_{k}^{\alpha}|\hat{H}|\tilde{\Psi}_{0}^{\alpha}}
=
\left\langle
\delta O_k(\mathbf{x}) E_{\rm L}(\mathbf{x})
\right\rangle_{\pi(\mathbf{x})}
=
-\cfrac{1}{2}f_k$, and
$
H_{0k}
=
\braket{\tilde{\Psi}_{0}^{\alpha}|\hat{H}|\tilde{\Psi}_{k}^{\alpha}}
=
\left\langle
W_k(\mathbf{x})
\right\rangle_{\pi(\mathbf{x})}
=
-\cfrac{1}{2}f_k$.
where $f_k$ (for $k \ge 1$) is the same as in Eq.~\ref{eq_general_force}.

\vspace{2mm}
The original LM optimizes all $P$ parameters simultaneously by solving the full $(P+1)\times(P+1)$ generalized eigenvalue problem~{\cite{2005_Sorella_LM, 2007_Umrigar_LM, 2007_Toulouse_LM}}. In practice, however, this can be numerically unstable and computationally expensive when many parameters with very different scales are present. For this reason, jQMC uses two practical variants based on the SR direction $g_k=(\bm{\mathcal{S}}^{-1}\mathbf{f})_k$, inspired by the TurboRVB's implementation~{\cite{2007_Sorella_adaptive_SR}}.

\subsubsection{Linear method with only the SR direction}
\label{subsec:practical_asr}
The first practical variant addresses a key limitation of the original LM formalism, which requires solving a generalized eigenvalue problem with all $P$ variational parameters. This variant keeps only the SR direction and removes all individual parameter directions. In this case, the LM update is constrained to
\begin{equation}
\delta\alpha_k=\gamma g_k,
\qquad
g_k=(\bm{\mathcal{S}}^{-1}\mathbf{f})_k,
\label{eq:asr_delta_alpha}
\end{equation}
so that the optimization reduces to a single scalar $\gamma$. The energy becomes
\begin{equation}
E_{\alpha+\delta\bm{\alpha}}
=
\frac{H_0+2\gamma H_1+\gamma^2 H_2}{1+\gamma^2 S_2},
\label{eq:asr_energy_gamma}
\end{equation}
with
\begin{equation}
H_0 = E_\alpha,
\qquad
H_1 = -\frac{1}{2}\mathbf{g}^T\mathbf{f},
\qquad
H_2 = \mathbf{g}^T (\mathbf{K} + \mathbf{B})\,\mathbf{g},
\qquad
S_2 = \mathbf{g}^T \bm{\mathcal{S}}\,\mathbf{g}.
\label{eq:asr_coefficients}
\end{equation}
Minimizing $E(\gamma)$ yields a quadratic equation,
\begin{equation}
H_1S_2\gamma^2 + (H_0S_2-H_2)\gamma - H_1 = 0,
\label{eq:asr_quadratic}
\end{equation}
whose solution is
\begin{equation}
\gamma
=
\frac{H_2-H_0S_2 \pm \sqrt{(H_2-H_0S_2)^2 + 4H_1^2S_2}}{2H_1S_2}.
\label{eq:asr_gamma_solution}
\end{equation}
In jQMC, the root with the smaller absolute value is selected in order to avoid spuriously large steps.
The final parameter update is simply
\begin{equation}
\delta\alpha_k=\gamma g_k.
\label{eq:asr_final_update}
\end{equation}

\subsubsection{Linear method with the SR collective variable and selected parameters}
\label{subsec:lm_sr_collective}
For a more efficient approach, we combine a single \emph{global SR collective mode}, which considers the contribution of all parameters through the SR matrix $\bm{\mathcal{S}}^{-1}$, combined with a small set of \emph{individually resolved directions} corresponding to the parameters whose generalized forces have large signal-to-noise ratios. Thus, the SR direction provides a global update for all parameters, while the selected directions based on the signal-to-ratios allow the LM to assign individually optimized steps to the few chosen parameters that should be far from their local minima. Specifically, we define the SR collective state
\begin{equation}
|\tilde{\Psi}_{\mathrm{SR}}\rangle
=
\sum_{k=1}^P g_k\,|\tilde{\Psi}_{k}^{\alpha}\rangle,
\qquad
g_k=(\bm{\mathcal{S}}^{-1}\mathbf{f})_k,
\label{eq:lm_sr_collective}
\end{equation}
where the sum runs over all variational parameters. The SR direction provides a well-conditioned collective mode that automatically incorporates the scale information encoded in the overlap matrix.
We then diagonalize the LM in the reduced subspace
\begin{equation}
\left\{
|\tilde{\Psi}_{0}^{\alpha}\rangle,\,
|\tilde{\Psi}_{\mathrm{SR}}\rangle,\,
|\tilde{\Psi}_{k_1}^{\alpha}\rangle,\,
\ldots,\,
|\tilde{\Psi}_{k_m}^{\alpha}\rangle
\right\},
\label{eq:lm_reduced_subspace}
\end{equation}
where $k_1,\ldots,k_m$ denote a small set of selected parameters. In jQMC, these are chosen as the parameters with the largest signal-to-noise ratio of the generalized force $f_k$, so that noisy directions are excluded from the explicit LM subspace.
The generalized eigenvalue problem retains the same form, as in Eq.~{\ref{eq:lm_generalized_evp}}:
\begin{equation}
\bar{H}\mathbf{v}=E\bar{\bm{\mathcal{S}}}\mathbf{v},
\label{eq:lm_reduced_evp}
\end{equation}
but now in a reduced $(m+2)$-dimensional basis. The matrix elements involving the SR collective variable follow from linearity:
\begin{equation}
\bar{H}_{0,\mathrm{SR}}=-\frac{1}{2}\mathbf{g}^T\mathbf{f},
\qquad
\bar{H}_{\mathrm{SR},\mathrm{SR}}=\mathbf{g}^T (\mathbf{K} + \mathbf{B})\,\mathbf{g},
\qquad
\bar{\mathcal{S}}_{\mathrm{SR},\mathrm{SR}}=\mathbf{g}^T \bm{\mathcal{S}}\,\mathbf{g},
\label{eq:lm_sr_matrix_elements}
\end{equation}
and similarly for the mixed SR-selected-parameter blocks.
The final update is written as
\begin{equation}
\delta\alpha_k
=
\frac{v_{\mathrm{SR}}}{v_0}g_k
+
\begin{cases}
v_{k_i}/v_0, & k=k_i\in\{k_1,\ldots,k_m\},\\
0, & \text{otherwise}.
\end{cases}
\label{eq:lm_sr_update}
\end{equation}
Thus, this variant supplements the global SR direction by a few individually optimized corrections.

\subsection{Optimization of molecular orbital coefficients with constraints}
\label{subsec:constrained_mo}

In jQMC, the nodal surface of the wave function can be optimized via the determinant part, namely, the geminal matrix $\Lambda_{l,l'}^{\uparrow,\downarrow}$ [Eq.~(\ref{eq:agp_pairing_expansion})]. The most straightforward strategy is an unconstrained optimization, that is, optimizing all independent variational parameters in the geminal matrix subject only to the required symmetry. jQMC implements this simplest scheme. In addition, jQMC also implements the molecular-orbital-constrained optimization introduced in TurboRVB~\cite{2009_Marchi_MOopt, 2015_Dupuy_MOopt, 2017BEC}. According to this scheme, one can optimize the geminal matrix $\Lambda^{\uparrow\downarrow}$ expanded over the atomic orbitals (Eq.~\ref{eq:paired_block_matrix_product}) by preserving a targeted rank of the matrix, chosen in input. For AGP wave functions, this approach makes it possible to introduce an effective coherence-length cutoff in the AGP matrix~\cite{2017BEC}, for both insulators and metals, because the AO representation bears information on the correlation length of the geminal, through the ionic centers of the AO basis. 
Moreover, the AGP expanded over atomic orbitals has an equivalent natural orbital representation in its AGPn or SD form, expanded over molecular orbitals. Therefore, by constraining the rank of $\Lambda$, the optimization can be carried out with a fixed number of molecular orbitals retained in the projection, denoted by $n$ ($n=N_e/2$ for SD and $n>N_e/2$ for AGPn), while exploiting the locality of the geminal built in the corresponding AO expansion. We describe the implementation of the fixed-rank optimization below. We notice that, following the convention stated at the beginning of Sec.~\ref{sec:methods}, all orbitals and the geminal matrix are taken to be real.

\vspace{2mm}
Below, we focus on the paired block $\boldsymbol{\Lambda}^{\uparrow,\downarrow}$ of the geminal matrix; the unpaired block $\boldsymbol{\Lambda}^{\uparrow}$ has no spin-down index and therefore no rank constraint, and is treated separately at the end of this Subsection. To keep the formulas simple, we specialize the derivation to the spin-unpolarized SD reference, where $n = N_e/2 = N_e^{\downarrow}$ and the $n$ retained MOs coincide with the occupied MOs; we therefore use the standard occupied / virtual terminology throughout the rest of the subsection instead of retained / discarded. The same formulas can be applied for AGPn ($n > N_e/2$) with occupied replaced with the retained $n$-dimensional MO subspace, and to the spin-polarized case via the substitutions for spin-dependent MO coefficients.

\vspace{2mm}
As shown in Eq.~(\ref{eq:agpn_pairing}), the AGP pairing function (the spin-$\uparrow,\downarrow$ part) can be represented in the MO basis as:
\begin{equation}
F^{\uparrow,\downarrow}(\mathbf r,\mathbf r')
=
\sum_{\mu=1}^{n}
\lambda_\mu
\Phi_\mu(\mathbf r)
\Phi_\mu(\mathbf r'),
\label{eq:agp_mo}
\end{equation}
where $\{\Phi_\mu\}_{\mu=1}^n$ are the $n$ occupied MOs and $\{\lambda_\mu\}$ the corresponding weights ($\lambda_\mu = 1$ for SD). Expanding the occupied MOs in the AO basis:
\begin{equation}
\Phi_\mu(\mathbf r)
=
\sum_l \tilde{C}_{l,\mu}\psi_l(\mathbf r),
\end{equation}
the pairing function takes the AO basis representation:
\begin{equation}
F^{\uparrow,\downarrow}(\mathbf r,\mathbf r')
=
\sum_{l,l'}
\psi_l(\mathbf r)
\Lambda^{\uparrow,\downarrow}_{l,l'}
\psi_{l'}(\mathbf r'),
\end{equation}
with
\begin{equation}
\boldsymbol{\Lambda}^{\uparrow,\downarrow}
=
\tilde{\mathbf{C}}
\boldsymbol{\lambda}
\tilde{\mathbf{C}}^T,
\label{eq:Lambda_factorization}
\end{equation}
where $\boldsymbol{\lambda} = \mathrm{diag}(\lambda_1,\ldots,\lambda_n)$ is the $(n,n)$ diagonal matrix of weights and $\tilde{\mathbf{C}}$ is the $(M_{ao},n)$ matrix of occupied MO coefficients; thus, $rank(\boldsymbol{\Lambda}^{\uparrow,\downarrow}) \le n$. The columns of $\tilde{\mathbf{C}}$ satisfy $\tilde{\mathbf{C}}^T\mathbf S\tilde{\mathbf{C}} = I_n$ with $\mathbf S$ being the AO overlap matrix. We notice that, in this Section, the row and column indices of the MO coefficients $\tilde{\mathbf{C}}$ are flipped compared with $\mathbf{C}$ in Eq.~{\ref{eq:mo_ao_relation}} to follow the same notation as Ref.~{\onlinecite{2017BEC}}.

{\vspace{2mm}}
A linearized rank-preserving variation is obtained by changing only the
quantities that appear in Eq.~(\ref{eq:Lambda_factorization}), namely
\(\tilde{\mathbf{C}}\) and \(\boldsymbol{\lambda}\). Under
\(\tilde{\mathbf{C}}\to\tilde{\mathbf{C}}+\delta\tilde{\mathbf{C}}\) and
\(\boldsymbol{\lambda}\to\boldsymbol{\lambda}+\delta\boldsymbol{\lambda}\),
the first-order variation of the geminal matrix is
\begin{equation}
(\delta\boldsymbol{\Lambda}^{\uparrow,\downarrow})^c
=
\delta\tilde{\mathbf{C}}\,\boldsymbol{\lambda}\,\tilde{\mathbf{C}}^T
+
\tilde{\mathbf{C}}\,\delta\boldsymbol{\lambda}\,\tilde{\mathbf{C}}^T
+
\tilde{\mathbf{C}}\,\boldsymbol{\lambda}\,\delta\tilde{\mathbf{C}}^T .
\label{eq:delta_lambda_matrix}
\end{equation}
Here, \(\delta\tilde{\mathbf{C}}\) denotes an arbitrary infinitesimal change of the
occupied MO coefficients.
We now introduce the AO-basis representations of the left and right projectors
associated with the occupied MO subspace:
\begin{align}
\mathbf R &= (\mathbf S\tilde{\mathbf C})\tilde{\mathbf C}^T,
\label{eq:R_projector}\\
\mathbf L &= \mathbf R^T
=
\tilde{\mathbf{C}}(\mathbf S\tilde{\mathbf{C}})^T
=
\tilde{\mathbf{C}}\tilde{\mathbf{C}}^T\mathbf S,
\label{eq:L_projector}
\end{align}
where \(\tilde{\mathbf{C}}\) is the same MO-coefficient matrix appearing in
Eq.~(\ref{eq:Lambda_factorization}), satisfying
\(\tilde{\mathbf{C}}^T\mathbf S\tilde{\mathbf{C}}=I_n\), and \(\mathbf S\) is the AO overlap
matrix.
\(\mathbf L\) acts on MO coefficient vectors from the left and is the
\(\mathbf S\)-orthogonal projector onto the occupied MO subspace, and thus, \((\mathbf I - \mathbf L)\) is the complementary projector onto the virtual MO subspace, in the AO-basis representation~{\cite{2009_Marchi_MOopt}}.
These matrices satisfy
\begin{equation}
\mathbf L^2=\mathbf L,\qquad
\mathbf R^2=\mathbf R,\qquad
(\mathbf I-\mathbf L)\tilde{\mathbf{C}}=0,\qquad
\tilde{\mathbf{C}}^T(\mathbf I-\mathbf R)=0 .
\label{eq:LR_projector_identities}
\end{equation}
The last two identities are the only ingredients needed to identify the
forbidden component of a rank-\(n\) variation. Applying the virtual projector
on both AO indices of Eq.~(\ref{eq:delta_lambda_matrix}) gives
\begin{align}
&
(\mathbf I-\mathbf L)
(\delta\boldsymbol{\Lambda}^{\uparrow,\downarrow})^c
(\mathbf I-\mathbf R)
\nonumber\\
&=
(\mathbf I-\mathbf L)\delta\tilde{\mathbf C}\,\boldsymbol{\lambda}\,
\underbrace{\tilde{\mathbf{C}}^T(\mathbf I-\mathbf R)}_{=0}
+
\underbrace{(\mathbf I-\mathbf L)\tilde{\mathbf{C}}}_{=0}\,
\delta\boldsymbol{\lambda}\,
\underbrace{\tilde{\mathbf{C}}^T(\mathbf I-\mathbf R)}_{=0}
+
\underbrace{(\mathbf I-\mathbf L)\tilde{\mathbf{C}}}_{=0}\,
\boldsymbol{\lambda}\,\delta\tilde{\mathbf{C}}^T(\mathbf I-\mathbf R)
\nonumber\\
&=0 .
\end{align}
Thus, every first-order variation generated within the factorized rank-\(n\) form has no
virtual--virtual component:
\begin{equation}
(\mathbf I-\mathbf L)
(\delta\boldsymbol{\Lambda}^{\uparrow,\downarrow})^c
(\mathbf I-\mathbf R)
=
0.
\label{eq:proj_constraint_matrix}
\end{equation}
Equivalently, the virtual--virtual block is absent from the tangent space of
the rank-\(n\) AGP/SD ansatz. Therefore, the virtual--virtual component is the
only component that must be removed from an arbitrary AO-basis variation in
order to impose this linearized rank-\(n\) constraint.

\vspace{2mm}
This observation gives a practical way to impose the rank constraint on an arbitrary optimization update. In other words, by eliminating the virtual--virtual component from a variation matrix, one can obtain a linearized rank-constrained update. The forbidden virtual--virtual component is
\begin{equation}
(\delta\boldsymbol{\alpha}^{\uparrow,\downarrow})_{vv}
=
(\mathbf I-\mathbf L)
\delta\boldsymbol{\alpha}^{\uparrow,\downarrow}
(\mathbf I-\mathbf R).
\end{equation}
Therefore, the constrained variation is obtained by subtracting only this forbidden component:
\begin{equation}
(\delta\boldsymbol{\alpha}^{\uparrow,\downarrow})^c
=
\delta\boldsymbol{\alpha}^{\uparrow,\downarrow}
-
(\mathbf I-\mathbf L)
\delta\boldsymbol{\alpha}^{\uparrow,\downarrow}
(\mathbf I-\mathbf R).
\label{proPf}
\end{equation}
This projection leaves the occupied--occupied, occupied--virtual, and virtual--occupied components unchanged, while it removes the virtual--virtual component. Therefore, $(\delta\boldsymbol{\alpha}^{\uparrow,\downarrow})^c$ satisfies Eq.~(\ref{eq:proj_constraint_matrix}) and keeps the rank-$n$ AGP ansatz.

\vspace{2mm}
In this Subsection, $\delta\boldsymbol{\alpha}^{\uparrow,\downarrow}$ denotes the matrix-valued variation of the paired geminal matrix $\boldsymbol{\Lambda}^{\uparrow,\downarrow}$ in the AO basis, with matrix elements $\alpha^{\uparrow,\downarrow}_{l,l'}$ (dimension $(M_{ao}, M_{ao})$). In addition, the per-sample logarithmic derivative for paired-block parameters is treated as a matrix $\mathbf{O}^{\rm paired}$ with entries $(\mathbf{O}^{\rm paired})_{l,l'} = \partial\log\Psi/\partial\alpha^{\uparrow,\downarrow}_{l,l'}$. This differs from the 1D flattened index $k$ used in Secs.~\ref{subsec:stochastic_reconfiguration_a_k_a_natural} and \ref{subsec:linear_method}, where $\alpha_k$ and $O_k$ are scalars; the two views are related by unfolding the 1D index $k$ into the matrix index pair $(l,l')$. The jQMC implementation matches this picture: the per-sample $O$-matrix is held with a 1D parameter axis, and only the lambda block is temporarily reshaped to 2D for the projection.

\vspace{2mm}
In practice, the rank constraint of Eq.~(\ref{proPf}) is enforced not only directly on the final parameter update but also upstream on the per-sample logarithmic-derivative matrix $\mathbf{O}^{\rm paired}$, from which the generalized force, the SR matrix, and the LM matrices are constructed~{\cite{2009_Marchi_MOopt, 2017BEC}}. Specifically, the paired observable is replaced by
\begin{equation}
\tilde{\mathbf{O}}^{\rm paired} \;=\; \mathbf{O}^{\rm paired} \;-\; (\mathbf{I}-\mathbf{L}^T)\,\mathbf{O}^{\rm paired}\,(\mathbf{I}-\mathbf{R}^T)
\label{eq:O_paired_projection}
\end{equation}
before any downstream optimization quantity is formed~{\cite{2009_Marchi_MOopt, 2017BEC}}.
After solving the SR/LM equations, the paired update is projected once more via Eq.~(\ref{proPf}) to exactly fulfill the rank constraint. 

\vspace{2mm}
For spin-polarized systems with unpaired electrons, the geminal matrix in the AO-basis representation is composed of the spin-paired and spin-unpaired blocks (Eq.~{\ref{eq:agp_augmented_pairing_matrix}}). Since the rank constraint originates from the paired block, $F^{\uparrow,\downarrow}$, the optimization constraint should be applied only to the paired geminal block $\boldsymbol{\Lambda}^{\uparrow,\downarrow}$. The unpaired block $\boldsymbol{\Lambda}^{\uparrow}$ carries no spin-down index and possesses no virtual--virtual subblock; neither the corresponding logarithmic derivative $\mathbf{O}^{\rm unpaired}$ nor the unpaired parameter update $\delta\boldsymbol{\alpha}^{\uparrow}$ is therefore subjected to any projection, and the unpaired part is optimized exactly as in the unconstrained SR/LM scheme of Secs.~\ref{subsec:stochastic_reconfiguration_a_k_a_natural} and~\ref{subsec:linear_method}.

\section{Implementation details}
\label{sec:implementations}
\begin{figure*}
    \centering
    \includegraphics[width=1.0\linewidth]{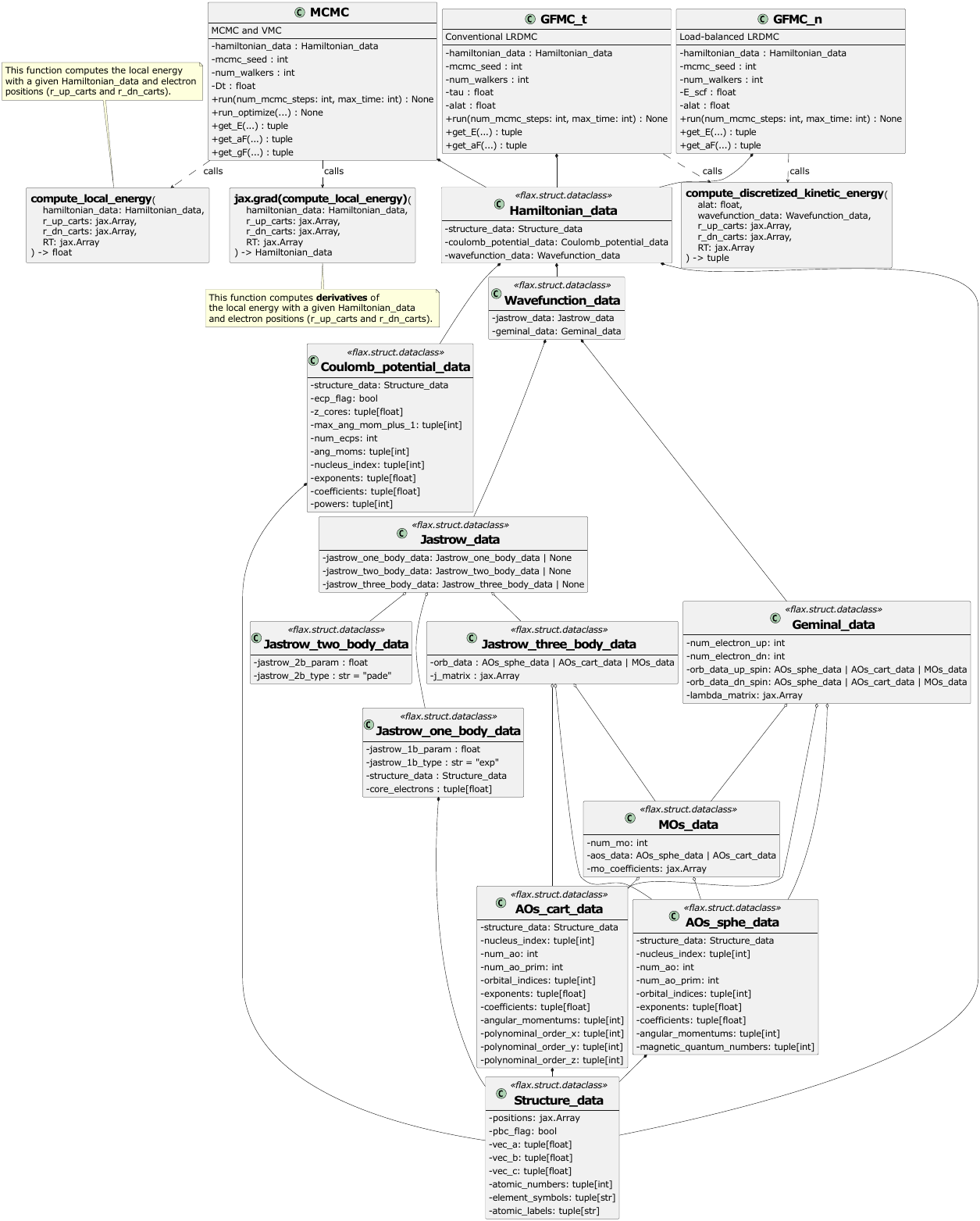}
    \caption[]{The diagram of the main jQMC data classes. Each box represents a data class. The solid arrows demonstrate relations among the data classes. The broken arrows link a data class with calling functions.}
    \label{fig:UML-diagram}
\end{figure*}

jQMC clearly separates \textit{data} and \textit{algorithms}. Data structures are implemented using Python \texttt{dataclass}. More specifically, we take the \texttt{struct} class from the \texttt{flax} library, which is compatible with JAX's \texttt{jit} and  \texttt{grad}. Algorithms are then defined as functions that take these data class instances, along with electron coordinates and control parameters.
An overview of the data classes and algorithms used in jQMC is provided in Table~{\ref{main_classes}}, and the dependency relationships among the data classes are depicted as a diagram in Figure~{\ref{fig:UML-diagram}}.

\begin{center}
\begin{table*}[hbtp!]
\caption{\label{main_classes} The main dataclasses and classes implemented in jQMC.}
\vspace{1mm}
\begin{tabular}{c|c|p{10cm}}
\Hline 
Type & Module/Class & Description \\ 
\Hline 
\multirow{11}{*}{Dataclass} 
& \texttt{Hamiltonian\_data} 
& Holds all information needed to evaluate the Hamiltonian, including structure, Coulomb potential, and wave function parameters. \\ 
\cline{2-3}
& \texttt{Structure\_data}
& Stores atomic positions in 3D space, atomic numbers, element symbols, and textual labels for each atom. \\ 
\cline{2-3}
& \texttt{AOs\_cart\_data} 
& Contains Cartesian atomic-orbital basis details such as the number of orbitals, contraction exponents and coefficients, angular momenta, and polynomial orders for each Cartesian component. \\ 
\cline{2-3}
& \texttt{AOs\_sphe\_data} 
& Contains spherical atomic-orbital basis details including the number of orbitals, contraction exponents and coefficients, angular momenta, and magnetic quantum numbers for each center. \\ 
\cline{2-3}
& \texttt{Coulomb\_potential\_data} 
& Stores Coulomb interaction parameters or effective core potential (ECP) data such as core charges, angular momentum cutoffs, radial exponents, coefficients, and powers for each nucleus. \\ 
\cline{2-3}
& \texttt{MOs\_data} 
& Describes molecular-orbital information by storing the total number of MOs, the underlying atomic-orbital basis reference, and the MO coefficient arrays. \\ 
\cline{2-3}
& \texttt{Geminal\_data}
& Represents the pairing (geminal) matrix and orbitals for spin-up and spin-down electrons along with their respective electron counts. \\ 
\cline{2-3}
& \texttt{Jastrow\_data} 
& Aggregates all Jastrow factor data by combining one-body, two-body, and three-body Jastrow parameter sets. \\ 
\cline{2-3}
& \texttt{wave function\_data}
& Encapsulates the complete wave function description by combining Jastrow data and geminal pairing information. \\ 
\hline 
\multirow{2}{*}{Class} 
& \texttt{MCMC} 
& Implements Markov-Chain Monte Carlo method and manages variational Monte Carlo (WF optimization). \\ 
\cline{2-3}
& \texttt{GFMC} 
& Implements Green Function Monte Carlo method for LRDMC calculations.\\ 
\Hline
\end{tabular}
\end{table*}
\end{center}

{\vspace{2mm}}
In jQMC, all computational kernels are JIT-compiled by JAX. These include the updates of electron positions in MCMC, the evaluation of $e_L$ and $\partial e_L$, and other observables, as well as the updates of electron positions in GFMC. In this sense, the numerically intensive kernels are executed on the GPU, while the CPU is used primarily for compilation and orchestration. The branching step in LRDMC, however, is still performed on CPUs. After the walker weights have been updated, the branching procedure based on these weights is executed on the CPU, because this step involves many conditional operations and also requires complex MPI communication associated with walker replication and deletion.

{\vspace{2mm}}
JAX transfers host data such as NumPy arrays from the CPU to the GPU when they are converted into JAX arrays or supplied to GPU-executed JAX computations. To avoid additional implementation complexity and to minimize repeated host--device transfers, we adopt a design in which the data required for the main numerical kernels is kept on the GPU whenever possible, while the memory demand on the GPU increases because most of the data required for the computation must reside in the GPU memory.

\begin{figure}
    \centering
    \includegraphics[width=0.5\linewidth]{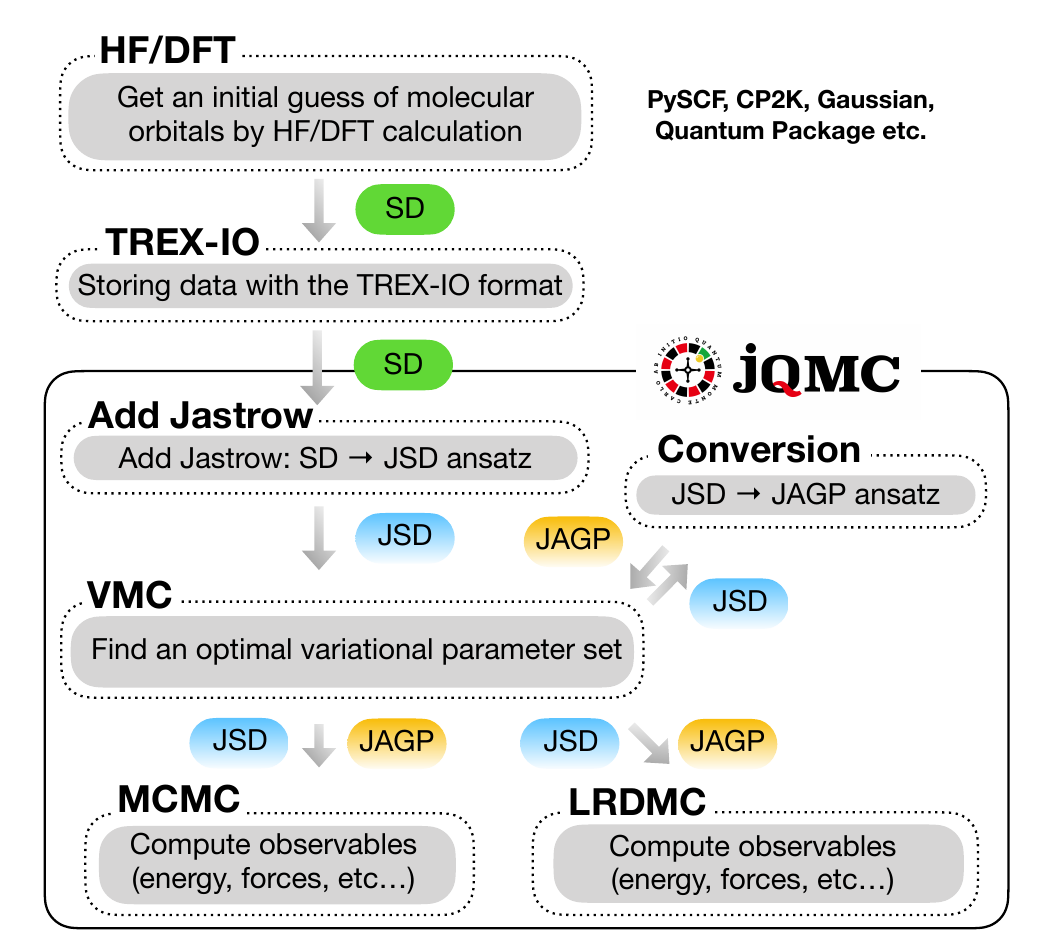}
    \caption[]{A typical workflow of jQMC to perform VMC, MCMC and LRDMC calculations starting from a HF/DFT calculation.}
    \label{fig:workflow}
\end{figure}

{\vspace{2mm}}
A typical workflow for performing LRDMC calculations with jQMC is illustrated in Figure~{\ref{fig:workflow}}. As it is standard in QMC simulations, the process begins by constructing a trial wave function, which typically requires input from HF or DFT calculations. jQMC supports the construction of trial wave functions via the TREX-IO interface, making it easy to import wave function data from many established \emph{ab initio} electronic structure packages, such as PySCF~\cite{2026_Qiming_pyscf}.
Once the trial wave function is imported, one may proceed to optimize the Jastrow factor and, if necessary, the determinant part. After the wave function optimization is completed, one can perform single-point MCMC or LRDMC calculations. In the case of LRDMC, simulations are carried out with several lattice discretizations, and an extrapolation to the zero-lattice-spacing limit (\( a \to 0 \)) is performed to obtain unbiased energy estimates.

\section{Mixed-precision support}
\label{sec:mixed-precision}
jQMC supports mixed-precision calculations. In jQMC, AO (except for gradients and Laplacian calculations) and Jastrow evaluations are performed using 32-bit floating-point (FP32). This works because the rounding errors in the AO evaluation from low-precision calculations are mitigated by the summation performed during MO calculations (in the case of SD) or geminal calculations (in the case of AGP), which are carried out in 64-bit floating-point (FP64). Furthermore, since the Jastrow term involves an exponential ($\exp$) function, its rounding errors from low-precision calculations are smoothed out by the $\exp$ function. In particular, using FP32 in AO evaluations yields a significant speed-up because many computationally intensive exponential calculations are required for GTO evaluations, and single-precision transcendental operations such as $\exp$ function are typically executed on special function units (SFUs), which are distinct from the main FP32/FP64 arithmetic pipelines~{\cite{NVIDIA_CUDA_Math_API_Single_2026}}.
Comparison of FP32/FP64 (mixed) and FP64 (full) calculations for water clusters is listed in Table~{\ref{tab:comparison-mixed-full-precisions}}. The table shows that both calculations yield energies that are consistent within the error bars ($3\sigma$) for all compounds (even when the number of electrons is increased up to 160), and that FP32/FP64 calculations are approximately twice as fast as the full-precision ones. We notice that users can switch between the full and mixed precisions via an input flag (the default is the full precision). In the following sections, full precision is used in Secs.~{\ref{sec:performance-analysis}}, {\ref{sec:weak-scaling-benchmark}}, {\ref{sec:verification}}, and{\ref{sec:demonstrations}}, whereas the mixed precision is used in Secs~{\ref{sec:vectorization}} and {\ref{sec:wall-time-benchmark}}.

\begin{table*}[t]
\centering
\caption{Comparison of MCMC total energies obtained with full (FP64) and mixed (FP32 + FP64) precision. $\Delta E = E_{\mathrm{mixed}} - E_{\mathrm{full}}$, and $\sigma_{\mathrm{c}} = \sqrt{\sigma_{\mathrm{full}}^2 + \sigma_{\mathrm{mixed}}^2}$. The speedup is $t_{\mathrm{full}}/t_{\mathrm{mixed}}$, where each $t$ is the net MCMC time summed over the restart chunks, divided by the total number of production steps. Numbers in parentheses indicate the statistical error on the last digits.}
\label{tab:comparison-mixed-full-precisions}
\begin{tabular}{l|c|cccc|c}
\hline
System & $N_{e}$ & $E_{\mathrm{full}}$ (Ha) & $E_{\mathrm{mixed}}$ (Ha) & $\Delta E$ (Ha) & $|\Delta E| / \sigma_{\mathrm{c}}$ & Speedup \\
\hline
H$_2$O & 8 & -17.21800(16) & -17.21795(16) & $+5.0 \times 10^{-5}$ & 0.22 & 1.22$\times$ \\
(H$_2$O)$_2$ & 16 & -34.43316(14) & -34.43297(14) & $+1.9 \times 10^{-4}$ & 0.96 & 1.40$\times$ \\
(H$_2$O)$_4$ & 32 & -68.86243(16) & -68.86253(16) & $-1.0 \times 10^{-4}$ & 0.44 & 1.22$\times$ \\
(H$_2$O)$_6$ & 48 & -103.28575(16) & -103.28556(16) & $+1.9 \times 10^{-4}$ & 0.84 & 1.30$\times$ \\
(H$_2$O)$_8$ & 64 & -137.70573(16) & -137.70516(16) & $+5.7 \times 10^{-4}$ & 2.52 & 1.46$\times$ \\
(H$_2$O)$_{20}$ & 160 & -343.92181(15) & -343.92162(16) & $+1.9 \times 10^{-4}$ & 0.87 & 1.30$\times$ \\
\hline
\end{tabular}
\end{table*}
%

\begin{sidewaystable*}                                                                                                       
\begin{center}
\caption{\label{machines} HPC facilities used in this study for benchmarking jQMC}
\vspace{1mm}
\begin{tabular}{l|l|l|l|l}
\Hline
\multicolumn{2}{l|}{Cluster name}  & Leonardo & Miyabi (GPU node) & Genkai (GPU node) \\
\Hline
\multicolumn{2}{l|}{Vendor and Model}  & BullSequana XH2000, Atos & PRIMERGY CX2550 M7, Fujitsu & PRIMERGY GX2560 M7, Fujitsu
\\
\hline
\multicolumn{2}{l|}{Operator and Location}
& CINECA in Italy                    & The University of Tokyo in Japan & Kyushu University in Japan \\
\hline
\multirow{7}{*}{CPU}
& Processor name               & Intel Xeon Platinum 8358          & NVIDIA Grace CPU C1      & Intel Xeon Platinum 8490H \\
& Number of processors         & 1 CPU                             & 1 CPU                    & 2 CPUs                   \\
& Number of cores              & 32 cores                          & 72 cores                 & 120 cores                \\
& Frequency                    & 2.6 GHz                           & 3.0 GHz                  & 1.9 GHz                  \\
& Memory                       & 512 GB                            & 120 GB                   & 1024 GB                  \\
& Memory BW (bandwidth)             & ---                                 & 512 GB/s                 & 614.4 GB/s               \\
\hline
\multirow{6}{*}{GPU}
& Processor name               & NVIDIA custom A100                & NVIDIA H100              & NVIDIA H100 NVL          \\
& Number of processors         & 4                                 & 1                        & 4                        \\
& Memory                       & 64 GB                             & 96 GB                    & 94 GB                    \\
& Memory BW (bandwidth)             & 461 GB/s                          & 4.02 TB/s                & 2.4 TB/s                 \\
& CPU--GPU connection           & NVLink 3.0 (200 GB/s)              & NVLink C2C (450 GB/s)    & PCIe Gen5 (128 GB/s)     \\
& Interconnect                 & InfiniBand 2$\times$dual-port HDR (400 Gbps) & InfiniBand NDR (200 Gbps) & InfiniBand 2$\times$NDR (400
Gbps) \\
\hline
\multicolumn{2}{l|}{Total nodes}  & 3456                              & 1120                     & 38                       \\
\hline
\multicolumn{2}{l|}{Total GPUs}   & 13824 GPUs                        & 1120 GPUs                & 152 GPUs                 \\
\Hline
\end{tabular}
\end{center}
\end{sidewaystable*}

\section{Vectorization strategy}
\label{sec:vectorization}
Within a single process, vectorization in \texttt{jQMC} is realized through \texttt{JAX}'s \texttt{vmap} function. In QMC algorithms, the top-level and most significant vectorization is over \textit{walkers}. Since walkers in QMC simulations evolve nearly independently, batching over walkers provides the most natural and efficient way to exploit vectorization on modern hardware. As a result, to maximize single-GPU performance, especially in QMC algorithms where matrix-matrix or matrix-vector operations inside the inner loops are not dominant in computational cost, it is generally most efficient to assign as many walkers as possible to each GPU, up to the point where the GPU's computational or memory throughput is saturated.

\begin{figure*}
    \centering
    \includegraphics[width=1.0\linewidth]{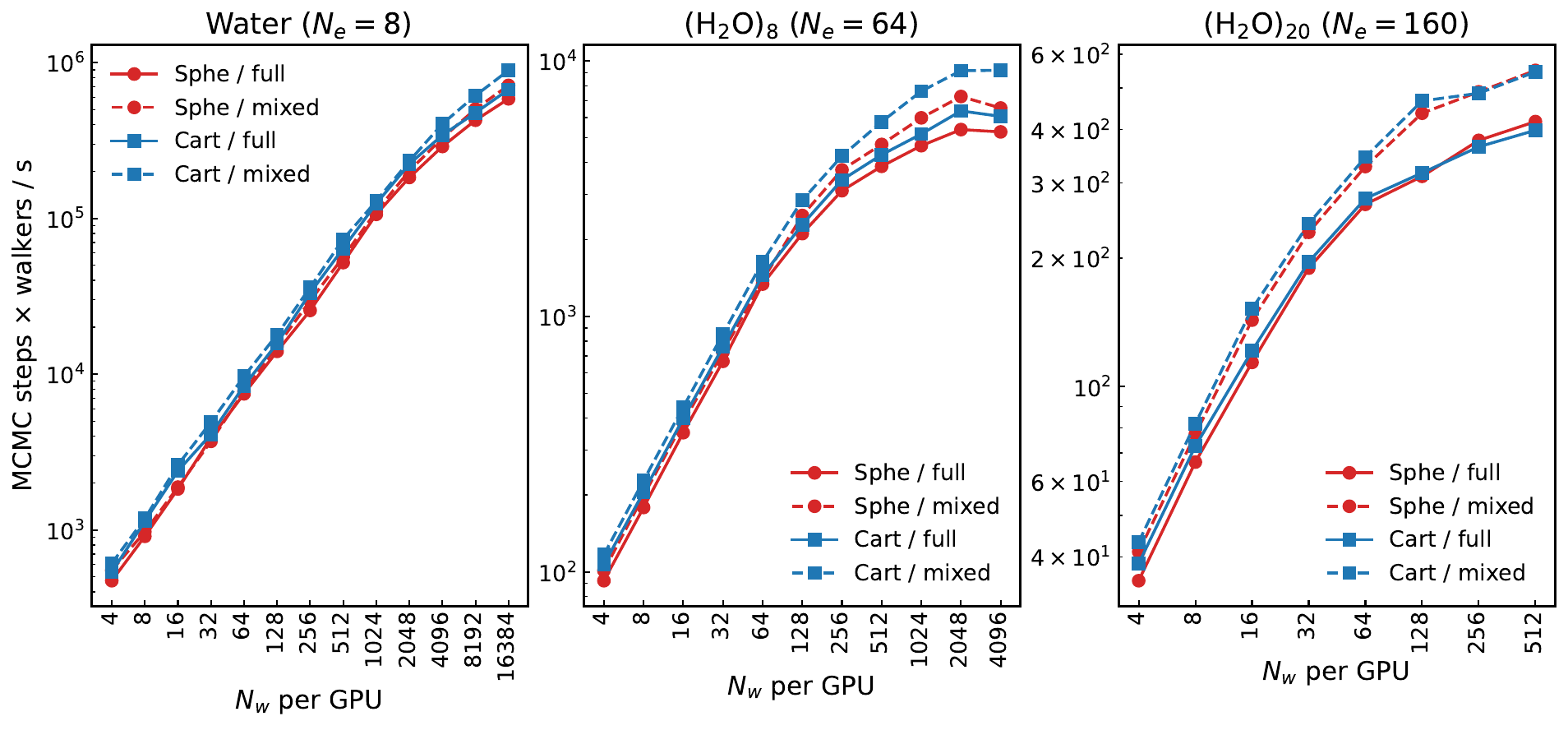}
    \caption[]{The MCMC throughput as a function of the number of walkers per GPU. The vertical axis represents MCMC steps $\times$ number of walkers per sec., which measures the actual computational power of the QMC runs. Each panel shows the four combinations of basis representation (spherical / cartesian GTOs) and precision (full / mixed precision). The benchmark was performed for six systems, but only three of them, \texttt{water} ($N_e$ = 8), \texttt{water cluster [8]} ($N_e$ = 64), and \texttt{water cluster [20]} ($N_e$ = 160), are plotted here. These benchmark tests were measured on a single Miyabi-G GPU node (NVIDIA H100) of the Miyabi supercomputer.
    }
    \label{fig:mcmc-vectorization-benchmark}
\end{figure*}

\begin{figure*}
    \centering
    \includegraphics[width=1.0\linewidth]{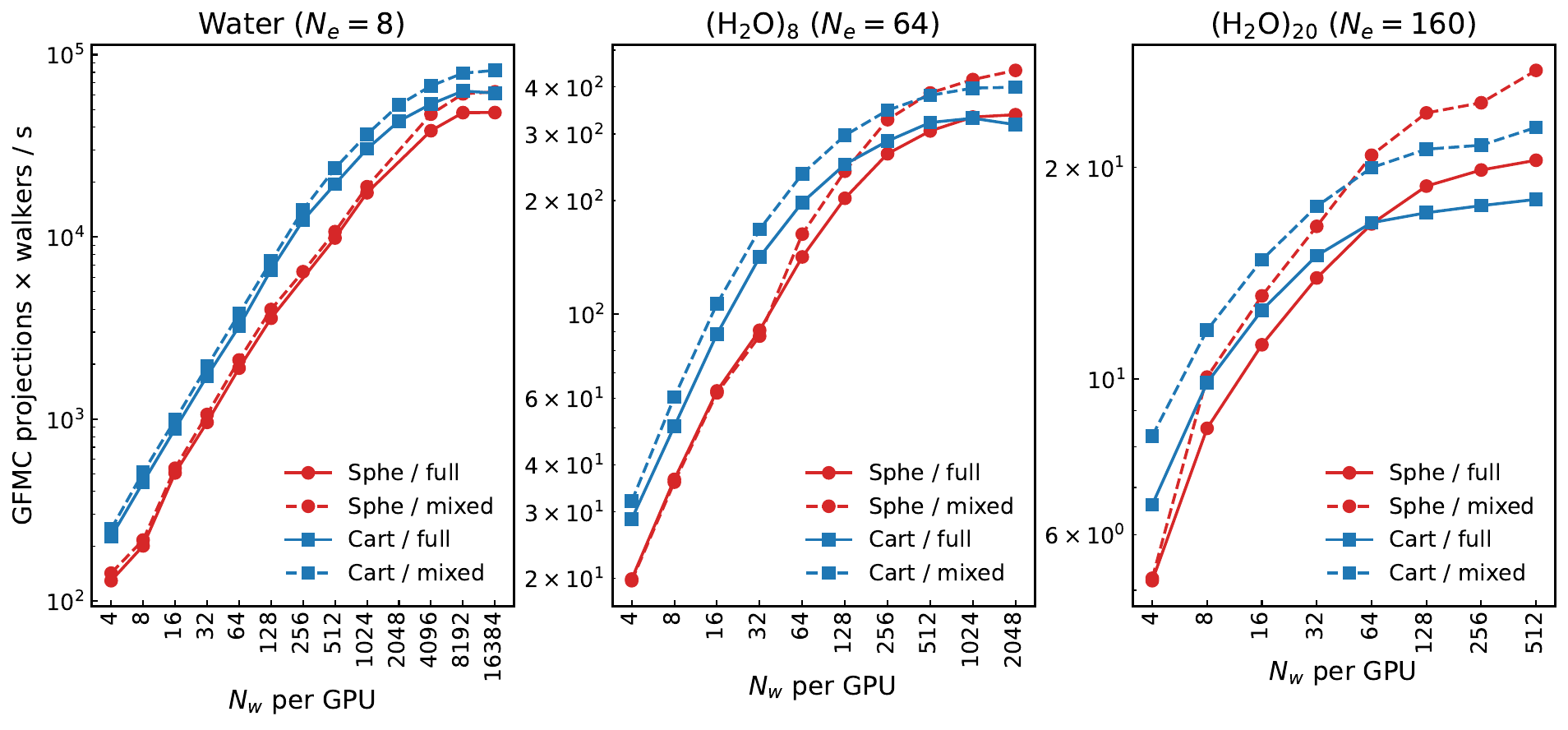}
    \caption[]{The LRDMC throughput as a function of the number of walkers per GPU. The vertical axis represents GFMC projections $\times$ number of walkers per sec., which measures the actual computational power of the QMC runs. Each panel shows the four combinations of basis representation (spherical / cartesian GTOs) and precision (full / mixed precision). The benchmark was performed for six systems, but only three of them, \texttt{water} ($N_e$ = 8), \texttt{water cluster [8]} ($N_e$ = 64), and \texttt{water cluster [20]} ($N_e$ = 160), are plotted here. These benchmark tests were measured on a single Miyabi-G GPU node (NVIDIA H100) of the Miyabi supercomputer.
    }
    \label{fig:lrdmc-vectorization-benchmark}
\end{figure*}

{\vspace{2mm}}
Figures~{\ref{fig:mcmc-vectorization-benchmark}} and {\ref{fig:lrdmc-vectorization-benchmark}} demonstrate the MCMC and LRDMC throughputs as the number of walkers per GPU is varied, for water clusters ranging from a single water molecule ($N_e = 8$) to a cluster of twenty water molecules ($N_e = 160$), using a single node of the \texttt{Miyabi} GPU cluster at the University of Tokyo. Notice that the throughput is the product of (i) the number of MCMC steps (or GFMC projections) executed per unit time by a single GPU multiplied by (ii) the number of walkers assigned to that GPU.
%
%

%
{\vspace{2mm}}
The benchmark calculations were performed using six water clusters: H$_2$O (\texttt{water}, $N_e$ = 8), (H$_2$O)$_2$ (\texttt{water dimer} $N_e$ = 16), (H$_2$O)$_4$ (\texttt{water cluster [4]}, $N_e$ = 32), (H$_2$O)$_6$ (\texttt{water cluster [6]}, $N_e$ = 48), (H$_2$O)$_{10}$ (\texttt{water cluster [8]}, $N_e$ = 64), and (H$_2$O)$_{20}$ (\texttt{water cluster [20]}, $N_e$ = 160). We used the \texttt{ccECP} pseudopotentials accompanied with the \texttt{cc-pVTZ} basis set.
As shown in Figs.~{\ref{fig:mcmc-vectorization-benchmark}} and {\ref{fig:lrdmc-vectorization-benchmark}}, performance improves 
monotonically with the number of walkers per GPU, up to a saturation point. This saturation reflects the GPU becoming fully utilized, as analyzed in the next Section, and the corresponding walker count represents the most efficient configuration for that system.
Since the optimal number of walkers per GPU depends on both the system size and hardware, users should perform preliminary benchmarking runs to identify the performance sweet spot before large-scale production simulations.

\section{Performance Analysis on NVIDIA GPU H100}
\label{sec:performance-analysis}
\begin{figure*}[h]
    \centering
    \includegraphics[width=1.0\linewidth]{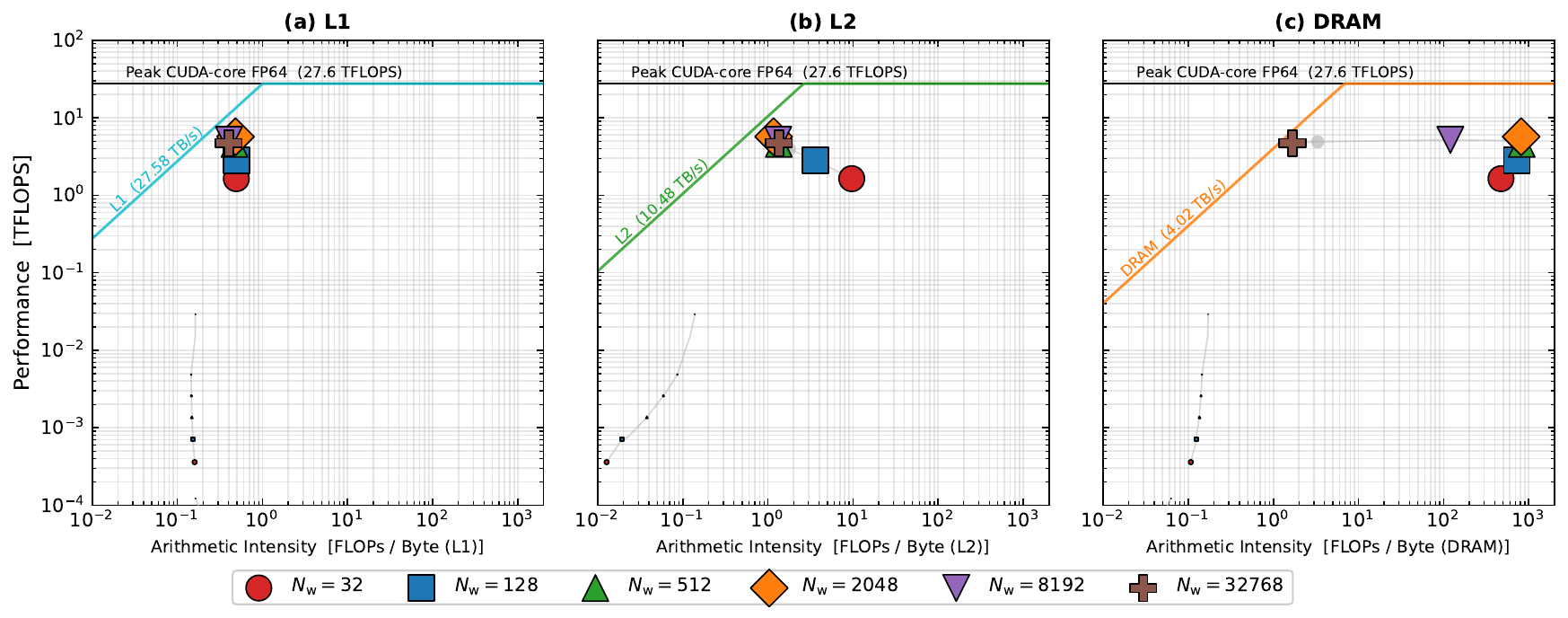}
    \caption[]{Hierarchical roofline of the function \texttt{compute\_AOs} (Cartesian, FP64) kernel on H100, swept over $N_w \in {32, 128, 1024, 8192, 32768}$, with the (a) L1, (b) L2, and (c) DRAM diagonals (peak BW $27.58$, $10.48$, and $4.02$ TB/s, respectively) and the $27.58$-TFLOPS CUDA-core FP64 ceiling. The L1, L2, and DRAM BWs and the FP64 peak performance
    are based on NVIDIA Nsight Compute measurements; the measured DRAM BW of 4.02~TB/s agrees well with the nominal Miyabi value of 4.02~TB/s.
    The FP64 horizontal line reflects the actual clock ($\sim$ 1.5 GHz) and is lower than the 33.5 TFLOPS H100 datasheet boost-clock peak.
    These benchmark tests were measured on a single Miyabi-G GPU node (NVIDIA H100) of the Miyabi supercomputer.
    }
    \label{fig:AO-roofline}
\end{figure*}

\begin{figure*}[h]
    \centering
    \includegraphics[width=0.5\linewidth]{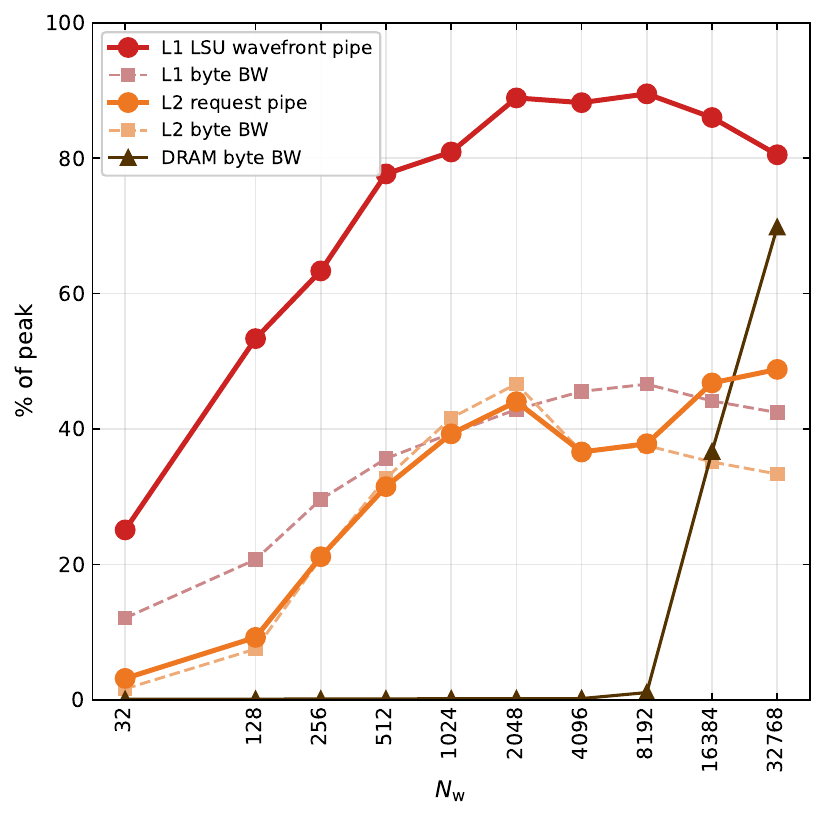}
    \caption[]{Achieved fraction of peak for the dominant \texttt{compute\_AOs} (Cartesian, FP64) kernel, as a function of walker count $N_w$ on a single H100 GPU. Dashed lines (L1/L2 byte BW and DRAM byte BW report the bytes-per-second utilization of each memory tier - the same quantities that set the ceilings of the hierarchical roofline. Solid lines (L1 LSU wavefront pipe, L2 request pipe) report instead the transaction-rate occupancy of each cache level's issue pipe (wavefronts/cycle at L1, request packets/cycle at L2 cache), which is the binding throughput for scatter/gather access patterns and is structurally absent from a byte-BW roofline. This benchmark test was measured on a single Miyabi-G GPU node (NVIDIA H100) of the Miyabi supercomputer.}
    \label{fig:AO-kernel-analysis}
\end{figure*}

\begin{figure}
    \centering
    \includegraphics[width=0.5\linewidth]{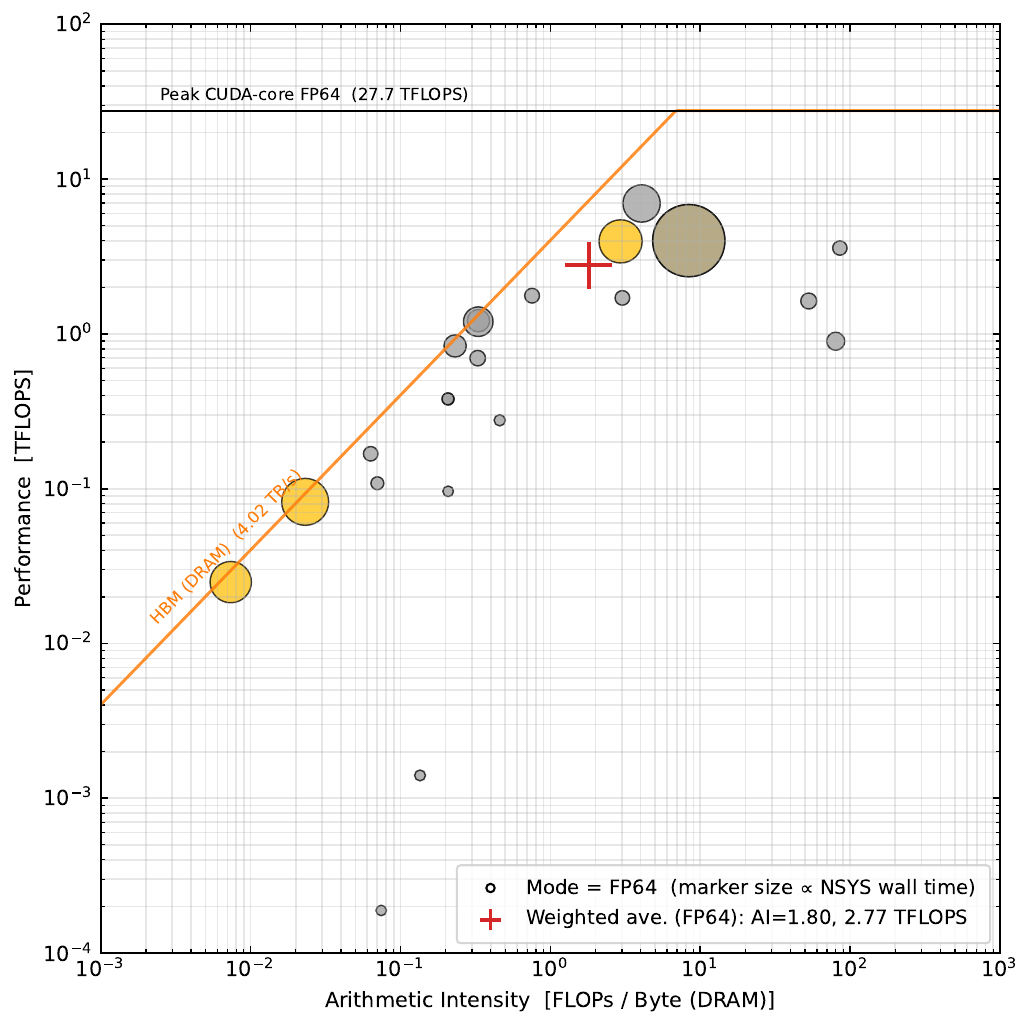}
    \caption[]{Roofline plot of the CUDA kernels involved in the discretized kinetic-energy computation, executed with 2048 walkers in FP64 on a single NVIDIA H100 GPU of the Miyabi supercomputer. The horizontal axis is the AI (FLOPs per byte of DRAM traffic) and the vertical axis is the achieved performance in TFLOPS. Each gray marker represents one CUDA kernel; its area is proportional to the NSYS-reported wall-clock contribution, so the dominant (fused) kernels appear as the largest disks.
    Yellow-filled markers indicate the five kernels with the largest wall time.
    The orange diagonal line is the HBM-DRAM BW ceiling (4.02 TB/s) and the black horizontal line is the peak CUDA-core FP64 throughput (27.7 TFLOPS); the red cross marks the wall-clock-weighted average over all kernels (AI $\approx 1.80$, $2.77$ TFLOPS). The DRAM BWs and the FP64 peak performance are based on NVIDIA Nsight Compute measurements; thus, the FP64 horizontal line reflects the actual clock ($\sim$ 1.5 GHz) and is lower than the 33.5 TFLOPS H100 datasheet boost-clock peak.
    }
    \label{fig:kinetic_discretized_roofline}
\end{figure}

In Section~{\ref{sec:vectorization}}, we observed the throughputs of MCMC and LRDMC as the number of walkers increased, but it was unclear to what extent the GPU was actually utilized. Therefore, we used NVIDIA Nsight Systems and NVIDIA Nsight Compute to analyze how jQMC's MCMC and LRDMC computations run on the GH200 chip of Miyabi Supercomputer. For the benchmark system, we used the water-cluster [8] ($N_e$=64) employed in Section~{\ref{sec:vectorization}}.
{\vspace{2mm}}
Figure~{\ref{fig:AO-roofline}} shows a roofline plot for the function that calculates atomic orbitals with Cartesian GTOs for a given batched electron positions (dim. of input: (number of walkers, $N_e$, 3) $\rightarrow$ dim. of output: (number of walkers, number of AOs, $N_e$)), which is the most basic computational function in jQMC. The x-axis plots the Arithmetic Intensity (AI), defined as floating-point operations (FLOPs) per byte, for the total data transfer involving (a) L1 cache, (b) L2 cache, and (c) DRAM (HBM3). In the roofline plot, each dot represents a (fused) CUDA kernel generated via XLA for different numbers of walkers, with wall time measured by NVIDIA Nsight Compute used to size the circles. The plot reveals that a single fused CUDA kernel dominates most of the AO computation. We focus on the kernel hereafter. Up to $N_w = 2048$, the achieved number of floating-point operations per second (FLOPS or FLOP/s) increases as the number of walkers, resulting in the achieved FLOPS of 5.65 and the AI of 817.32. The achieved FLOPS is about 22\% of the peak FLOPS of the CUDA 64-bit kernel. However, the achieved FLOPS are saturated after $N_w = 2048$, and the kernel point just shifts to the left (i.e., the AI decreases) without increasing FLOPS. Finally, none of the CUDA kernels are on the L1, L2, DRAM rooflines, even with $N_w = 32768$. NVIDIA Nsight Compute reveals that the bottleneck is neither L1, L2, nor DRAM BW (i.e., the bytes/s sustained between the Streaming Multiprocessors and each memory tier, as drawn by the roofline ceilings), but the L1 wavefront pipe (i.e., the transactions/cycle throughput of the L1 load/store unit, which is distinct from its byte BW when warp threads access non-contiguous addresses~{\cite{nvidia_ncu_guide}}).
Figure~\ref{fig:AO-kernel-analysis} plots, as a function of $N_w$, the achieved fraction of the peak of five quantities. The dashed lines (L1 byte BW, L2 byte BW) and the DRAM byte BW curve report the bytes/s delivered by the kernel normalized by the device-peak BW of each level, which correspond to the ceilings of the hierarchical roofline of Fig.~\ref{fig:AO-roofline}. The two solid lines, in contrast, report a transaction-rate occupancy: L1 LSU (Load Store Unit) wavefront pipe is the number of wavefronts/cycle issued through the L1 load/store data pipe normalized by its peak, and L2 request pipe is the analogous quantity at the L2 boundary. Figure~\ref{fig:AO-kernel-analysis} reveals that the L1 LSU wavefront pipe reaches $\sim 90\%$ of peak for $N_w = 2048$, while L1 byte BW shows plateaus near $\sim 50\%$. We notice that the $\sim 40$-point gap between the solid and the dashed red curves is invisible on the byte-only roofline yet identifies the kernel as L1 transaction-throughput bound rather than L1 BW bound. The L2 request pipe stays at a lower absolute level (peaking near $\sim 48\%$), indicating that the L1 absorbing is most of the gather traffic. DRAM byte BW only becomes appreciable at the largest $N_w = 32768$, where the working set finally exceeds the on-chip L1 and L2 caches. These results suggest that cache optimizations will be necessary for future improvements in FLOPS and AI performance. 
In addition, we analyzed each component, such as continuum kinetic energy, discretized kinetic energy, and non-local ECP computations, which are dominant in the computation of the local energy in MCMC and the projection in LRDMC. For example, Figure~{\ref{fig:kinetic_discretized_roofline}} shows the roofline plot for the discretized kinetic energy computation (Eq.~{\ref{eq:kinetic_ratio}}) with $N_w = 2048$. The plot shows that the dominant (fused) CUDA kernels are all in DRAM-bound regions, suggesting that QMC calculations as a whole are also dominated by memory-bound operations. The bottleneck of each kernel depends on whether it is L1-, L2-, or DRAM-limited, according to our Compute analysis with NVIDIA Nsight Compute, and the optimization must be tailored to the specific circumstances.
The CUDA-kernel-level optimization (i.e., via high-level operations (HLO) generated by JAX) for jQMC is an interesting future work.

\section{Wall-time benchmark}
\label{sec:wall-time-benchmark}
The walltime required in practice for MCMC and LRDMC calculations depends strongly on the quality of the employed wave function, because the ultimate target of QMC simulations is the statistical error bar, and the efficiency for reaching a given error bar is significantly affected by the wave-function quality. Since different QMC packages adopt different wave-function parameterizations and different functional forms of the Jastrow factor, a fair comparison between packages is generally difficult. However, TurboRVB and jQMC use the same Jastrow factor, enabling a genuinely fair comparison. In this benchmark, we prepared exactly the same Jastrow--Slater ansatz in both TurboRVB and jQMC, and compared the absolute wall-clock time required to achieve a target statistical accuracy of 1 kcal/mol in the total energy for a given ansatz. Parameters related to computational efficiency, such as the electron hopping distance used in the MCMC proposal, the frequency of observable calculations in MCMC, and the lattice discretization mesh and projection time in LRDMC, were set to the  same values for both TurboRVB and jQMC.

\vspace{2mm}
Benchmark calculations were carried out for six systems of increasing size: \texttt{water} ($N_e = 8$), \texttt{water dimer} ($N_e = 16$), \texttt{water cluster [4]} ($N_e = 32$), \texttt{water cluster [6]} ($N_e = 48$), \texttt{water cluster [8]} ($N_e = 64$), and \texttt{water cluster [20]} ($N_e = 160$). All calculations employed the \texttt{ccECP} pseudopotentials together with the \texttt{cc-pVTZ} basis set. For the determinant part, we directly used the orbitals obtained from an LDA-PZ DFT calculation with PySCF-forge. TurboRVB employs spherical GTOs, whereas jQMC supports both spherical and Cartesian GTOs. Since the Cartesian representation is faster than the Spherical one in jQMC, we transformed the orbitals from spherical to Cartesian form for the jQMC calculations; the physical energy is, of course, unchanged by this transformation. For the Jastrow factor, we used $J = J_2 + J_3$, where $J_2$ was of Pad\'e type and $J_3$ was expanded in localized atomic orbitals, using H:[3s1p] and O:[4s2p1d]. The variational parameters of the Jastrow factor were optimized in TurboRVB, and the resulting values were copied into the \texttt{Jastrow\_data} of jQMC. Using these wave functions, we computed the total energy and variance with both TurboRVB and jQMC by MCMC, and confirmed that the results agreed within the statistical error bars for all six systems.

\vspace{2mm}
For these wave functions, we prepared one node (equipped with two multi-core CPUs with four GPUs) on the Genkai supercomputer, and measured the time required to achieve a statistical error bar of 1 kcal/mol in two cases: using CPUs only and using CPU+GPUs (TurboRVB) or only GPUs (jQMC). In the CPU-only calculations, both TurboRVB and jQMC used 120 MPI processes, corresponding to the maximum number available of cores on a single node, with one walker assigned to each MPI process. In the GPU calculations with jQMC, one MPI process was bound to each GPU, and the number of walkers per MPI process was chosen to maximize the MCMC and LRDMC throughput on the Genkai GPU node. In contrast, TurboRVB does not implement walker-level vectorization, unlike jQMC; instead, NVIDIA CUDA MPS was used~{\cite{2026_Bellentani_NvidiaMPS}}, and 30 MPI processes (i.e., all cores in one NUMA domain, which is the optimal setting for the CPU-bound part, despite possibly oversubscribing a single GPU) were assigned to each GPU. The compilation and runtime environments are summarized in Table~\ref{tab:compile_env}.

\begin{table*}[t]
\caption{Compilation and runtime environments used in the benchmark calculations.}
\label{tab:compile_env}
\centering
\begin{tabular}{lll}
\hline
Code / platform & Environment \\
\hline
TurboRVB-CPU & Intel compiler (ifort, icc) 2021.10.0; Intel MKL 2023.2; Intel MPI 2021.10 \\
TurboRVB-GPU & nvfortran 23.9; CUDA 12.2.2; Open MPI 4.1.5; NVIDIA MPS (\texttt{nvidia-cuda-mps-control}) \\
jQMC-CPU & Python 3.12.7; JAX 0.7.2; Open MPI 4.1.6; mpi4py 4.1.1 \\
jQMC-GPU & Python 3.12.7; JAX 0.7.2; Open MPI 4.1.6; mpi4py 4.1.1; CUDA 12.9 \\
\hline
\end{tabular}
\end{table*}

\begin{figure*}[h]
    \centering
    \includegraphics[width=1.0\linewidth]{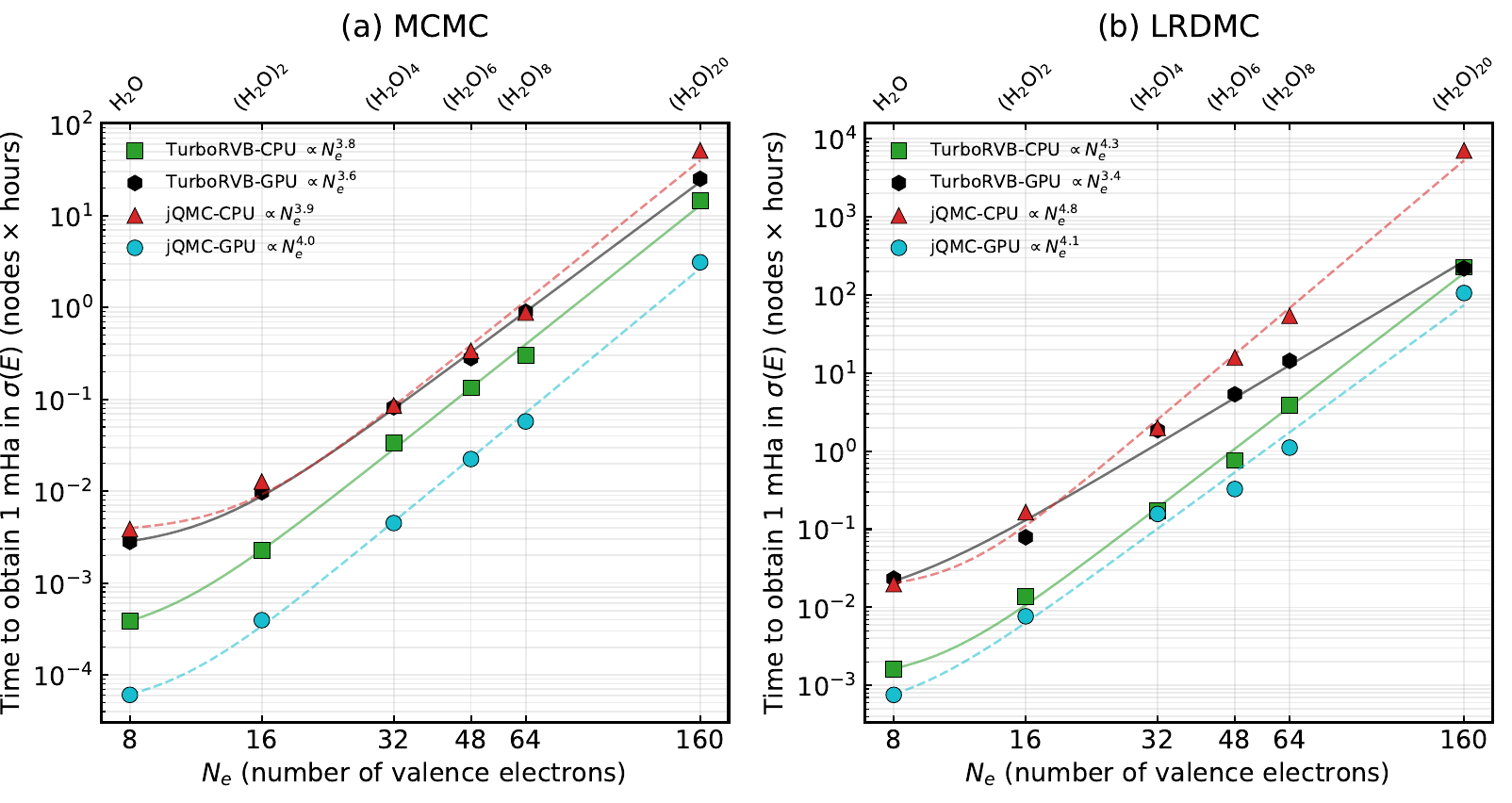}
    \caption[]{Wall-clock time (nodes $\cdot$ hours) required to obtain the target statistical error of $\sigma(E) = 1$ mHa, as a function of the number of valence electrons $N_e$, for (a) MCMC and (b) LRDMC sampling. The wall-clock time is estimated from a reference run via $t_\text{target} = (\sigma_\text{run}/\sigma_\text{target})^2 \cdot t_\text{net}$, where $t_\text{net}$ refers to the net elapsed time of the reference run excluding initialization and equilibration. jQMC runs use cartesian GTOs in mixed precision; TurboRVB uses spherical GTOs in full precision. Solid lines are power-law fits $t \propto N_e^{p}$ with the exponent $p$ given in the legend. In the LRDMC calculations, a lattice discretization $a = 0.3$ Bohr was employed. These benchmark tests were measured on a single node (2 $\times$ Intel Xeon Platinum 8490H + 4 $\times$ NVIDIA H100) of the Genkai supercomputer. In the CPU-only calculations, both TurboRVB and jQMC used 120 MPI processes with one walker assigned to each MPI process. In the GPU calculations with jQMC, one MPI process was bound to each GPU and the number of walkers per MPI process was chosen to maximize the MCMC and LRDMC throughput. In the TurboRVB calculation on GPUs, 30 MPI processes were assigned to each GPU via NVIDIA CUDA MPS.
    \label{fig:comparision-turborvb-jqmc}
    }
\end{figure*}

\vspace{2mm}
Figure~\ref{fig:comparision-turborvb-jqmc} shows the absolute node-seconds required to reach a statistical error bar of 1 kcal/mol in the total energy for the systems listed above. Focusing first on jQMC, one can clearly see that GPU acceleration yields speedups exceeding two orders of magnitude, if compared with the same calculations run on CPUs. This is consistent with expectations, since JAX has been developed with GPU execution in mind. By contrast, in TurboRVB, the GPU calculations are slower than the CPU calculations for the system sizes and hardware configurations tested here. This is because TurboRVB does not parallelize at the walker level, as noted above, and only partially offloads the computation to the GPU by replacing some matrix and vector operations (BLAS/LAPACK) with cuBLAS and cuSolver routines. In TurboRVB, the GPU version becomes more advantageous as the system size increases because kernels replaced with cuBLAS and cuSolver are more dominant; Fig.~\ref{fig:comparision-turborvb-jqmc} demonstrates that for systems with roughly $N_e \sim 100$, GPU acceleration becomes clearly beneficial, particularly in the LRDMC implementation. We expect that for even larger systems, the GPU version of TurboRVB will eventually be faster than the CPU one.

\vspace{2mm}
Figure~\ref{fig:comparision-turborvb-jqmc} also shows that the current CPU version of jQMC is about 10 times slower than TurboRVB on the same CPU (Intel Xeon Platinum 8490H), even though jQMC implements essentially the same algorithm to the best of our knowledge. This indicates that the current jQMC is suboptimal on CPU via XLA. For instance, using hardware-counter measurements, we found that JAX/XLA emits mostly scalar instructions in the AO kernel, with only a small amount of AVX2 and almost no AVX-512, even though the CPU supports them; forcing a 512-bit preferred vector width does not change this, pointing to a code-structure rather than a flag issue. After rewriting the relevant part of the AO kernel, we obtained a speedup of a factor of 2 for the AO evaluation (c.f., We did not merge this change because it made the GPU version slightly slower). In contrast, a minimal Fortran/C implementation of the same operation, compiled with the Intel compiler, is almost fully vectorized with AVX-512 and is much faster (about 6 times over the baseline) than the JAX AO kernel. Beyond the AO kernel, there are most likely further CPU-side bottlenecks in other kernels of jQMC, which we leave for future identification. The current jQMC layout is well suited to GPU but remains suboptimal for CPU; a unified implementation that performs well on both is left for future work.

\vspace{2mm}
Comparing the two faster cases in each package, jQMC on GPUs with TurboRVB on CPUs, jQMC is faster for both MCMC and LRDMC in the wall times, thanks to GPUs. The jQMC MCMC and LRDMC are faster than the corresponding TurboRVB calculations by 10 times and 3 times, respectively, for the system sizes taken into account here. In the LRDMC calculations, the speed-up offered by jQMC is not as drastic as that found in MCMC. This is because, in LRDMC, the computational bottleneck involves a significant amount of AO evaluations, needed in the discretized kinetic energy and in the non-local ECP calculations, resulting in a performance gain that are not as significant as those achieved with MCMC. Further acceleration is expected if AO calculations can be made more optimal in jQMC. Nevertheless, the present benchmark demonstrates that jQMC has already been tuned to deliver wall-clock performance comparable to, or even better than, TurboRVB, despite the fact that TurboRVB is a long-optimized Fortran90 code benefiting from Intel Fortran compiler optimizations.

\section{Parallelization strategy: Weak scaling benchmarks}
\label{sec:weak-scaling-benchmark}
Although the vectorization scheme described above improves the throughput on a single GPU, production QMC calculations often require simulations that exceed the capacity of a single device. In \texttt{jQMC}, multi-GPU calculations are therefore supported by explicit MPI-based parallelization via \texttt{mpi4py} for CPUs or by the sharding mechanisms provided by \texttt{JAX} for GPUs.
We examined the weak-scaling performance of \texttt{jQMC} on the Miyabi and Leonardo supercomputers. The benchmark target is the water cluster (H$_2$O)$_{20}$ ($N_e = 160$). In these benchmarks, the number of walkers per GPU is fixed at 100, while the number of GPUs was increased, allowing us to assess the implementation's parallel efficiency in realistic multi-GPU settings.

\begin{figure*}[h]
    \centering
    \includegraphics[width=1.0\linewidth]{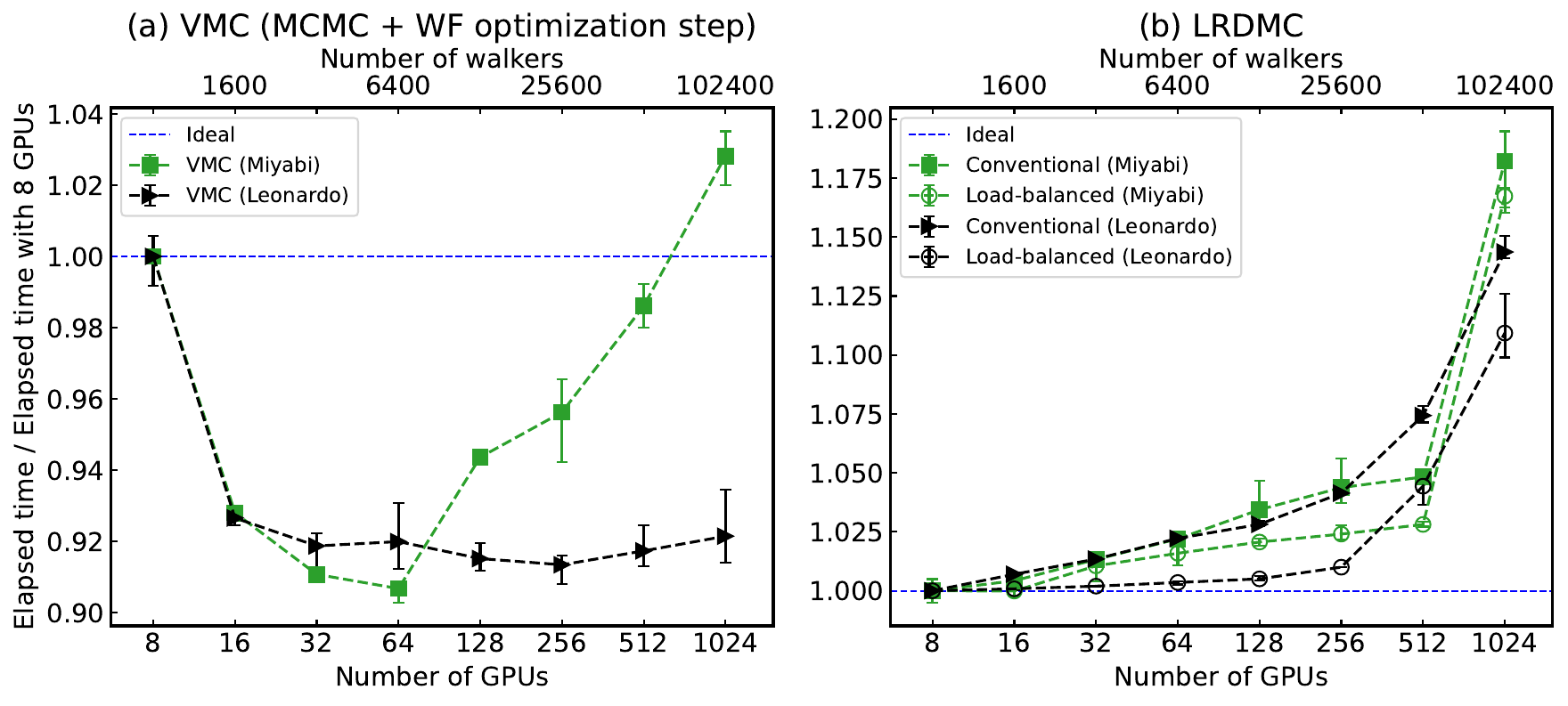}
    \caption[]{Weak-scaling benchmark. Elapsed time per (a) VMC step (MCMC sampling + one step of WF optimization) and (b) LRDMC branching step, normalized to the value obtained with 8 GPUs on each machine. Error bars show the min--max range across 2-5 independent runs on different node allocations. The benchmark target is a water cluster (H$_2$O)$_{20}$ ($N_e = 160$). The number of walkers per GPU is fixed at 100, so the total walker count (top axis) scales linearly with the number of GPUs (bottom axis); this corresponds to the standard weak-scaling protocol in which the workload per GPU is held constant. For LRDMC, two branching schemes are compared: the conventional (\texttt{GFMC\_tau} class) algorithm (filled markers) and the load-balanced (\texttt{GFMC\_bra} class) algorithm (open markers), where the latter applies a fixed number of projections per measurement chosen from the conventional LRDMC with $\tau = 0.1$ a.u.$^{-1}$. The horizontal dashed line indicates ideal weak scaling (= 1.0). Miyabi (green squares; NVIDIA H100 nodes at the University of Tokyo) was scanned from 8 to 1024 GPUs, and Leonardo (black triangles; NVIDIA A100 Booster module at CINECA) from 8 up to 1024 GPUs. Both VMC and LRDMC retain near-ideal weak scaling across the full range tested, demonstrating that jQMC sustains its parallel efficiency up to $10^3$ GPUs.}
    \label{fig:weak-scailing-benchmark}
\end{figure*}

{\vspace{2mm}}
Figure~\ref{fig:weak-scailing-benchmark} shows the weak-scaling benchmark results for (a) VMC (MCMC sampling + one step of wave function optimization ) and (b) LRDMC. As seen, the weak-scaling behavior differs between VMC and LRDMC. In the VMC case [Fig.~\ref{fig:weak-scailing-benchmark}(a)], the performance initially improves as the number of walkers increases, both on Miyabi and Leonardo. This improvement is attributed to the faster convergence of the CG solver when the number of walkers per GPU, and hence the total number of MCMC samples, is increased. With an even larger number of walkers, a slight degradation is observed on Miyabi, whereas the good weak-scaling performance is maintained up to 1024 GPUs on Leonardo. The slowdown on Miyabi is likely due to network latency and/or hardware-induced GPU-to-GPU performance variability across different GPUs, which has not yet been identified. Nevertheless, the weak-scaling efficiency remains close to unity on Miyabi (1.02), while Leonardo maintains good performance (0.92) up to 1024 GPUs, corresponding to 102,400 walkers.
In the LRDMC case [Fig.~\ref{fig:weak-scailing-benchmark}(b)], the slowdown is observed as the number of walkers increases on both Miyabi and Leonardo. This behavior is expected since LRDMC requires inter-walker communication between two consecutive branching steps, as described in Sec.~\ref{subsubsec:lrdmc_implementation}. Here, we compare two branching schemes for LRDMC: the conventional algorithm~\cite{2005CAS} and the load-balanced algorithm~\cite{2025_NAK_LRDMC_load_balanced}. As expected, the load-balanced algorithm shows better weak-scaling performance than the conventional one on both Miyabi and Leonardo, because it is specifically designed to balance the computational workload among walkers~\cite{2025_NAK_LRDMC_load_balanced}.
Both LRDMC algorithms maintain performance close to unity up to 512 GPUs on Miyabi and 256 GPUs on Leonardo. A more pronounced degradation is observed at 1024 GPUs on Miyabi and 512 GPUs on Leonardo, likely again due to network latency and/or hardware-induced GPU-to-GPU performance variability. Nevertheless, the load-balanced LRDMC algorithm still achieves good weak-scaling performance, with values of $\sim$ 1.2 on Miyabi and $\sim$ 1.1 on Leonardo, for the largest calculations considered, corresponding to 102,400 walkers.
These benchmarks demonstrate that the MPI- and GPU-parallel implementation introduces only a small overhead over the present range of system sizes, both for VMC and LRDMC.

\section{Verification of the code}
\label{sec:verification}
\subsection{Intra-software tests}
\label{subsec:intra_software_tests}
The \textit{intra-software tests} in jQMC are designed according to the following principle. Achieving high computational performance with JAX requires \texttt{jit} compilation and \texttt{vmap} vectorization. They may obscure the underlying algorithmic flow and make it more difficult to debug the code than conventional Python implementations.
To address the trade-off between performance and maintainability, jQMC adopts a dual-implementation strategy. More specifically, for the main functions, two versions are implemented in parallel: a production version without a suffix and a corresponding debug version with the suffix \texttt{\_debug}. The production version is used for production calculations and is highly optimized for performance. By contrast, the \texttt{\_debug} version prioritizes readability by using standard Python constructs, allowing users and developers to follow the logic step by step.
Then, to ensure correctness, we implement consistency tests for as many function pairs as possible. These tests verify that the production and debug implementations produce equivalent results. The CD/CI framework on GitHub triggers \texttt{pytest} every time a change is pushed or nightly.  
Every Python module is accompanied by a corresponding test Python module. For example, \texttt{test\_wavefunction.py} contains comprehensive tests for the dataclasses and associated functions defined in \texttt{wavefunction.py}.
These tests help reduce the risk of introducing bugs, thereby improving the maintainability and reliability of jQMC.

\subsection{Inter-software tests}
\label{subsec:inter_software_tests}
\begin{figure*}
    \centering
    \includegraphics[width=1.0\linewidth]{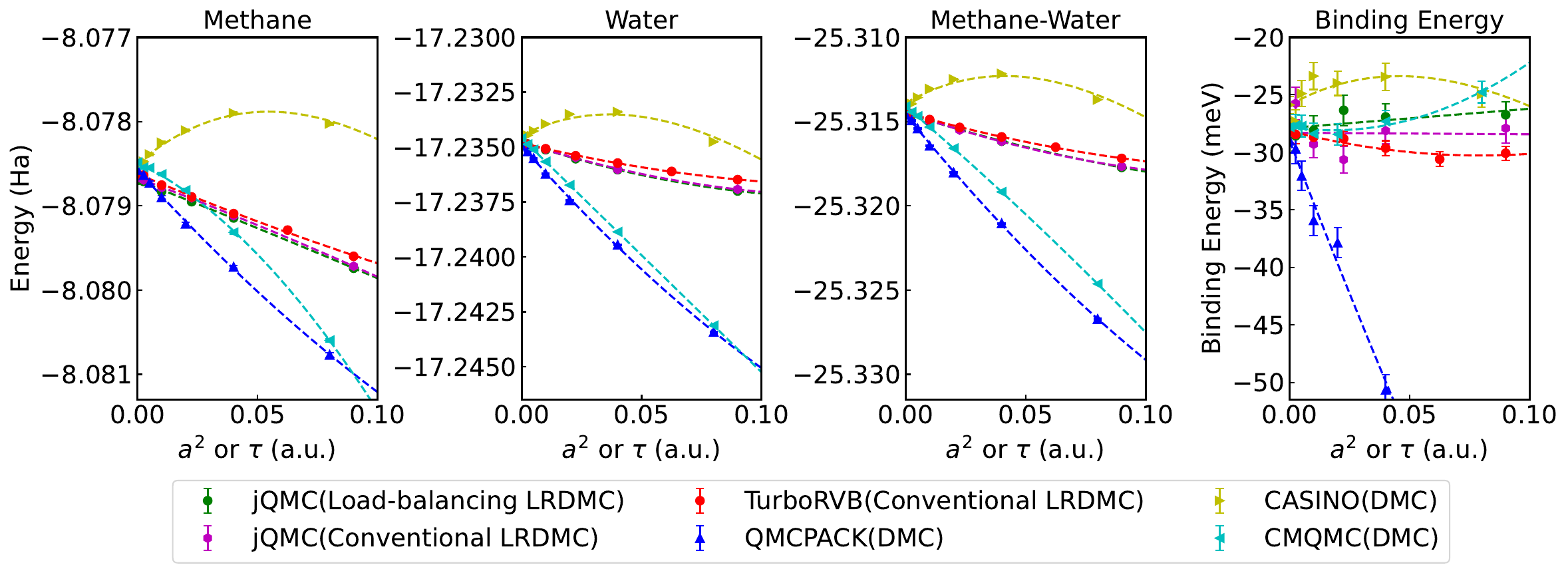}
    \caption[]{The total energy of Methane, Water, and Methane-Water dimer, and the binding energy. The CASINO, QMCPACK, and CMQMC results are taken from Ref.~{\onlinecite{2025_DellaPia_allQMC}}. The TurboRVB values are taken from Ref.~{\onlinecite{nakano2025-loadbalance-lrdmc}}. The corresponding numbers are shown in Table~{\ref{tab:inter-software-comparison}}.
    }
    \label{fig:inter-software-comparison}
\end{figure*}

In the QMC community, recent efforts have been made to systematically examine reproducibility across multiple codes~{\cite{2025_DellaPia_allQMC}}. One notable study involved 11 different community-developed QMC codes and focused on computing the total energy and binding energy of the water--methane dimer. These calculations were performed using a common basis set (\texttt{cc-pVTZ}) and a common pseudopotential (\texttt{ccECP}), allowing a controlled investigation of reproducibility across implementations.
In DMC calculations employing ECPs, one of the technical challenges is how to deal with the non-local part of the ECP. Approaches such as the Locality Approximation (LA) and T-move (TM) are known to affect the zero-time-step (\( t \rightarrow 0 \)) limit. This is due not only to the non-local part treatment, but also to the amplitude of the wave function, set by the Jastrow factor. As a result, differences in Jastrow implementations among codes can lead to discrepancies in energies.
On the other hand, the Determinant Locality Approximation (DLA) and DLA combined with TM (DTM) yield significantly better agreement among codes, as it minimizes sensitivity to the specific form of the Jastrow factor in the small-\( t \) or small-\( a \) limits.
We carry out calculations analogous to the consistency test in the aforementioned reproducibility study corresponding to the DTM locality approximation (i.e., we used LRDMC with DLA) to assess the consistency of our implementation. We present our validation results in Fig.~{\ref{fig:inter-software-comparison}} and Table.~{\ref{tab:inter-software-comparison}}.

\section{Demonstrations: VMC and LRDMC atomic forces with optimized JSD ansatz}
\label{sec:demonstrations}
Recently, the quantitative disagreement between CCSD(T)~\cite{1989_Raghavachari_CCSDT} and FN-DMC with a Jastrow--Slater determinant ansatz has been actively discussed. In particular, since the report of discrepancies of up to about 8 kcal/mol for the binding energies of dispersion-dominated systems, such as C$_{60}$ encapsulated in a [6]cycloparaphenyleneacetylene ring (C$_{60}$+5PPA), reported by Al-Hamdani and co-workers~\cite{2021Yasmin_C60}, the origin of these discrepancies has been debated from both the CCSD(T) and DMC perspectives~\cite{2025_Fishman_CCSDT_vdw, 2025_benjamin_S66, 2026_Nakano_AGPn}. Beyond total energies, first derivatives, namely atomic forces, have also been examined. For the same molecular structures, Slootman \emph{et al.} computed the differences between CCSD(T) and FN-DMC forces for ethanol~{\cite{slootman2024accurate}}. They reported that, for ethanol, DMC forces agree with all-electron CCSD(T) within a mean absolute error (MAE) of 1.5 kcal mol$^{-1}$ \AA$^{-1}$, even when the Reynolds approximation is employed (Eq.~\ref{eq:force_lrdmc_mixed}). In the present work, we extend this comparison by performing LRDMC force calculations for ethanol and malonaldehyde, both included in the rMD17 set~\cite{Christensen2020_MLST,Christensen2020_rMD17_Figshare}, and assessing their agreement with CCSD(T).

\vspace{2mm}
As in Ref.~\onlinecite{2026_Nakano_JCP_CP2K}, we selected ethanol and malonaldehyde from the rMD17 dataset~\cite{Christensen2020_MLST,Christensen2020_rMD17_Figshare}, and computed (i) VMC forces with fully optimized wave functions, (ii) LRDMC forces using trial wave functions fully optimized at the VMC level, (iii) Hartree--Fock (HF), second-order M{\o}ller--Plesset perturbation theory (MP2), CCSD, and DFT forces, and (iv) CCSD(T) forces as references. We extracted 50 structures for each molecule from the rMD17 dataset~\cite{Christensen2020_MLST,Christensen2020_rMD17_Figshare}. We employed the cc-pVQZ basis set for the HF, MP2~\cite{1934MOL_MP2}, and CCSD~\cite{Purvis1982_CCSD}, and CCSD(T) calculations. We employed def2-QZVPPD~\cite{Weigend2005_def2} basis set for DFT calculations with the SVWN~\cite{Vosko1980_VWN}, PBE~\cite{1996PER_PBE}, B3LYP~{\cite{1994_Stephens_B3LYP}}, PBE0+D3BJ~\cite{Adamo1999_PBE0}, or $\omega$B97M+D3BJ~\cite{Mardirossian2016_wB97M_V,Najibi2018_wB97X_wB97M_D3BJ} functionals. In the MP2, CCSD, and CCSD(T) calculations, the 1$s$ core orbitals were kept frozen. The HF, DFT, MP2, CCSD, and CCSD(T) calculations were performed using the Psi4 software suite (version 1.9.1)~\cite{Smith2020_PSI4_1_4}. For the VMC and LRDMC calculations, we used ccECP~\cite{2017BEN_ccECP} together with the corresponding cc-pVTZ basis sets. The MO coefficients in the Slater determinant and the parameters in the Jastrow factor were optimized variationally at the VMC level. The Jastrow factor included inhomogeneous one-body (Eq.~\ref{onebody_J_inhom}), two-body (Eq.~\ref{twobody_jastrow}), and three-body (Eq.~\ref{threebody_jastrow}) terms. For the basis sets of the inhomogeneous one-body and three-body Jastrow terms, we used uncontracted primitive GTOs with [3s1p] for H and [4s2p1d] for C, N, and O. The Gaussian exponents were optimized for each configuration during the wave-function optimization. The total numbers of optimized variational parameters were 38233 and 54749 for ethanol and malonaldehyde, respectively. The trial wave functions were generated with PySCF-forge~\cite{2026_Qiming_pyscf}.

\begin{figure*}[h]
    \centering    \includegraphics[width=1.00\linewidth]{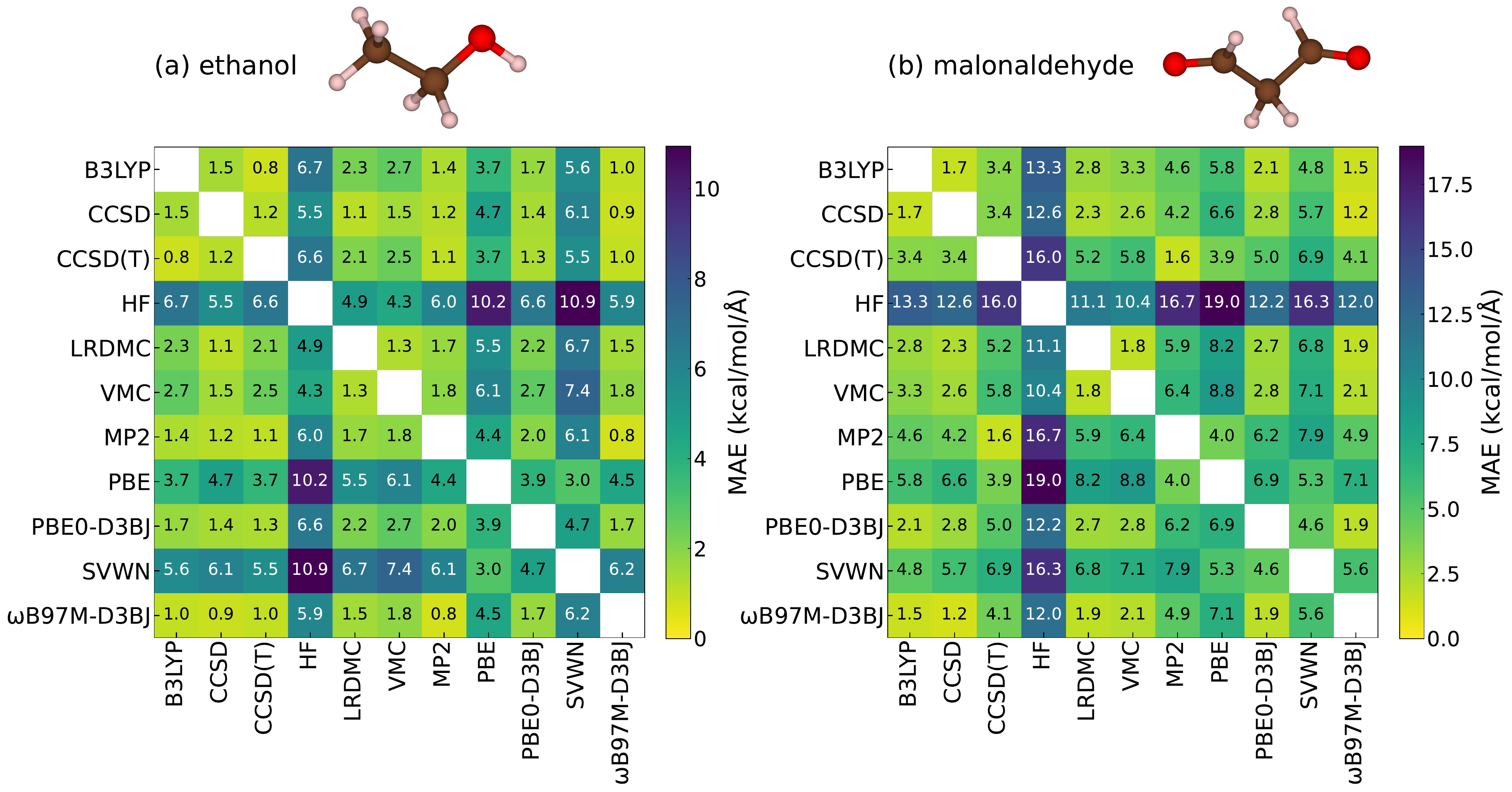}
    \caption[]{Pairwise MAEs of atomic forces (kcal mol$^{-1}$ \AA$^{-1}$) between all pairs of the benchmarked methods for (a) Ethanol and (b) Malonaldehyde. The numbers indicate the corresponding MAEs for each pair.
    The molecular structures were depicted using VESTA~{\cite{2011MOM_VESTA}}.
    }
    \label{fig:pairwise-rmse-heatmap}
\end{figure*}

\begin{table*}[t]
  \centering
  \caption{\label{tab:inter-software-comparison} The total energies (Ha) of the Methane, Water, and Methane-Water, and its binding energy (meV), obtained with the extrapolation to $\tau \rightarrow 0$ (DMC) and $a \rightarrow 0$ (LRDMC). The CASINO, QMCPACK, and CMQMC results are taken from Ref.{\onlinecite{2025_DellaPia_allQMC}}. The TurboRVB values are taken from Ref.~{\onlinecite{nakano2025-loadbalance-lrdmc}}.}
  \begin{tabular}{l|ccc|c}
    \hline
    Package                            & Methane (Ha)   & Water (Ha)     & Methane--Water (Ha) & Binding energy (meV) \\
    \hline
    CASINO (DMC)                      & -8.07856(1)    & -17.23473(2)   & -25.31432(3)     & -26.8(1.0)         \\
    QMCPACK (DMC)                     & -8.07858(2)    & -17.23482(3)   & -25.31443(7)     & -29.0(1.1)         \\
    CMQMC (DMC)                       & -8.07848(2)    & -17.23460(1)   & -25.31411(3)     & -27.8(3)           \\
    \hline
    TurboRVB (Conventional LRDMC)     & -8.07860(1)    & -17.23479(1)   & -25.31445(1)     & -28.1(3)           \\
    TurboRVB (Load-balanced LRDMC)    & -8.07862(2)    & -17.23482(2)   & -25.31447(2)     & -29.0(7)           \\
    \hline
    jQMC (Conventional LRDMC)         & -8.07867(2)    & -17.23478(2)   & -25.31444(4)     & -28.3(8)           \\
    jQMC (Load-balancing LRDMC)       & -8.07870(2)    & -17.23477(2)   & -25.31452(4)     & -27.9(8)           \\
    \hline
  \end{tabular}
\end{table*}

\vspace{2mm}
To compare the methods considered here, we computed the pairwise MAEs between their force predictions and displayed them as a heat map in Fig.~\ref{fig:pairwise-rmse-heatmap}. 
The computed VMC and LRDMC forces carry statistical errors. Suppose that the error bar of each force component is an independent Gaussian random variable, 
$\varepsilon_{i,I} \sim\mathcal{N} (0,\sigma_{i,I}^{2})$, where $i$ and $I$ denote sample index and force component. 
Since the MAE is computed using $|d_{i,I}| \equiv |\Delta F_{i,I} + \varepsilon_{i,I}|$, the measured MAE is always biased by the statistical errors. Indeed, even when the two methods agree exactly ($\Delta F_{i,I} = 0$), the obtained MAE does not vanish but reaches a floor,
\begin{equation}
\mathrm{MAE}_{\mathrm{floor}} = \mathbb{E}[|\varepsilon_{i,I}|] = \sqrt{\frac{2}{\pi}}\,\langle \sigma_{i,I} \rangle ,
\end{equation}
where the brackets denote the average over all samples and force components. Notice that $\sigma_{i,I} = [(\sigma^{A}_{i,I})^{2} + (\sigma^{B}_{i,I})^{2}]^{1/2}$ when both methods are stochastic and independent.
In the present calculations, the MAE floors are 0.35 and 0.72 kcal mol$^{-1}$ \AA$^{-1}$ for VMC and LRDMC in ethanol, respectively (0.82 kcal mol$^{-1}$ \AA$^{-1}$ when VMC is compared with LRDMC), and 0.33 and 1.13 kcal mol$^{-1}$ \AA$^{-1}$ in malonaldehyde, repectively (1.20 kcal mol$^{-1}$ \AA$^{-1}$ for VMC vs LRDMC). Unless the measured MAE clearly exceeds this floor, we cannot conclude that there is a genuine difference between the two methods.

\vspace{2mm}
For ethanol, Slootman \emph{et al.}~\cite{slootman2024accurate} reported a benchmark result for 200 configurations. Their calculation yielded an MAE of 3.055 kcal mol$^{-1}$ \AA$^{-1}$ for VMC forces, with respect to the CCSD(T) values taken as reference. Our MAE on VMC forces for ethanol, (2.5 kcal mol$^{-1}$) in MAE, is lower than theirs, likely because of the more flexible Jastrow factor used in the present study. We note, however, that our ethanol configurations are not identical to theirs and MAEs have the uncertainty as noted above, so this is not a strict one-to-one comparison. Slootman \emph{et al.} also reported an MAE of 2.0 kcal mol$^{-1}$ \AA$^{-1}$ for DMC forces evaluated with the Reynolds approximation. In our calculations, we obtained a similar MAE of 2.1 kcal mol$^{-1}$ \AA$^{-1}$ for ethanol, which is close to the value reported by Slootman \emph{et al.}

\vspace{2mm}
For malonaldehyde, one of the authors of the present study previously reported a large discrepancy between VMC and CCSD(T) forces in Ref.~\onlinecite{2026_Nakano_JCP_CP2K}, amounting to 7.2 kcal mol$^{-1}$ \AA$^{-1}$ in RMSE (4.8 kcal mol$^{-1}$ \AA$^{-1}$ in MAE). We reproduce the same inconsistency in the present study, obtaining an MAE of 5.8 kcal mol$^{-1}$ \AA$^{-1}$ for the MCMC forces. The obtained MAE differs from the value reported in Ref.~\onlinecite{2026_Nakano_JCP_CP2K} by 1~kcal mol$^{-1}$ \AA$^{-1}$. We attribute the difference mainly to variations in the trial wave functions employed in the two calculations. In this study, jQMC uses the exponential form for the two-body Jastrow factor, whereas the previous work uses the Pad\'e form. In addition, the trial wave functions were generated using different XC functionals, namely B3LYP and LDA-PZ in the present and previous works, respectively. Finally, the GTOs used in the three-body Jastrow factor also differ between the two calculations, namely, Cartesian and Spherical GTOs, respectively (i.e., the number of $d$ orbitals are different).

\vspace{2mm}
Interestingly, even LRDMC shows a large discrepancy relative to CCSD(T), with an MAE of 5.2 kcal mol$^{-1}$ \AA$^{-1}$. Since the trial wave function, including its nodal surface, was optimized at the VMC level, the mean-field nodal surface is unlikely to be the primary origin of the discrepancy. The reason for the significant deviation in malonaldehyde is not yet fully understood and will require further investigation from both the DMC and CCSD(T) perspectives.

\section{Conclusions and perspectives}
\label{sec:conclusions_and_perspectives}
In this work, we presented the implementation details and benchmark results of jQMC, a Python-based \emph{ab initio} quantum Monte Carlo (QMC) package built on top of JAX. jQMC is designed with modern heterogeneous high-performance computing (HPC) architectures in mind, particularly those combining CPUs and GPUs, which are rapidly becoming mainstream. The code is parallelized at the top level of QMC algorithms, i.e., based on a multi-walker strategy. By packing as many walkers as GPU memory permits, jQMC fully leverages GPU computational power, leading to significant acceleration of QMC simulations. Benchmark results confirm the effectiveness of this design.

\vspace{2mm}
jQMC remains under active development, and several important directions can be envisaged. From a physics perspective, the implementation of periodic boundary conditions is a major milestone for future work, which is in progress and will be most likely supported in the next release. From a performance point of view, improving cache and memory efficiencies is an urgent challenge. While future hardware advancements may alleviate this issue, software-level solutions will also be essential.

\section{Code and Data availability}
\label{sec:code_and_data_availability}
The code and data that support the findings of this study are available from GitHub repositories [\url{https://github.com/jqmc-project/jQMC}] and [\url{https://github.com/jqmc-project/jQMC-data}], respectively. All calculations in this work were performed using jQMC version corresponding to v0.2.2.

\section*{Acknowledgments}
K.N. is grateful for computational resources from the Numerical Materials Simulator at National Institute for Materials Science (NIMS), from Research Institute for Information Technology at Kyushu University under the category of General Projects on the supercomputer Genkai, from Information technology center at the University of Tokyo under the category of Regular Use Projects on the supercomputer Miyabi. 
The weak-scaling benchmark on Miyabi with more than 512 GPUs was conducted through "Large-scale HPC Challenge" Project, Joint Center for Advanced High Performance Computing~(JCAHPC).
M.C. and K.N. acknowledge EuroHPC for the computational grant EUHPC\_B35\_064 allocated on the Leonardo booster partition through the EuroHPC Benchmark Access Call.

K.N. acknowledges financial support from the MEXT Leading Initiative for Excellent Young Researchers (Grant No.~JPMXS0320220025) and from Japan Science and Technology Agency (JST), PRESTO (Grant No.~JPMJPR24J9). M.C. acknowledges financial support from the European High Performance Computing Joint Undertaking (JU) through the "EU-Japan Alliance in HPC" HANAMI project (Hpc AlliaNce for Applications and supercoMputing Innovation: the Europe - Japan collaboration). 

The authors acknowledge technical support of CINECA, especially grateful to Dr.~Tommaso Gorni~(CINECA) and Dr.~Laura Bellentani~(CINECA) for their assistance.
%
The authors gratefully acknowledge fruitful discussions with our collaborators, especially Prof. Giuseppe Carleo (EPFL) and Prof. Filippo Vicentini (Ecole Polytechnique), on the JAX implementation and its application to \emph{ab initio} QMC.
%
The authors thank Dr. Tan Naruhiko and Dr. Shinnosuke Furuya of NVIDIA for carefully reading Sec.~{\ref{sec:performance-analysis}} and providing valuable suggestions for improving the clarity and wording.
%
K.N. acknowledges Dr. Atsushi Togo (NIMS) for fruitful discussions on the high-level design and software architecture of the package.

Claude-code (Anthropic) and Copilot (Microsoft) were used to support code development, specifically for performance tuning and memory-efficiency optimization on GPUs. The AI-generated parts were critically tested and reviewed by K.N.

\bibliographystyle{apsrev4-1}
\bibliography{./references.bib}

\end{document}